# Nominals, Inverses, Counting, and Conjunctive Queries or: Why Infinity is your Friend!

**Sebastian Rudolph** RUDOLPH@KIT.EDU
*AIFB, Karlsruhe Institute of Technology, DE*

**Birte Glimm** BIRTE.GLIMM@COMLAB.OX.AC.UK
*Oxford University Computing Laboratory, UK*

## Abstract

Description Logics are knowledge representation formalisms that provide, for example, the logical underpinning of the W3C OWL standards. Conjunctive queries, the standard query language in databases, have recently gained significant attention as an expressive formalism for querying Description Logic knowledge bases. Several different techniques for deciding conjunctive query entailment are available for a wide range of DLs. Nevertheless, the combination of nominals, inverse roles, and number restrictions in OWL 1 and OWL 2 DL causes unsolvable problems for the techniques hitherto available. We tackle this problem and present a decidability result for entailment of unions of conjunctive queries in the DL $\mathcal{ALCHOIQ}b$ that contains all three problematic constructors simultaneously. Provided that queries contain only simple roles, our result also shows decidability of entailment of (unions of) conjunctive queries in the logic that underpins OWL 1 DL and we believe that the presented results will pave the way for further progress towards conjunctive query entailment decision procedures for the Description Logics underlying the OWL standards.

## 1. Introduction

We present a decidability result for entailment of unions of conjunctive queries in the very expressive Description Logic $\mathcal{ALCHOIQ}b$. The article is an extended version of the conference paper *Status $\mathcal{QIO}$: Conjunctive Query Entailment is Decidable*, Proceedings of the 12th International Conference on the Principles of Knowledge Representation and Reasoning (KR 2010), May 09–13, 2010 (Glimm & Rudolph, 2010).

Description Logics (DLs) are a family of logic based knowledge representation formalisms (Baader, Calvanese, McGuinness, Nardi, & Patel-Schneider, 2003). Most DLs correspond to the function-free two variable fragment of First-Order Logic (FOL) often extended with counting quantifiers (e.g., $\forall x \exists_{\leq n} y(R(x, y))$) and DLs are also closely related to the (2-variable) guarded fragment since DL formulae naturally result in guarded formulae when translated into FOL. In line with the restriction to 2 variables, DL formulae contain only unary and binary predicates, which are called concepts and roles in DLs. The constructors for building complex expressions are usually chosen such that the key inference problems, such as concept satisfiability, are decidable. A DL knowledge base (KB) consists of a TBox, which contains intensional knowledge such as concept definitions and general background knowledge (essentially a FOL theory), and an ABox, which contains extensional knowledge and is used to describe individuals (a set of ground facts). Using a database metaphor, the TBox corresponds to the schema, and the ABox corresponds to the data. In contrast to databases, however, DL knowledge bases, as FOL in general, adopt an open world semantics, i.e., they represent information about the domain in an incomplete way.

Standard DL reasoning services include testing concepts for satisfiability and retrieving certain instances of a given concept. The latter retrieves, for a knowledge base consisting of an ABox $\mathcal{A}$ and a TBox $\mathcal{T}$, all (ABox) individuals that are instances of the given (possibly complex) concept expression $C$, i.e., all those individuals $a$ such that $\mathcal{T}$ and $\mathcal{A}$ entail that $a$ is an instance of $C$. The underlying reasoning problems are well-understood, and the computational complexity of the standard reasoning tasks given a knowledge base as input range from PTime-complete for DLs with limited expresivity such as DL-Lite (Calvanese, De Giacomo, Lembo, Lenzerini, & Rosati, 2005), $\mathcal{EL}$ (Baader, 2003), and ELP (Krötzsch, Rudolph, & Hitzler, 2008) to 2-NExpTime-complete for very expressive DLs such as $\mathcal{SROIQ}$ (Kazakov, 2008).

Despite the high worst case complexity of the standard reasoning problems for very expressive DLs such as $\mathcal{SROIQ}$, there are highly optimized implementations available, e.g., FaCT++ (Tsarkov & Horrocks, 2006), Pellet (Sirin, Parsia, Cuenca Grau, Kalyanpur, & Katz, 2007), and HermiT (Motik, Shearer, & Horrocks, 2009). These systems are used in a wide range of applications, e.g., biology (Sidhu, Dillon, Chang, & Sidhu, 2005), bio informatics (Wolstencroft, Brass, Horrocks, Lord, Sattler, Turi, & Stevens, 2005), medicine (Golbreich, Zhang, & Bodenreider, 2006), information integration (Calvanese, De Giacomo, Lenzerini, Nardi, & Rosati, 1998b), geography (Goodwin, 2005), geology (Jet Propulsion Laboratory, 2006), defense (Lacy, Aviles, Fraser, Gerber, Mulvehill, & Gaskill, 2005), and configuration (McGuinness & Wright, 1998). Most prominently, DLs are known for their use as a logical underpinning of ontology languages, e.g., OIL, DAML+OIL, the W3C standard OWL 1 (Bechhofer, van Harmelen, Hendler, Horrocks, McGuinness, Patel-Schneider, & Stein, 2004), and its successor OWL 2 (W3C OWL Working Group, 2009). There are three species of OWL 1: OWL Lite, OWL DL, and OWL Full. OWL 2 extends OWL 1 and adds three further sublanguages (called OWL 2 profiles): OWL EL, OWL QL, and OWL RL. OWL Lite corresponds to the DL $\mathcal{SHIF}$ in which the standard reasoning tasks are ExpTime-complete, OWL 1 DL corresponds to the DL $\mathcal{SHOIN}$, in which the standard reasoning tasks are NExpTime-complete, and OWL 2 DL extends this to the DL $\mathcal{SROIQ}$. For OWL Full the standard reasoning tasks are no longer decidable. The new QL, EL, and RL profiles are more restrictive than OWL DL and each of the profiles trades off different aspects of OWL's expressive power in return for different computational and/or implementational benefits. OWL EL corresponds to the DL $\mathcal{EL}$ ++ (Baader, Brandt, & Lutz, 2005) and the basic reasoning problems can be performed in time that is polynomial with respect to the size of the input knowledge base. OWL 2 QL is based on the DL-Lite family of Description Logics, where the data complexity of conjunctive query entailment is in $AC^0$. Thus, conjunctive query answering can be implemented using standard relational database technology. OWL 2 RL enables the implementation of polynomial time reasoning algorithms using rule-extended database technologies.

In data-intensive applications, querying KBs plays a central role. Instance retrieval is, in some aspects, a rather weak form of querying: although possibly complex concept expressions are used as queries, we can only query for tree-like relational structures, as a DL concept cannot express arbitrary cyclic structures. This property is known as the tree model property and is considered an important reason for the decidability of most Modal and Description Logics (Grädel, 2001; Vardi, 1997) and we also heavily exploit a variant of this property to establish our decidability result. Conjunctive queries (CQs) and unions of conjunctive queries (UCQs) are well known in the database community and constitute an expressive query language with capabilities that go well beyond standard instance retrieval. In FOL terms, CQs and UCQs are formulae from the positive existential fragment. Free variables in a query (not bound by an existential quantifier) are also called answer variables or distinguished variables, whereas existentially quantified variables are called non-distinguished.

If the query contains no distinguished variables, the query answer is just *true* or *false* and the query is called a Boolean query. Given a knowledge base $\mathcal{K}$ and a Boolean UCQ $q$, the query entailment problem is deciding whether $q$ is *true* or *false* w.r.t. $\mathcal{K}$, i.e., we have to decide whether each model of $\mathcal{K}$ provides for a suitable assignment for the variables in $q$. For a query with distinguished variables, the answers to the query are those tuples of individual names (constants) for which the knowledge base entails the query that is obtained by replacing the free variables with the individual names in the answer tuple. These answers are also called *certain answers*. The problem of finding all answer tuples is known as query answering. We present a decidability result for query entailment, which is a decision problem, but this is no restriction since query answering can easily be reduced to query entailment as we illustrate in more detail in Section 3.

### 1.1 Related Work

Conjunctive queries have been first mentioned in the context of Description Logics (DLs) by Levy and Rousset (1996). The first account of conjunctive queries as main topic is given by Calvanese, De Giacomo, and Lenzerini (1998a). In particular in recent years, the problem of decidability of conjunctive query entailment and the complexity of the problem in different logics has gained significant attention. For the DLs $\mathcal{SHIQ}$ and $\mathcal{SHOQ}$ decidability and 2-ExpTime-completeness of the problem is known (Glimm, Horrocks, Lutz, & Sattler, 2008a; Glimm, Horrocks, & Sattler, 2008b; Lutz, 2008; Eiter, Lutz, Ortiz, & Simkus, 2009). Conjunctive query entailment is already 2-ExpTime-hard in the relatively weak DL $\mathcal{ALCI}$ (Lutz, 2008), which was initially attributed to inverse roles. Recently, it was shown, however, that also transitive roles together with role hierarchies as in the DL $\mathcal{SH}$ make conjunctive query entailment 2-ExpTime-hard (Eiter et al., 2009). The techniques by Glimm et al. for $\mathcal{SHIQ}$ and $\mathcal{SHOQ}$ (Glimm et al., 2008a, 2008b) reduce query entailment to the standard reasoning task of knowledge base satisfiability checking in the DL extended with role conjunctions. An alternative technique is the so-called knots technique (Ortiz, Simkus, & Eiter, 2008b), which is an instance of the mosaic technique originating in Modal Logic. This technique also gives worst-case optimal algorithms for $\mathcal{SHIQ}$ and several of its sub-logics. Further, there are automata-based decision procedures for positive existential path queries (Calvanese, Eiter, & Ortiz, 2007, 2009). Positive existential path queries generalize unions of conjunctive queries and, therefore, decision procedures for this kind of query also provides decision procedures for unions of conjunctive queries. In particular the most recent extension (Calvanese et al., 2009) is very close to a conjunctive query entailment decision procedure for OWL 2, which corresponds to the DL $\mathcal{SROIQ}$, because it covers $\mathcal{SRIQ}$, $\mathcal{SROQ}$, and $\mathcal{SROI}$. The use of the three problematic constructors for nominals, inverses, and number restrictions is, however, not covered.

Regarding data complexity, i.e., the complexity with respect to the ABox (the data) only, CQ entailment is usually coNP-complete for expressive logics. For example, for DLs from $\mathcal{ALE}$ up to $\mathcal{SHIQ}$ this is the case (Glimm et al., 2008a) and this holds also for CQ entailment in the two variable guarded fragment with counting (Pratt-Hartmann, 2009). The latter work is quite closely related since many Description Logics can be translated into the two variable guarded fragment with counting, i.e., the results of Pratt-Hartmann also hold for $\mathcal{SHIQ}$ with only simple roles (roles that are not transitive and have no transitive subrole) in the query. Given the same restriction on the query, also $\mathcal{SHOQ}$ and $\mathcal{SHOI}$ were shown to have coNP-complete data complexity w.r.t. conjunctive query entailment (Ortiz, Calvanese, & Eiter, 2008a).

Query entailment and answering have also been studied in the context of databases with incomplete information (Rosati, 2006b; van der Meyden, 1998; Grahne, 1991). In this setting, DLs can be used as schema languages, but the expressivity of the considered DLs is usually much lower than the expressivity of the DL $\mathcal{ALCHOIQ}b$ that we consider here and reasoning in them is usually tractable. For example, the constructors provided by logics of the DL-Lite family (Calvanese, De Giacomo, Lembo, Lenzerini, & Rosati, 2007) are chosen such that the standard reasoning tasks are in PTime regarding combined complexity and query entailment is in $\mathrm{AC}^0$ with respect to data complexity. Thus, TBox reasoning can be done independently of the ABox and the ABox can be stored and accessed using a standard database SQL engine. Another tractable DL is $\mathcal{EL}$ (Baader, 2003). Conjunctive query entailment in $\mathcal{EL}$ is, however, not tractable as the complexity increases to coNP-complete (Rosati, 2007b). Moreover for $\mathcal{EL}^{++}$ (Baader et al., 2005), a still tractable extension of $\mathcal{EL}$, query entailment is even undecidable (Krötzsch, Rudolph, & Hitzler, 2007). This is mainly because in $\mathcal{EL}^{++}$, one can use unrestricted role compositions. This allows for encoding context-free languages, and conjunctive queries can then be used to check the intersection of such languages, which is known to be an undecidable problem. Since the logics used in databases with incomplete information are considerable less expressive than $\mathcal{ALCHOIQ}b$, the techniques developed in that area do not transfer to our setting.

Given that query entailment is a (computationally) harder task than, for example, knowledge base satisfiability, it is not very surprising that decidability of the latter task does not necessarily transfer to the problem of CQ entailment. Most of the undecidability results can be transferred from FOL since many DLs can directly be translated into an equivalent FOL theory. For example, it is known that conjunctive query entailment is undecidable in the two variable fragment of First-Order Logic $\mathcal{L}_2$ (Rosati, 2007a), and Rosati identifies a relatively small set of constructors that cause the undecidability (most notably role negation axioms, i.e., axioms of the form $\forall x, y\,(\neg R(x, y) \to P(x, y))$ for $R, P$ binary predicates). Pratt-Hartmann (2009) recently established decidability for CQ entailment in the two variable guarded fragment with counting ($\mathcal{GC}_2$). It is worth noting that Pratt-Hartmann assumes that the background theory (that is the knowledge base in our case) is constant free and formulae of the form $\exists_{=1}x(P(x))$, which can be used to simulate constants/nominals, are not considered guarded. His result covers, therefore, only the DL $\mathcal{ALCHIQ}b$ and is not applicable to the case, when the input knowledge base (the background theory) contains nominals (individual constants).

Most of the implemented DL reasoners, e.g., KAON2,[1] Pellet, and RacerPro,[2] provide an interface for conjunctive query answering, although KAON2 and RacerPro consider only named individuals in the ABox for the assignments of variables. Under that restriction queries do no longer have the standard FOL semantics and decidability is obviously not an issue since conjunctive query answering with this restriction can be reduced to standard instance retrieval by replacing the variables with individual names from the ABox and then testing entailment of each conjunct separately. Pellet goes beyond that and also provides an interface for conjunctive queries with FOL semantics under the restriction that the queries have a kind of tree shape. Under this restriction decidability is known since CQs can then be expressed as normal concepts (possibly by adding role conjunctions).

### 1.2 Contributions and Overview

Given all these results, which show a great interest in the problem of conjunctive query entailment over expressive DLs, it is very interesting that for the DLs $\mathcal{SHIF}$, $\mathcal{SHOIN}$, and $\mathcal{SROIQ}$ that underpin the widely adopted standards OWL Lite, OWL 1 DL, and OWL 2 DL, respectively, decidability of conjunctive query entailment has only been established for OWL Lite. The main obstacle in devising a decision procedure is the combination of inverse roles ($\mathcal{I}$), nominals ($\mathcal{O}$), and number restrictions/counting quantifiers ($\mathcal{F}$ stands for functionality, $\mathcal{N}$ for unqualified number restrictions, and $\mathcal{Q}$ for qualified number restrictions). The complications arising from the combination of these constructors caused also a major hurdle in the development of implementable algorithms for knowledge base satisfiability in $\mathcal{SHOIN}$ and extensions thereof, but Horrocks and Sattler (2005) devised a tableau-based decision procedure that has since been extended to $\mathcal{SROIQ}$. Meanwhile also alternative approaches such as resolution (Kazakov & Motik, 2008), and hypertableau-based procedures (Motik et al., 2009) are available and implemented.

The key obstacle in establishing a decision procedure is the existence of potentially infinitely many *new nominals*, i.e., elements that are uniquely identifiable in any model of a KB. For an example, consider the KB $\mathcal{K}$ given in Fig. 1. A concept of the form $\{o\}$ has to be interpreted as a singleton set, containing only the interpretation of the constant $o$. For simplicity, we assume for now that a constant is always interpreted as itself, e.g., the interpretation of $o$ is $o$. An axiom of the form $\{o_1\} \sqsubseteq \exists f.\exists s.\exists f^-.\{o_2\}$ can then be understood as follows: For the constant $o_1$, there must be two elements, say $d_1$ and $d_2$, such that $f(o_1, d_1)$, $s(d_1, d_2)$, and $f(o_2, d_2)$ holds. Note that $o_2$ occurs as the first element in $f(o_2, d_2)$ since an inverse role ($f^-$) is used. Thus, an interpretation for the KB must contain the three elements $o_1$, $o_2$, and $o_3$, which must be interconnected in the following way: paths of the shape $\xrightarrow{f} \cdot \xrightarrow{s} \cdot \xleftarrow{f}$ have to lead from $o_1$ to $o_2$ as well as from $o_2$ to $o_3$ and from $o_3$ to $o_1$. Moreover, the role $f$ is defined to be functional, meaning that every element can have at most one $f$-successor. This also applies to all individuals $o_i$, which forces the existence of an $s$-cycle. Observe that a cyclic Boolean query such as $\{s(x, y), s(y, z), s(z, x)\}$ that checks for the existence of such a cycle cannot be answered by applying standard techniques such as replacing variables with individual names ($o_i$) or rewriting the query into an equivalent

1. http://kaon2.semanticweb.org
2. http://www.racer-systems.com

$$\{o_1\} \sqsubseteq \exists f.\exists s.\exists f^-.\{o_2\}$$
$$\{o_2\} \sqsubseteq \exists f.\exists s.\exists f^-.\{o_3\}$$
$$\{o_3\} \sqsubseteq \exists f.\exists s.\exists f^-.\{o_1\}$$
$$\mathsf{func}(f)$$

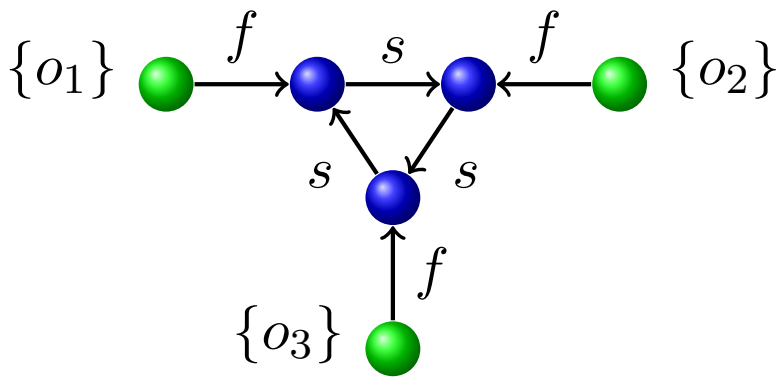

Figure 1: Example knowledge base $\mathcal{K}$ and a representation for a model, where the three elements in the $s$-cycle are so-called *new nominals*.

tree-shaped query. The elements in the cycle behave as if they were nominals, but we do not have names for them.

We tackle the problem of conjunctive query entailment in a very expressive DL that contains all the three problematic constructors simultaneously and prove decidability of (unions of) conjunctive queries. The most challenging part is to establish finite representability of countermodels in case the query given as input is not entailed by the knowledge base. Our results also hold for $\mathcal{SHOIQ}$ knowledge bases, i.e., with some roles declared as transitive, provided that the queries contain only simple roles (roles that are neither transitive nor have a transitive subrole). This is essentially the same restriction that is placed on roles that can occur in number restrictions since otherwise the standard reasoning tasks become undecidable. Under this restriction, we can use standard techniques for eliminating transitivity (Kazakov & Motik, 2008). Hence, we also show decidability of conjunctive query entailment in OWL DL, for queries with only simple roles.

We believe that our work is also valuable for understanding, in general, the structure of models in DLs that contain nominals, inverse roles, and number restrictions. Furthermore, we devise non-trivial extensions of standard techniques such as unraveling, which we believe will prove useful when working with such expressive DLs.

The paper is organized as follows: in Section 2, we give a bird's-eye view of the techniques and ideas used to establish decidability. In Section 3, we give the necessary definitions and introduce standard notations. In Sections 4, 5, and 6 we present the main results that we then use in Section 7 to show how models that do not satisfy the query can be finitely represented before we conclude in Section 8.

## 2. The Big Picture

Before going into the technical details, we will describe our overall line of argumentation establishing decidability of conjunctive query entailment in $\mathcal{ALCHOIQ}b$.

### 2.1 Decidability via Finitely Representable Countermodels

Let $\mathcal{K}$ be an $\mathcal{ALCHOIQ}b$ knowledge base and let $q$ be the conjunctive query in question, i.e., we aim to determine whether

$$\mathcal{K} \models q.$$

Clearly, as $\mathcal{ALCHOIQ}b$ is a fragment of first-order predicate logic with equality, $\mathcal{K}$ can be translated into a FOL sentence $FOL(\mathcal{K})$. Likewise we find a FOL sentence $FOL(q)$ for $q$ being just an existentially quantified formula. Hence, checking the above entailment is equivalent to determining whether the first-order theory $FOL(\mathcal{K})$ entails $FOL(q)$. As a result of the completeness theorem for FOL (Gödel, 1929), the consequences of a finite FOL theory are recursively enumerable, which provides us with a procedure that terminates if $\mathcal{K} \models q$. Hence, we can establish decidability by providing another algorithm that terminates iff the entailment above does not hold – i.e., if there is a so-called *countermodel* being a model $\mathcal{I}$ of $\mathcal{K}$ for which $\mathcal{I} \not\models q$.

We will provide such an algorithm by showing that, whenever such a countermodel $\mathcal{I}$ exists at all, there is also a countermodel $\overline{\mathcal{I}}$ that is finitely representable. More precisely, $\overline{\mathcal{I}}$ can be encoded into a word $Rep(\mathcal{I})$ of finite length over a finite alphabet, whereby the encoding *Rep* has the property that for every such finite word it can be effectively checked whether it represents a countermodel for a given knowledge base and query.

As a consequence thereof, we can create the desired algorithm that enumerates all words, checks each for being a countermodel, and terminates as soon as it has found one.

### 2.2 Finite Representability by Bounding Nominals and Blocking

We now outline how we are going to show that there is always a finitely representable countermodel, if there is one at all. We do this by taking an arbitrary countermodel and cautiously transforming it into a countermodel that is finitely representable. Cautiously means that we have to make sure that the transformation does preserve the two properties of 1) being a model of the underlying knowledge base $\mathcal{K}$ and 2) not entailing the considered query $q$.

The result of the overall transformation is going to be a *regular model*, i.e., a structure where substructures are being in a certain sense periodically repeated. It is common practice in DL theory to construct this kind of models from arbitrary ones by blocking techniques, whereby certain element configurations occurring twice in the original model are detected and the new model is generated by infinitely stringing together the same finite substructure that is delimited by those two configurations.

In the case we consider, this technique cannot be applied directly to the original countermodel. This is due to an intricate interplay of nominals, inverse roles and cardinality constraints by which an arbitrary – even an infinite – number of domain elements can be forced to "behave" like nominals; this is why those elements are usually referred to as *new nominals* in a DL setting. In FOL, nominals are often called kings and the new nominals are called the court. In our case, the presence of infinitely many new nominals in the model may prevent the existence of repeated configurations needed for blocking.

We overcome this difficulty by first applying a transformation by means of which the original countermodel is converted into a countermodel with only finitely many new nominals. This guarantees that the subsequent blocking-based transformation is applicable and will yield the desired regular (and thus finitely representable) model.

### 2.3 Bounding Nominals by Transformations of Forest Quasi-Models

For our argumentation, we introduce the notion of *forest quasi-models*. These are structures not satisfying the originally considered knowledge base but a weakened form of it. In return to this concession, they exhibit a proper forest structure that is easier to handle and manipulate.

We employ two techniques to turn "proper" models into forest quasi-models and vice versa: a model can be *unraveled* yielding a forest quasi-model. A forest quasi-model can be *collapsed* to obtain a "proper" model. Both techniques preserve certain structural properties.

Our strategy to construct a countermodel with finitely many nominals consists of the following three steps:

- Take an arbitrary countermodel and unravel it.
- Transform the obtained forest quasi-model by substituting critical parts by well-behaved ones,
- Collapse the obtained structure into a (proper) model.

The mentioned "critical parts" are those giving rise to new nominals. They have to be – at least largely – avoided (we do not care about a finite set of those critical parts remaining).

The central question is: where do these mysterious well-behaved substitutes come from? Fortunately, the plethora of critical parts brings about its own remedy. We can use infinite sets of critical parts to construct well-behaved ones in an infinite approximation process (this is why infinity is your friend). We thereby obtain parts which have not been present in our structure before, but are well compatible with it and can hence be used for its reorganization.

After having informally introduced our main line of argumentation, we now move on to the technical details.

## 3. Preliminaries

We first define the syntax and semantics of roles, and then go on to $\mathcal{SHOIQ}b$-concepts, individuals, and knowledge bases. We do not actually use the full expressivity of $\mathcal{SHOIQ}b$, but it is a convenient umbrella for all DLs we are working with and we can define less expressive DLs of interest as restrictions of $\mathcal{SHOIQ}b$.

**Definition 1** (Syntax of $\mathcal{SHOIQ}b$)**.** Let $N_C$, $N_R$, and $N_I$ be countable, infinite, and pairwise disjoint sets of *concept names*, *role names*, and *individual names*, respectively. We call $\mathcal{S} = (N_C, N_R, N_I)$ a *signature*. The set $\mathsf{rol}(\mathcal{S})$ of *$\mathcal{SHOIQ}b$-roles* over $\mathcal{S}$ (or roles for short) is $N_R \cup \{r^- \mid r \in N_R\}$, where roles of the form $r^-$ are called *inverse roles*. A *role inclusion axiom* is of the form $r \sqsubseteq s$ with $r, s$ roles. A *transitivity axiom* is of the form $\mathsf{trans}(r)$ for $r$ a role. A *role hierarchy* $\mathcal{H}$ is a finite set of role inclusion and transitivity axioms.

For a role hierarchy $\mathcal{H}$, we define the function $\mathsf{inv}$ over roles as $\mathsf{inv}(r) := r^-$ if $r \in N_R$ and $\mathsf{inv}(r) := s$ if $r = s^-$ for a role name $s \in N_R$. Further, we define $\sqsubseteq_\mathcal{H}$ as the smallest transitive reflexive relation on roles such that $r \sqsubseteq s \in \mathcal{H}$ implies $r \sqsubseteq_\mathcal{H} s$ and $\mathsf{inv}(r) \sqsubseteq_\mathcal{H} \mathsf{inv}(s)$. We write $r \equiv_\mathcal{H} s$ if $r \sqsubseteq_\mathcal{H} s$ and $s \sqsubseteq_\mathcal{H} r$. A role $r$ is *transitive w.r.t. $\mathcal{H}$* (notation $r^+ \sqsubseteq_\mathcal{H} r$) if a role $s$ exists such that $r \sqsubseteq_{\mathcal{H}} s$, $s \sqsubseteq_{\mathcal{H}} r$, and $\mathsf{trans}(s) \in \mathcal{H}$ or $\mathsf{trans}(\mathsf{inv}(s)) \in \mathcal{H}$. A role $s$ is called *simple w.r.t.* $\mathcal{H}$ if there is no role $r$ such that $r$ is transitive w.r.t. $\mathcal{H}$ and $r \sqsubseteq_{\mathcal{H}} s$.

For $r \in \mathsf{rol}(\mathcal{S})$ a simple role, a *Boolean role expressions* $U$ is defined as follows:

$$U ::= r \mid \neg U \mid U \sqcap U \mid U \sqcup U.$$

We use $\vdash$ to denote standard Boolean entailment between a set of roles $\mathcal{R} \subseteq \mathsf{rol}(\mathcal{S})$ and role expressions. Let $r \in \mathsf{rol}(\mathcal{S})$, and $U$ a Boolean role expression over $\mathcal{R}$. We inductively define:

- $\mathcal{R} \vdash r$ if $r \in \mathcal{R}$, and $\mathcal{R} \not\vdash r$ otherwise,
- $\mathcal{R} \vdash \neg U$ if $\mathcal{R} \not\vdash U$, and $\mathcal{R} \not\vdash \neg U$ otherwise,
- $\mathcal{R} \vdash U \sqcap V$ if $\mathcal{R} \vdash U$ and $\mathcal{R} \vdash V$, and $\mathcal{R} \not\vdash U \sqcap V$ otherwise,
- $\mathcal{R} \vdash U \sqcup V$ if $\mathcal{R} \vdash U$ or $\mathcal{R} \vdash V$, and $\mathcal{R} \not\vdash U \sqcup V$ otherwise.

A Boolean role expression $U$ is *safe* if $\emptyset \not\vdash U$.

Given a signature $\mathcal{S} = (N_C,\ N_R,\ N_I)$, the set of $\mathcal{SHOIQ}b$*-concepts* (or concepts for short) over $\mathcal{S}$ is the smallest set built inductively over symbols from $\mathcal{S}$ using the following grammar, where $o \in N_I, A \in N_C, n \in \mathbb{N}_0$, $s$ is a simple role, and $U$ is a role or a safe Boolean role expression:

$$\begin{aligned} C ::= \ & \top \mid \bot \mid \{o\} \mid A \mid \neg C \mid C_1 \sqcap C_2 \mid C_1 \sqcup C_2 \mid \\ & \forall U.C \mid \exists U.C \mid \leqslant n\ s.C \mid \geqslant n\ s.C. \end{aligned}$$

△

Alternatively, safeness can be characterized as follows: a Boolean role expression U is safe if, after transforming it into disjunctive normal form, each disjunct contains at least one non-negated role. Intuitively, this implies that a safe role expression can never relate individuals that are not in a direct role relation with each other.

**Definition 2** (Semantics of $\mathcal{SHOIQ}b$-concepts)**.** An *interpretation* $\mathcal{I} = (\Delta^{\mathcal{I}}, \cdot^{\mathcal{I}})$ consists of a non-empty set $\Delta^{\mathcal{I}}$, the *domain* of $\mathcal{I}$, and a function $\cdot^{\mathcal{I}}$, which maps every concept name $A \in N_C$ to a subset $A^{\mathcal{I}} \subseteq \Delta^{\mathcal{I}}$, every role name $r \in N_R$ to a binary relation $r^{\mathcal{I}} \subseteq \Delta^{\mathcal{I}} \times \Delta^{\mathcal{I}}$, and every individual name $a \in N_I$ to an element $a^{\mathcal{I}} \in \Delta^{\mathcal{I}}$. For each role name $r \in N_R$, the interpretation of its inverse role $(r^-)^{\mathcal{I}}$ consists of all pairs $\langle \delta, \delta' \rangle \in \Delta^{\mathcal{I}} \times \Delta^{\mathcal{I}}$ for which $\langle \delta', \delta \rangle \in r^{\mathcal{I}}$.

The semantics of $\mathcal{SHOIQ}b$-concepts over a signature $\mathcal{S}$ is defined as follows:

$$\begin{array}{rlrlrl} (\neg r)^{\mathcal{I}} &= \Delta^{\mathcal{I}} \times \Delta^{\mathcal{I}} \setminus r^{\mathcal{I}} & (r_1 \sqcap r_2)^{\mathcal{I}} &= r_1^{\mathcal{I}} \cap r_2^{\mathcal{I}} & (r_1 \sqcup r_2)^{\mathcal{I}} &= r_1^{\mathcal{I}} \cup r_2^{\mathcal{I}} \\ \top^{\mathcal{I}} &= \Delta^{\mathcal{I}} & \bot^{\mathcal{I}} &= \emptyset & (\{o\})^{\mathcal{I}} &= \{o^{\mathcal{I}}\} \\ (\neg C)^{\mathcal{I}} &= \Delta^{\mathcal{I}} \setminus C^{\mathcal{I}} & (C \sqcap D)^{\mathcal{I}} &= C^{\mathcal{I}} \cap D^{\mathcal{I}} & (C \sqcup D)^{\mathcal{I}} &= C^{\mathcal{I}} \cup D^{\mathcal{I}} \end{array}$$

$$\begin{aligned} (\forall U.C)^{\mathcal{I}} &= \{\delta \in \Delta^{\mathcal{I}} \mid \text{if } \langle \delta, \delta' \rangle \in U^{\mathcal{I}}, \text{ then } \delta' \in C^{\mathcal{I}}\} \\ (\exists U.C)^{\mathcal{I}} &= \{\delta \in \Delta^{\mathcal{I}} \mid \text{there is a } \langle \delta, \delta' \rangle \in U^{\mathcal{I}} \text{ with } \delta' \in C^{\mathcal{I}}\} \\ (\leqslant n\ s.C)^{\mathcal{I}} &= \{\delta \in \Delta^{\mathcal{I}} \mid \sharp(s^{\mathcal{I}}(\delta, C)) \leq n\} \\ (\geqslant n\ s.C)^{\mathcal{I}} &= \{\delta \in \Delta^{\mathcal{I}} \mid \sharp(s^{\mathcal{I}}(\delta, C)) \geq n\} \end{aligned}$$

where $\sharp(M)$ denotes the cardinality of the set $M$ and $s^{\mathcal{I}}(\delta, C)$ is defined as

$$\{\delta' \in \Delta^{\mathcal{I}} \mid \langle \delta, \delta' \rangle \in s^{\mathcal{I}} \text{ and } \delta' \in C^{\mathcal{I}}\}.$$

A concept $C$ is in *negation normal form (NNF)* if negation occurs only in front of concept names and we use $\mathsf{nnf}(C)$ to denote the negation normal form of a concept $C$. △

Any concept can be transformed in linear time into an equivalent one in NNF by pushing negation inwards, making use of de Morgan's laws and the duality between existential and universal restrictions, and between at-most and at-least number restrictions of the form $\leqslant n\ r.C$ and $\geqslant n\ r.C$ respectively (Horrocks, Sattler, & Tobies, 2000).

**Definition 3** (Syntax and Semantics of Axioms and Knowledge Bases)**.** A *functionality restriction* is an expression $\mathsf{func}(f)$ for $f$ a role. For $C, D$ concepts, a *general concept inclusion* (GCI) is an expression $C \sqsubseteq D$. We introduce $C \equiv D$ as an abbreviation for $C \sqsubseteq D$ and $D \sqsubseteq C$. A finite set of GCIs and functionality restrictions is called a *TBox*. An (ABox) *assertion* is an expression of the form $C(a)$, $r(a,b)$, $\neg r(a,b)$, $a \doteq b$, or $a \not\doteq b$, where $C$ is a concept, $r$ is a role, and $a, b \in N_I$ are individual names. An *ABox* is a finite set of assertions. A *knowledge base* $\mathcal{K}$ is a triple $(\mathcal{T}, \mathcal{H}, \mathcal{A})$ with $\mathcal{T}$ a TBox, $\mathcal{H}$ a role hierarchy, and $\mathcal{A}$ an ABox.

We use $\mathsf{con}(\mathcal{K})$, $\mathsf{rol}(\mathcal{K})$, and $\mathsf{nom}(\mathcal{K})$ to denote, respectively, the set of concept names, roles (including inverses), and individual names occurring in $\mathcal{K}$. The *closure* $\mathsf{cl}(\mathcal{K})$ *of* $\mathcal{K}$ is the smallest set containing $\mathsf{nnf}(\neg C \sqcup D)$ if $C \sqsubseteq D \in \mathcal{T}$; $D$ if $D$ is a sub-concept of $C$ and $C \in \mathsf{cl}(\mathcal{K})$; and $\mathsf{nnf}(\neg C)$ if $C \in \mathsf{cl}(\mathcal{K})$. A role $f$ is *functional in* $\mathcal{K}$ if $\mathcal{K}$ contains the functionality axiom $\mathsf{func}(f)$ and it is *inverse functional in* $\mathcal{K}$ if $\mathcal{K}$ contains the functionality axiom $\mathsf{func}(\mathsf{inv}(f))$.

Let $\mathcal{I} = (\Delta^\mathcal{I}, \cdot^\mathcal{I})$ be an interpretation. Then $\mathcal{I}$ *satisfies* a role inclusion axiom $r \sqsubseteq s$ if $r^\mathcal{I} \subseteq s^\mathcal{I}$, $\mathcal{I}$ satisfies a transitivity axiom $\mathsf{trans}(r)$ if $r^\mathcal{I}$ is a transitive binary relation, and a role hierarchy $\mathcal{H}$ if it satisfies all role inclusion and transitivity axioms in $\mathcal{H}$. The interpretation $\mathcal{I}$ *satisfies* a functionality restriction $\mathsf{func}(f)$ if, for each $\delta \in \Delta^\mathcal{I}$, $\sharp(\{\delta' \mid \langle \delta, \delta' \rangle \in f^\mathcal{I}\}) \leq 1$; $\mathcal{I}$ *satisfies* a GCI $C \sqsubseteq D$ if $C^\mathcal{I} \subseteq D^\mathcal{I}$; and $\mathcal{I}$ *satisfies* a TBox $\mathcal{T}$ if it satisfies each functionality restriction and each GCI in $\mathcal{T}$. The interpretation $\mathcal{I}$ *satisfies* an assertion $C(a)$ if $a^\mathcal{I} \in C^\mathcal{I}$, $r(a,b)$ if $\langle a^\mathcal{I}, b^\mathcal{I} \rangle \in r^\mathcal{I}$, $\neg r(a,b)$ if $\langle a^\mathcal{I}, b^\mathcal{I} \rangle \notin r^\mathcal{I}$, $a \doteq b$ if $a^\mathcal{I} = b^\mathcal{I}$, and $a \not\doteq b$ if $a^\mathcal{I} \neq b^\mathcal{I}$; $\mathcal{I}$ *satisfies* an ABox if it satisfies each assertion in $\mathcal{A}$. We say that $\mathcal{I}$ *satisfies* $\mathcal{K}$ if $\mathcal{I}$ satisfies $\mathcal{T}$, $\mathcal{H}$, and $\mathcal{A}$. In this case, we say that $\mathcal{I}$ is a *model* of $\mathcal{K}$ and write $\mathcal{I} \models \mathcal{K}$. We say that $\mathcal{K}$ *is consistent* if $\mathcal{K}$ has a model. △

If the knowledge base $\mathcal{K}$ is clear from the context, we simply say that a role $f$ is (inverse) functional instead of saying $f$ is (inverse) functional in $\mathcal{K}$.

The names of DLs indicate which constructors are supported. The basic DL $\mathcal{ALC}$ supports Boolean concept constructors and GCIs, but no role hierarchies, functionality restrictions et cetera. If transitivity axioms are added, we use $\mathcal{S}$ instead of $\mathcal{ALC}$. Inverse roles are indicated by the letter $\mathcal{I}$, role inclusion axioms by $\mathcal{H}$, nominals, i.e., concepts of the form $\{o\}$ for $o \in N_I$, by $\mathcal{O}$, functionality restrictions by $\mathcal{F}$, qualified number restrictions, i.e., concepts of the form $\leqslant n\ s.C$ and $\geqslant n\ s.C$, by $\mathcal{Q}$, and safe Boolean role expressions by $b$. If number restrictions are limited to concepts of the form $\leqslant n\ s.\top$ and $\geqslant n\ s.\top$, we use the letter $\mathcal{N}$.

We mostly refer to a few particular DLs in this paper: the DL $\mathcal{SHOIQ}$ is obtained from $\mathcal{SHOIQ}b$ by disallowing Boolean role expressions. The DLs $\mathcal{SHIQ}$, $\mathcal{SHOQ}$, and $\mathcal{SHOI}$ are obtained from $\mathcal{SHOIQ}$ by disallowing nominals, inverse roles, and number restrictions (incl. functionality restrictions), respectively. Finally, the DL $\mathcal{ALCOIF}b$ is obtained from $\mathcal{SHOIQ}b$ by disallowing transitivity axioms (we use $\mathcal{ALC}$ instead of $\mathcal{S}$ in the name of the DL to indicate this), role inclusion axioms, and concepts of the form $\leqslant n\ s.C$ and $\geqslant n\ s.C$.

### 3.1 Conjunctive Queries and Unions of Conjunctive Queries

We now introduce Boolean conjunctive queries since they are the basic form of queries we are concerned with. We later also define non-Boolean queries and show how they can be reduced to Boolean queries. Finally, unions of conjunctive queries are just a disjunction of conjunctive queries.

**Definition 4** (Syntax and Semantics of Conjunctive Queries)**.** Let $\mathcal{S} = (N_C, N_R, N_I)$ be a signature and $N_V$ a countably infinite set of variables disjoint from $N_C$, $N_R$, and $N_I$. A *term* $t$ is an element from $N_V \cup N_I$. Let $A \in N_C$ be a concept name, $r \in N_R$ a role name, and $t, t'$ terms. An *atom* is an expression $A(t)$ or $r(t, t')$ and we refer to these two types of atoms as *concept atoms* and *role atoms* respectively. A *Boolean conjunctive query* $q$ is a non-empty set of atoms. We use $\mathsf{var}(q)$ to denote the set of (existentially quantified) variables occurring in $q$ and $\mathsf{term}(q)$ to denote the set of variables and individual names occurring in $q$. As usual, we use $\sharp(q)$ to denote the cardinality of $q$, which is simply the number of atoms in $q$, and we use $|q|$ for the size of $q$, i.e., the number of symbols necessary to write $q$.

Let $\mathcal{I} = (\Delta^\mathcal{I}, \cdot^\mathcal{I})$ be an interpretation. A total function $\pi\colon \mathsf{term}(q) \to \Delta^\mathcal{I}$ is an *evaluation* if $\pi(a) = a^\mathcal{I}$ for each individual name $a$ occurring in $q$. For $A(t), r(t, t')$ atoms, we write

- $\mathcal{I} \models^\pi A(t)$ if $\pi(t) \in A^\mathcal{I}$;
- $\mathcal{I} \models^\pi r(t, t')$ if $(\pi(t), \pi(t')) \in r^\mathcal{I}$.

If, for an evaluation $\pi$, $\mathcal{I} \models^\pi \mathsf{At}$ for all atoms $\mathsf{At} \in q$, we write $\mathcal{I} \models^\pi q$. We say that $\mathcal{I}$ *satisfies* $q$ and write $\mathcal{I} \models q$ if there exists an evaluation $\pi$ such that $\mathcal{I} \models^\pi q$. We call such a $\pi$ a *match* for $q$ in $\mathcal{I}$.

Let $\mathcal{K}$ be a knowledge base and $q$ a conjunctive query. If $\mathcal{I} \models \mathcal{K}$ implies $\mathcal{I} \models q$, we say that $\mathcal{K}$ *entails* $q$ and write $\mathcal{K} \models q$. △

The *query entailment problem* is defined as follows: given a knowledge base $\mathcal{K}$ and a query $q$, decide whether $\mathcal{K} \models q$.

**Definition 5** (Unions of Conjunctive Queries)**.** A *union of Boolean conjunctive queries* is a formula $q_1 \vee \ldots \vee q_n$, where each disjunct $q_i$ is a Boolean conjunctive query.

A knowledge base $\mathcal{K}$ *entails* a union of Boolean conjunctive queries $q_1 \vee \ldots \vee q_n$, written as $\mathcal{K} \models q_1 \vee \ldots \vee q_n$, if, for each interpretation $\mathcal{I}$ such that $\mathcal{I} \models \mathcal{K}$, there is some $i$ such that $\mathcal{I} \models q_i$ and $1 \leq i \leq n$. △

We now clarify the connection between query entailment and query answering. For query answering, let the variables of a conjunctive query be typed: each variable can either be existentially quantified (also called *non-distinguished*) or free (also called *distinguished* or *answer variables*). Let $q$ be a query in $n$ variables (i.e., $\sharp(\mathsf{var}(q)) = n$), of which $v_1, \ldots, v_m$ ($m \leq n$) are answer variables. The *answers* of $\mathcal{K}$ to $q$ are those $m$-tuples $(a_1, \ldots, a_m)$ of individual names such that, for all models $\mathcal{I}$ of $\mathcal{K}$, $\mathcal{I} \models^\pi q$ for some $\pi$ that satisfies $\pi(v_i) = a_i^\mathcal{I}$ for all $i$ with $1 \leq i \leq m$. Recall that we use $\mathsf{nom}(\mathcal{K})$ to denote the set of individual names occurring in $\mathcal{K}$ (in the form of nominals or ABox individuals). It is not hard to see (cf. Chandra & Merlin, 1977) that the answers of $\mathcal{K}$ to $q$ can be computed by testing, for each $(a_1, \ldots, a_m) \in \mathsf{nom}(\mathcal{K})^m$, whether the query $q_{[v_1,\ldots,v_m/a_1,\ldots,a_m]}$ obtained from $q$ by replacing each occurrence of $v_i$ with $a_i$ for $1 \leq i \leq m$ is entailed by $\mathcal{K}$. The set of *certain answers* to $q$ is then the set of all $m$-tuples $(a_1, \ldots, a_m)$ for which $\mathcal{K} \models q_{[v_1,\ldots,v_m/a_1,\ldots,a_m]}$. Let $k = \sharp(\mathsf{nom}(\mathcal{K}))$ be the number of individual names occurring in $\mathcal{K}$. Since $\mathcal{K}$ is finite, clearly $k$ is finite. Hence, deciding which tuples belong to the set of answers can be checked with at most $k^m$ entailment tests.

The algorithm that we present in this paper decides query entailment. The reasons for devising a decision procedure for query entailment instead of query answering are two-fold: first, query answering can be reduced to query entailment as shown above; second, in contrast to query answering, query entailment is a decision problem and can be studied in terms of complexity theory.

### 3.2 Simplifying Assumptions

In the following, we make several assumptions that are without loss of generality, but simplify the presentation of the decision procedure.

#### 3.2.1 From $\mathcal{SHOIQ}$ and $\mathcal{ALCHOIQ}b$ to simplified $\mathcal{ALCOIF}b$ Knowledge Bases

In the following, we only work with $\mathcal{ALCOIF}b$ knowledge bases. Nevertheless, our results also hold for $\mathcal{SHOIQ}$ knowledge bases and queries with only simple roles in the query and for $\mathcal{ALCHOIQ}b$ knowledge bases, i.e., when the knowledge base contains safe Boolean role expressions, but no transitivity. The restriction to $\mathcal{ALCOIF}b$ is without loss of generality, as we show now.

Provided the query contains only simple roles, we can use the elimination techniques for transitivity (Kazakov & Motik, 2008) to reduce a $\mathcal{SHOIQ}$ knowledge base to an $\mathcal{ALCHOIQ}$ knowledge base with extended signature. We can further eliminate qualified number restrictions and role inclusion axioms by transforming an $\mathcal{ALCHOIQ}b$ knowledge base into an $\mathcal{ALCOIF}b$ knowledge base that is equivalent to the original one up to an extension of the signature (Rudolph, Krötzsch, & Hitzler, 2008). We do not repeat a formal proof here, but rather give an informal argument as to how this reduction works.

We assume that the knowledge base is in *negation normal form*, i.e., all GCIs are of the form $\top \sqsubseteq C$ with $C$ a concept in NNF. Now, consider a concept expression of the form $\geqslant n\ r.C$ with $r$ a role and $C$ a concept. This means that there are at least $n$ distinct $r$-neighbors satisfying $C$. However, this situation can be enforced by introducing $n$ new roles $r_1, \ldots, r_n$ each of which is deemed to have $r$ as a superrole ($r_i \sqsubseteq r$) and which are pairwise disjoint ($\top \sqsubseteq \forall(r_i \sqcap r_j).\bot$). Under those "side conditions", the above concept expression can be replaced by $\exists r_1.C \sqcap \ldots \sqcap \exists r_n.C$.

A somewhat dual argumentation is possible for concept expressions of the form $\leqslant n\ r.C$ restricting the number of $r$-neighbors satisfying $C$ to at most $n$. Again we extend the signature by introducing new roles $r_1, \ldots, r_n$, but this time, we let them "cover" all outgoing $r$-links in the following sense: whenever an $r$-link leads to some domain element $\delta$ which satisfies $C$, then one of the roles $r_1, \ldots, r_n$ also leads there. Indeed, safe Boolean role expressions allow for expressing this correspondence via the concept description $\forall(r \sqcap \neg r_1 \sqcap \ldots \sqcap \neg r_n).\neg C$. It is now easy to see, that this concept expression can replace the above if we additionally demand all roles $r_1, \ldots, r_n$ to be functional.

$$\{o\} \sqsubseteq \exists r.A \qquad A \sqsubseteq \exists r.A \qquad A \sqsubseteq \exists s.B$$
$$\mathsf{func}(f^-) \qquad \mathsf{func}(g^-) \qquad B \sqsubseteq C \sqcup D$$
$$C \sqsubseteq \exists f.E \qquad D \sqsubseteq \exists g.E \qquad E \sqsubseteq B \sqcup \{o\}$$

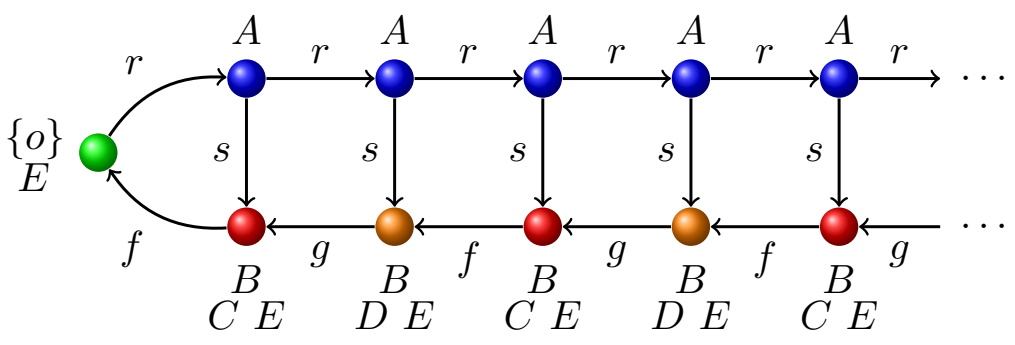

Figure 2: Knowledge base for our running example and a representation of a model for the knowledge base.

Finally consider a role hierarchy statement $r \sqsubseteq s$, stating that whenever two domain elements $\delta_1$ and $\delta_2$ are connected by role $r$, they are also interconnected via $s$. Clearly, this statement can be reformulated as: there are no two domain elements connected by $r$ and by $\neg s$. This, in turn, can be equivalently rephrased by saying that no domain element has an $r \sqcap \neg s$-neighbor or, expressed as GCI, $\top \sqsubseteq \forall(r \sqcap \neg s).\bot$.

These transformations can be applied to an $\mathcal{ALCHOIQ}b$ knowledge base, whereby all cardinality constraints and role inclusion axioms are eliminated. This leaves us with an equivalent $\mathcal{ALCOIF}b$ knowledge base up to an extension of the signature.

Figure 2 displays an $\mathcal{ALCOIF}b$ knowledge base and an according model, which we will refer to as a running example throughout the paper.

Furthermore, we assume that the ABox is internalized (e.g., $C(a)$ is replaced by the equivalent GCI $\{a\} \sqsubseteq C, r(a, b)$ by $\{a\} \sqsubseteq \exists r.\{b\}$, etc.). Thus, we effectively decide query entailment with respect to a TBox only since knowledge bases in this setting have an empty ABox.

For $\mathcal{T}$ an $\mathcal{ALCOIF}b$ TBox, it is always possible to transform $\mathcal{T}$ into an equivalent TBox $\mathcal{T}'$ up to signature extension such that all GCIs in $\mathcal{T}'$ have one of the following *simplified forms*:

$$\bigsqcap A_i \sqsubseteq \bigsqcup B_j \;\mid\; A \equiv \{o\} \;\mid\; A \sqsubseteq \forall U.B \;\mid\; A \sqsubseteq \exists U.B \;\mid\; \mathsf{func}(f), \tag{1}$$

where $A_{(i)}$ and $B_{(j)}$ are concept names, $o$ is an individual name, $U$ is a safe Boolean role expression, and $f$ is a role. If $i = 0$, we interpret $\bigsqcap A_i$ as $\top$ and if $j = 0$, we interpret $\bigsqcup B_j$ as $\bot$. An $\mathcal{ALCOIF}b$ knowledge base $\mathcal{K} = (\mathcal{T}, \mathcal{A})$ is *simplified* if $\mathcal{T}$ is simplified and $\mathcal{A}$ is empty. Every $\mathcal{ALCOIF}b$ knowledge base, which is not in this form, can be transformed in polynomial time into the desired form by using the standard structural transformation, which iteratively introduces definitions for compound sub-concepts (Kazakov & Motik, 2008). Thus, we assume in the remainder that any knowledge base is rewritten into a simplified $\mathcal{ALCOIF}b$ knowledge base.

#### 3.2.2 Connected and Constant-free Queries

We assume that queries are connected. More precisely, let $q$ be a conjunctive query. We say that $q$ is *connected* if, for all $t, t' \in \mathsf{term}(q)$, there exists a sequence $t_1, \ldots, t_n$ such that $t_1 = t$, $t_n = t'$ and, for all $1 \leq i < n$, there exists a role name $r$ such that $r(t_i, t_{i+1}) \in q$ or $r(t_{i+1}, t_i) \in q$. A collection $q_1, \ldots, q_n$ of queries is a *partitioning* of $q$ if $q = q_1 \cup \ldots \cup q_n$, $\mathsf{term}(q_i) \cap \mathsf{term}(q_j) = \emptyset$ for $1 \leq i < j \leq n$, and each $q_i$ is connected.

**Lemma 6.** *Let $\mathcal{K}$ be a knowledge base, $q$ a conjunctive query, and $q_1, \ldots, q_n$ a partitioning of $q$. Then $\mathcal{K} \models q$ iff $\mathcal{K} \models q_i$ for each $i$ with $1 \leq i \leq n$.*

A proof is given by Tessaris (2001) and, with this lemma, it is clear that the restriction to connected queries is indeed without loss of generality since entailment of $q$ can be decided by checking entailment of each $q_i$ at a time. In what follows, we therefore assume queries to be connected without further notice.

In unions of conjunctive queries, we assume that the variable names in each disjunct are different from the variable names in the other disjuncts. This can always be achieved by naming variables apart. We further assume that each disjunct in a UCQ is a connected conjunctive query. This is without loss of generality since a UCQ which contains unconnected disjuncts can always be transformed into conjunctive normal form; we can then decide entailment for each resulting conjunct separately and each conjunct is a union of connected conjunctive queries (Glimm et al., 2008a). Note that, due to the transformation into conjunctive normal form, the resulting number of unions of connected conjunctive queries for which we have to test entailment can be exponential in the size of the original query.

We further assume that queries do not contain constants (individual names) to occur in the position of variables. In the presence of nominals this is without loss of generality: for each individual name $a$ occurring in $q$, we extend the knowledge base $\mathcal{K}$ with the axioms $\{a\} \equiv N_a$ for $N_a \in N_C$ a fresh concept name, and replace each occurrence of $a$ in $q$ with a fresh variable $x_a \in N_V$ and add a concept atom $N_a(x_a)$ to $q$.

#### 3.2.3 General Notation

Throughout this paper, concept names and role expressions are written in upper case, while roles and individual names are written in lower case. Unless stated otherwise, we use $A$ and $B$ for concept names; $C$ and $D$ for possibly complex concepts; $r$ and $s$ for roles, $f$ for functional or inverse functional roles; $U$ and $V$ for safe Boolean role expressions; and $o$ for nominals that are used in TBox axioms or that occur in complex concepts. Sub- and superscripts might be appended if necessary. If not stated otherwise, we use $q$ (possibly with subscripts) for a connected Boolean conjunctive query, $\mathcal{K}$ for a simplified $\mathcal{ALCOIF}b$ knowledge base, $\mathcal{I}$ for an interpretation $(\Delta^{\mathcal{I}}, \cdot^{\mathcal{I}})$, and $\pi, \mu$ for evaluations.

## 4. Model Construction

In this section, we introduce interpretations and models that have a kind of forest shape. The main notion of a forest is, however, very weak since we do also allow for arbitrary relations between tree elements and roots. Without such relations, we call the result a strict forest. We exploit the nice properties of trees and forests in the following sections, when we replace parts in interpretations that give rise to an infinite number of new nominals. Since even models of an $\mathcal{ALCOIF}b$ knowledge base that have a kind of forest shape are not really forests, we also introduce "approximations" of models in which nominals are no longer interpreted as singleton sets. We call these structures quasi-interpretations or quasi-models and such interpretations can have the form of real forests. Further, we provide a way of "unraveling" an arbitrary model into a forest that is a quasi-model for the knowledge base and a way of "collapsing" such forest quasi-models back into real models of the knowledge base that still have a kind of forest shape.

**Definition 7** (Forest (Quasi-)Interpretations and (Quasi-)Models)**.** A *tree* $T$ is a non-empty, prefix-closed subset of $\mathbb{N}^*$. For $w, w' \in T$, we call $w'$ a *successor* of $w$ if $w' = w \cdot c$ for some $c \in \mathbb{N}$, where "$\cdot$" denotes concatenation. We call $w'$ a *predecessor* of $w$ if $w = w' \cdot c$ for some $c \in \mathbb{N}$, and $w'$ is a *neighbor* of $w$ if $w'$ is a successor of $w$ or vice versa. The empty word $\varepsilon$ is called the root of the tree. We use $|w|$ to denote the *length* of $w$.

A *forest* $F$ is a subset of $R \times \mathbb{N}^*$, where $R$ is a countable, possibly infinite set of elements such that, for each $\rho \in R$, the set $\{w \mid (\rho, w) \in F\}$ is a tree. Each pair $(\rho, \varepsilon) \in F$ is called a *root* of $F$. For $(\rho, w), (\rho', w') \in F$, we call $(\rho', w')$ a *successor* of $(\rho, w)$ if $\rho' = \rho$ and $w'$ is a successor of $w$; $(\rho', w')$ is a *predecessor* of $(\rho, w)$ if $\rho' = \rho$ and $w'$ is a predecessor of $w$; $(\rho', w')$ is a *neighbor* of $(\rho, w)$ if $(\rho', w')$ is a successor of $(\rho, w)$ or vice versa. A node $(\rho, w)$ is an *ancestor* of a node $(\rho', w')$ if $\rho = \rho'$ and $w$ is a prefix of $w'$ and it is a *descendant* if $\rho = \rho'$ and $w'$ is a prefix of $w$.

A *forest interpretation* of a knowledge base $\mathcal{K}$ is an interpretation $\mathcal{I} = (\Delta^\mathcal{I}, \cdot^\mathcal{I})$ that satisfies the following conditions:

FI1 $\Delta^\mathcal{I}$ is a forest with roots $R$;

FI2 there is a total and surjective function $\lambda \colon \mathsf{nom}(\mathcal{K}) \to R \times \{\varepsilon\}$ such that $\lambda(o) = (\rho, \varepsilon)$ iff $o^\mathcal{I} = (\rho, \varepsilon)$;

FI3 for each role $r \in \mathsf{rol}(\mathcal{K})$, if $\langle (\rho, w), (\rho', w') \rangle \in r^\mathcal{I}$, then either

(a) $w = \varepsilon$ or $w' = \varepsilon$, or
(b) $(\rho, w)$ is a neighbor of $(\rho', w')$.

If $\mathcal{I} \models \mathcal{K}$, we say that $\mathcal{I}$ is a *forest model for* $\mathcal{K}$. If $\Delta^\mathcal{I}$ has a single root, we call $\mathcal{I}$ a *tree interpretation* and a *tree model for* $\mathcal{K}$, respectively.

Let $\mathcal{K}$ be an $\mathcal{ALCOIF}b$ knowledge base. With $\mathsf{nomFree}(\mathcal{K})$, we denote the $\mathcal{ALCIF}b$ knowledge base obtained from $\mathcal{K}$ by replacing each nominal concept $\{o\}$ with $o \in \mathsf{nom}(\mathcal{K})$ with a fresh concept name $N_o$. A *forest quasi-interpretation* for $\mathcal{K}$ is an interpretation $\mathcal{J} = (\Delta^\mathcal{J}, \cdot^\mathcal{J})$ of $\mathsf{nomFree}(\mathcal{K})$ that satisfies the following properties:

FQ1 $\Delta^\mathcal{J}$ is a forest with roots $R$;

FQ2 there is a total and surjective function $\lambda \colon \mathsf{nom}(\mathcal{K}) \to R \times \{\varepsilon\}$ such that $\lambda(o) = (\rho, \varepsilon)$ iff $(\rho, \varepsilon) \in N_o^\mathcal{J}$

FQ3 for each role $r \in \mathsf{rol}(\mathcal{K})$, if $\langle (\rho, w), (\rho', w') \rangle \in r^\mathcal{I}$, then either

(a) $w = \varepsilon$ or $w' = \varepsilon$, or

(b) $(\rho, w)$ is a neighbor of $(\rho', w')$.

Note that condition FQ2 allows for elements $(\rho, w) \in \Delta^{\mathcal{J}}$ with $w \neq \varepsilon$ such that $(\rho, w) \in N_o^{\mathcal{J}}$. We call $\mathcal{J}$ *strict* if in condition FQ3, only FQ3(b) is allowed. If $\mathcal{J} \models \mathsf{nomFree}(\mathcal{K})$ we say that $\mathcal{J}$ is a *forest quasi-model for* $\mathcal{K}$.

The *branching degree* $d(w)$ of a node $w$ in a tree $T$ is the number of successors of $w$. Let $\mathcal{I} = (\Delta^{\mathcal{I}}, \cdot^{\mathcal{I}})$ be a forest (quasi) interpretation for $\mathcal{K}$. If there is a $k$ such that $d(w) \leq k$ for each $(\rho, w) \in \Delta^{\mathcal{I}}$, then we say that $\mathcal{I}$ has branching degree $k$. △

In the remainder, when we use the concept name $N_o$, we mean the fresh concept name that was introduced in $\mathsf{nomFree}(\mathcal{K})$ for the nominal concept $\{o\}$ with $o \in \mathsf{nom}(\mathcal{K})$. Elements in the extension of a concept $N_o$ are called nominal placeholders. Please note that, in a forest quasi-interpretations $\mathcal{J}$, we can have several elements $(\rho, w)$ with $w \neq \varepsilon$ such that $(\rho, w) \in N_o^{\mathcal{J}}$.

In the following, we define a notion of isomorphism between forest interpretations. Note that we demand not only structural identity w.r.t. concepts and roles but also w.r.t. the successor relation.

**Definition 8** (Isomorphism between Forest Interpretations)**.** Let $\mathcal{I}, \mathcal{I}'$ be two forest interpretations of $\mathcal{K}$ with $\delta_1, \delta_2 \in \Delta^{\mathcal{I}}, \delta_1', \delta_2' \in \Delta^{\mathcal{I}'}$. The pairs $\langle \delta_1, \delta_2 \rangle, \langle \delta_1', \delta_2' \rangle$ are isomorphic w.r.t. $\mathcal{K}$, written $\langle \delta_1, \delta_2 \rangle \cong_{\mathcal{K}} \langle \delta_1', \delta_2' \rangle$ iff

1. $\langle \delta_1, \delta_2 \rangle \in r^{\mathcal{I}}$ iff $\langle \delta_1', \delta_2' \rangle \in r^{\mathcal{I}'}$ for each $r \in \mathsf{rol}(\mathcal{K})$,
2. $\delta_i \in A^{\mathcal{I}}$ iff $\delta_i' \in A^{\mathcal{I}'}$ for $i \in \{1, 2\}$ and each $A \in \mathsf{con}(\mathcal{K})$,
3. $\delta_i = o^{\mathcal{I}}$ iff $\delta_i' = o^{\mathcal{I}'}$ for $i \in \{1, 2\}$ and each $o \in \mathsf{nom}(\mathcal{K})$.

We say that $\mathcal{I}$ and $\mathcal{I}'$ are *isomorphic* w.r.t. $\mathcal{K}$, written: $\mathcal{I} \cong_{\mathcal{K}} \mathcal{I}'$, if there is a bijection $\varphi : \Delta^{\mathcal{I}} \to \Delta^{\mathcal{I}'}$ such that, for each $\delta_1, \delta_2 \in \Delta^{\mathcal{I}}$, $\langle \delta_1, \delta_2 \rangle \cong_{\mathcal{K}} \langle \varphi(\delta_1), \varphi(\delta_2) \rangle$ and $\delta_1$ is a successor of $\delta_2$ iff $\varphi(\delta_1)$ is a successor of $\varphi(\delta_2)$. △

If clear from the context, we omit the subscript $\mathcal{K}$ of $\cong_{\mathcal{K}}$. We extend the above definition in the obvious way to forest quasi-interpretations, i.e., by omitting condition 3 and defining the isomorphism with respect to $\mathcal{K}' = \mathsf{nomFree}(\mathcal{K})$.

Forest quasi-models have, intuitively, the purpose of an intermediate step between arbitrary models of $\mathcal{K}$ and forest models of $\mathcal{K}$. When identifying each $\delta$ in the interpretation of a concept $N_o$ in the knowledge base $\mathcal{K}'$ with a root that is in the interpretation of $N_o$, we obtain an interpretation that would be a model for $\mathcal{K}$ apart from functionality restrictions for some nominals that might be violated. We show later how we can eliminate those relations from the forest back to the roots that violate functionality restrictions and how we can eventually obtain a forest model from a forest quasi-model.

Another useful property of quasi-interpretations is that, for simplified $\mathcal{ALCIF}b$ knowledge bases, it can be checked locally whether an interpretation $\mathcal{I}$ is actually a model of $\mathcal{K}$.

**Definition 9** (Local $\mathcal{K}$-consistency)**.** Let $\mathcal{I} = (\Delta^{\mathcal{I}}, \cdot^{\mathcal{I}})$ be an interpretation for a simplified $\mathcal{ALCIF}b$ knowledge base $\mathcal{K}$ with $\delta \in \Delta^{\mathcal{I}}$. We define local satisfaction for $\delta$ and concepts that can occur in simplified $\mathcal{ALCIF}b$ axioms as follows:

1. for $A_1, \ldots, A_n \in \mathsf{con}(\mathcal{K})$:

   (a) $\mathcal{I}, \delta \models \bigsqcap A_i$ if $\delta \in A_i^{\mathcal{I}}$ for each $i$ with $1 \leq i \leq n$; $\mathcal{I}, \delta \not\models \bigsqcap A_i$ otherwise;

   (b) $\mathcal{I}, \delta \models \bigsqcup A_i$ if $\delta \in A_i^{\mathcal{I}}$ for some $i$ with $1 \leq i \leq n$; $\mathcal{I}, \delta \not\models \bigsqcup A_i$ otherwise;

2. for $U$ a safe Boolean role expression over $\mathsf{rol}(\mathcal{K})$, $A \in \mathsf{con}(\mathcal{K})$:

   (a) $\mathcal{I}, \delta \models \exists U.A$ if there is some $\delta' \in \Delta^{\mathcal{I}}$ such that $\langle \delta, \delta' \rangle \in U^{\mathcal{I}}$ and $\mathcal{I}, \delta' \models A$; $\mathcal{I}, \delta \not\models \exists U.A$ otherwise;

   (b) $\mathcal{I}, \delta \models \forall U.A$ if, for each $\delta' \in \Delta^{\mathcal{I}}$ such that $\langle \delta, \delta' \rangle \in U^{\mathcal{I}}$, $\mathcal{I}, \delta' \models A$; $\mathcal{I}, \delta \not\models \forall U.A$ otherwise;

3. for $f \in \mathsf{rol}(\mathcal{K})$, $\mathcal{I}, \delta \models \mathsf{func}(f)$ if $\sharp(\{\delta' \in \Delta^{\mathcal{I}} \mid \langle \delta, \delta' \rangle \in f^{\mathcal{I}}\}) \leq 1$; $\mathcal{I}, \delta \not\models \mathsf{func}(f)$ otherwise.

An element $\delta \in \Delta^{\mathcal{I}}$ *locally satisfies* a GCI $C \sqsubseteq D$ with $C, D$ $\mathcal{ALCIF}b$-concepts if $\mathcal{I}, \delta \models C$ implies $\mathcal{I}, \delta \models D$. It locally satisfies a functionality restriction $\mathsf{func}(f)$ if $\mathcal{I}, \delta \models \mathsf{func}(f)$. An element $\delta \in \Delta^{\mathcal{I}}$ is *locally $\mathcal{K}$-consistent* if it locally satisfies each axiom in $\mathcal{K}$. △

**Lemma 10.** *Let $\mathcal{K}$ be a simplified $\mathcal{ALCIF}b$ knowledge base and $\mathcal{I} = (\Delta^{\mathcal{I}}, \cdot^{\mathcal{I}})$ an interpretation for $\mathcal{K}$. Then $\mathcal{I}$ is a model for $\mathcal{K}$ iff each element $\delta \in \Delta^{\mathcal{I}}$ is locally $\mathcal{K}$-consistent.*

*Proof.* For simplified $\mathcal{ALCIF}b$ knowledge bases, only axioms of the form $A \sqsubseteq \forall U.B$ and $A \sqsubseteq \exists U.B$ involve checking neighbors of an element $\delta$ and, since $B$ is a concept name in simplified knowledge bases, it is immediate that satisfaction of $B$ can be checked locally for the neighbor of $\delta$ in question. □

For a knowledge base $\mathcal{K}$ with nominals, we can also use local $\mathcal{K}$-consistency, but we need an additional global condition that ensures that nominals are interpreted as singleton sets. The following is an immediate consequence of Lemma 10 and the extra condition 2 for nominals:

**Proposition 11.** *Let $\mathcal{K}$ be a simplified $\mathcal{ALCOIF}b$ knowledge base and $\mathcal{I} = (\Delta^{\mathcal{I}}, \cdot^{\mathcal{I}})$ an interpretation for $\mathcal{K}$. Then $\mathcal{I}$ is a model for $\mathcal{K}$ iff*

1. *each element $\delta \in \Delta^{\mathcal{I}}$ is locally $\mathcal{K}$-consistent and,*

2. *for each $o \in \mathsf{nom}(\mathcal{K})$, there is exactly one element $\delta \in \Delta^{\mathcal{I}}$ such that $o^{\mathcal{I}} = \delta$.*

We now show how we can obtain a forest quasi-model from a model of $\mathcal{K}$ by using an adapted version of unraveling.

**Definition 12** (Unraveling)**.** Let $\mathcal{K}$ be a consistent $\mathcal{ALCOIF}b$ knowledge base and $\mathcal{I} = (\Delta^{\mathcal{I}}, \cdot^{\mathcal{I}})$ a model for $\mathcal{K}$. Let $\mathsf{choose}$ be a function that returns, for a concept $C = \exists U.B \in \mathsf{cl}(\mathcal{K})$ and an element $\delta \in (\exists U.B)^{\mathcal{I}}$ an element $\delta_{C,\delta} \in \Delta^{\mathcal{I}}$ such that $\langle \delta, \delta_{C,\delta} \rangle \in U^{\mathcal{I}}$ and $\delta_{C,\delta} \in B^{\mathcal{I}}$.

Without loss of generality, we assume that, for all $\delta \in \Delta^{\mathcal{I}}$ and concepts $C_1 = \exists U_1.B_1, C_2 = \exists U_2.B_2 \in \mathsf{cl}(\mathcal{K})$ such that $\delta \in C_1^{\mathcal{I}} \cap C_2^{\mathcal{I}}$, if $\mathsf{choose}(C_1, \delta) = \delta_1, \mathsf{choose}(C_2, \delta) = \delta_2$, and $\langle \delta, \delta_1 \rangle \cong \langle \delta, \delta_2 \rangle$, then $\delta_1 = \delta_2$.

An *unraveling* for some element $\delta \in \Delta^{\mathcal{I}}$, denoted as $\downarrow(\mathcal{I}, \delta)$, is an interpretation that is obtained from $\mathcal{I}$ and $\delta$ as follows: we define the set $S \subseteq (\Delta^{\mathcal{I}})^*$ of *sequences* to be the smallest set such that

- $\delta$ is a sequence;
- $\delta_1 \cdots \delta_n \cdot \delta_{n+1}$ is a sequence, if
  - $\delta_1 \cdots \delta_n$ is a sequence,
  - if $n > 2$ and $\langle \delta_n, \delta_{n-1} \rangle \in f^{\mathcal{I}}$ for some functional role $f$, then $\delta_{n+1} \neq \delta_{n-1}$,
  - $\delta_{n+1} = \mathsf{choose}(C, \delta_n)$ for some $C = \exists U.B \in \mathsf{cl}(\mathcal{K})$.

Now fix a set $F \subseteq \{\delta\} \times \mathbb{N}^*$ and a bijection $\lambda \colon F \to S$ such that

(i) $F$ is a forest,

(ii) $\lambda(\delta, \varepsilon) = \delta$,

(iii) if $(\delta, w), (\delta, w \cdot c) \in F$ with $w \cdot c$ a successor of $w$, then $\lambda(\delta, w \cdot c) = \lambda(\delta, w) \cdot \delta_{n+1}$ for some $\delta_{n+1} \in \Delta^{\mathcal{I}}$.

Such a forest $F$ and bijection $\lambda$ exist because $S$ is a prefix-closed set with root $\delta$. Thus, we just map from the notion of sequences to that of forests.

For each $o \in \mathsf{nom}(\mathcal{K})$, let $N_o \in N_C$ be a fresh concept name. For each $(\delta, w) \in F$, set $\mathsf{Tail}(\delta, w) = \delta_n$ if $\lambda(\delta, w) = \delta_1 \cdots \delta_n$. Now, we define the *unraveling* for $\delta$ as the interpretation $\mathcal{J} = (\Delta^{\mathcal{J}}, \cdot^{\mathcal{J}})$ with $\Delta^{\mathcal{J}} = F$ and, for each $(\delta, w) \in \Delta^{\mathcal{J}}$, we define the interpretation of concept and role names as follows:

(a) for each $o \in \mathsf{nom}(\mathcal{K}), N_o^{\mathcal{J}} = \{(\delta, w) \in \Delta^{\mathcal{J}} \mid \mathsf{Tail}(\delta, w) \in o^{\mathcal{I}}\}$;

(b) for each concept name $A \in \mathsf{con}(\mathcal{K})$, $A^{\mathcal{J}} = \{(\delta, w) \in \Delta^{\mathcal{J}} \mid \mathsf{Tail}(\delta, w) \in A^{\mathcal{I}}\}$;

(c) for each role name $r \in \mathsf{rol}(\mathcal{K})$, $\langle (\delta, w), (\delta, w') \rangle \in r^{\mathcal{J}}$ iff $w'$ is a neighbor of $w$, and $\langle \mathsf{Tail}(\delta, w), \mathsf{Tail}(\delta, w') \rangle \in r^{\mathcal{I}}$.

Let $R$ be the subset of $\Delta^{\mathcal{I}}$ that contains exactly those $\delta \in \Delta^{\mathcal{I}}$ such that $o^{\mathcal{I}} = \delta$ for some $o \in \mathsf{nom}(\mathcal{K})$. Let $U$ be a set containing an unraveling of $\mathcal{I}$ starting from each $\delta \in R$. The union of all interpretations from $U$ is called an *unraveling* for $\mathcal{I}$, denoted as $\downarrow(\mathcal{I})$, where unions of interpretations are defined in the natural way. △

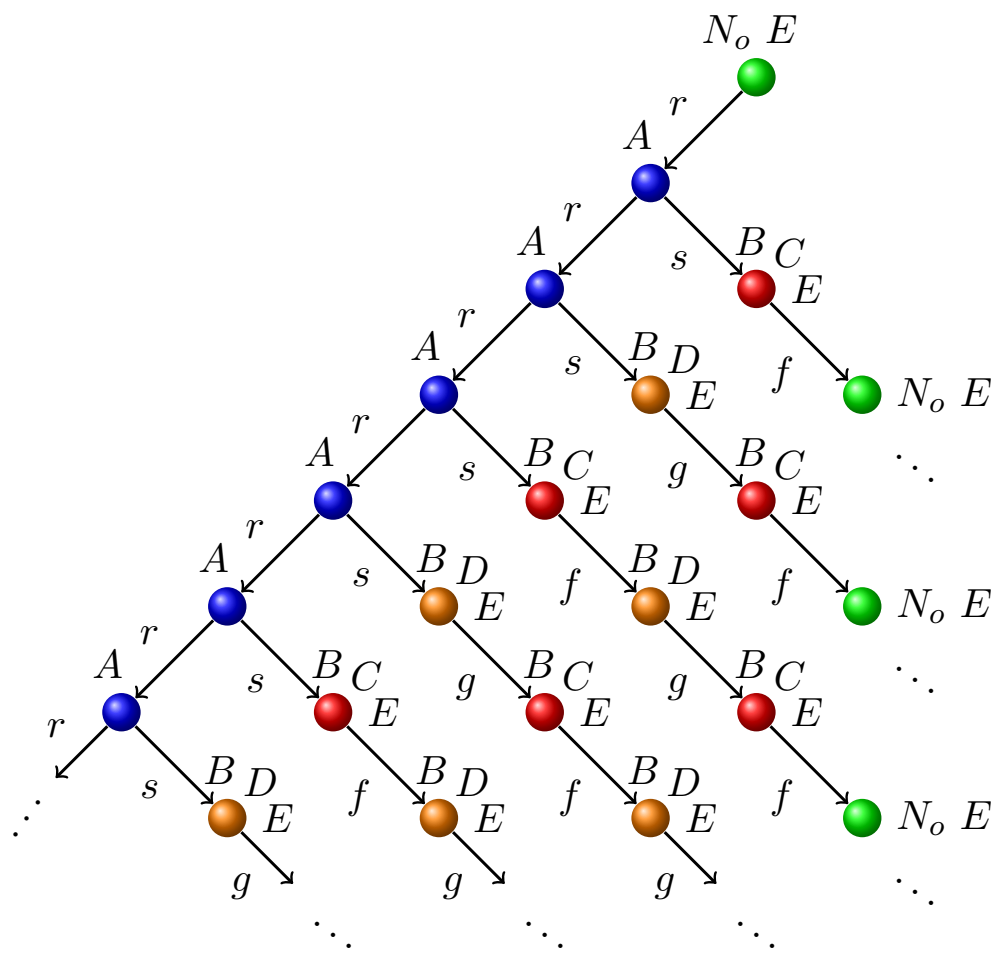

Figure 3: Unraveling of the model displayed in Figure 2.

Figure 3 shows the unraveling for our example knowledge base and model. The dotted lines under the non-root elements labeled $N_o$ indicate that a copy of the whole tree should be appended since we do not stop unraveling at nominal placeholders.

It might be helpful to think of the function $\mathsf{Tail}$ as a homomorphism (up to signature extension) from the elements in the unraveling $\mathcal{J}$ to elements in the original model $\mathcal{I}$. Indeed, $\mathsf{Tail}$ satisfies the following properties: For each $(\delta, w), (\delta', w') \in \Delta^{\mathcal{J}}$,

- $\mathsf{Tail}(\delta, w) = o^{\mathcal{I}}$ iff $(\delta, w) \in N_o^{\mathcal{J}}$, for all $o \in \mathsf{nom}(\mathcal{K})$,
- $\mathsf{Tail}(\delta, w) \in A^{\mathcal{I}}$ iff $(\delta, w) \in A^{\mathcal{J}}$, for all $A \in \mathsf{con}(\mathcal{K})$, and
- $\langle \mathsf{Tail}(\delta, w), \mathsf{Tail}(\delta', w') \rangle \in r^{\mathcal{I}}$ iff $\langle (\delta, w), (\delta', w') \rangle \in r^{\mathcal{J}}$, for all $r \in \mathsf{rol}(\mathcal{K})$.

Unravelings are the first step in the process of transforming an arbitrary model of $\mathcal{K}$ into a forest model since the resulting model is a forest quasi-model of $\mathcal{K}$, as we show in the next lemma.

**Lemma 13.** *Let $\mathcal{K}$ be a consistent $\mathcal{ALCOIF}b$ knowledge base and $\mathcal{I} = (\Delta^{\mathcal{I}}, \cdot^{\mathcal{I}})$ a model of $\mathcal{K}$. Then $\mathcal{J} = (\Delta^{\mathcal{J}}, \cdot^{\mathcal{J}}) = \downarrow(\mathcal{I})$ is a strict forest quasi-model for $\mathcal{K}$.*

*Proof.* Let $\mathcal{K}' = \mathsf{nomFree}(\mathcal{K})$. By construction, $\mathcal{J}$ satisfies conditions $\mathsf{FQ1}$ and $\mathsf{FQ3}$ of forest quasi-models and the strictness condition. Since $\mathcal{J}$ is obtained from a model $\mathcal{I}$ of $\mathcal{K}$, by definition of unravelings as starting from each $\delta \in \Delta^{\mathcal{I}}$ such that $o^{\mathcal{I}} = \delta$ for some $o \in \mathsf{nom}(\mathcal{K})$, and by condition (a) of unravelings, there is, for each $o \in \mathsf{nom}(\mathcal{K})$, one root $(\delta, \varepsilon) \in \Delta^{\mathcal{J}}$ such that $(\delta, \varepsilon) \in N_o^{\mathcal{J}}$. Thus, $\mathcal{J}$ satisfies also property $\mathsf{FQ2}$ and $\mathcal{J}$ is a forest quasi-interpretation for $\mathcal{K}$. We show that $\mathcal{J}$ is a model of $\mathcal{K}'$ by demonstrating that each $(\delta, w) \in \Delta^{\mathcal{J}}$ is locally $\mathcal{K}'$-consistent. Since we assume all knowledge bases to be simplified, we only have to consider axioms of form (1).

Let $\mathsf{Ax}$ be an axiom of the form $\bigsqcap A_i \sqsubseteq \bigsqcup B_j$ and assume that $(\delta, w) \in (\bigsqcap A_i)^{\mathcal{J}}$. By condition (b) of unravelings, we have $\delta_w = \mathsf{Tail}(\delta, w) \in (\bigsqcap A_i)^{\mathcal{I}}$ and, since $\mathcal{I} \models \mathcal{K}$, we have $\delta_w \in B_j^{\mathcal{I}}$ for some $j$. Again by condition (b) of unravelings, we then have $(\delta, w) \in B_j^{\mathcal{J}}$ as required.

Axioms of the form $A \equiv \{o\}$ in $\mathcal{K}$ are rewritten into $A \equiv N_o$ in $\mathcal{K}'$. We consider $A \sqsubseteq N_o$ and $N_o \sqsubseteq A$ separately. Let $\mathsf{Ax}$ be of the form $A \sqsubseteq N_o$ for $o \in \mathsf{nom}(\mathcal{K})$ and assume that $(\delta, w) \in A^{\mathcal{J}}$. By condition (b), we have that $\delta_w = \mathsf{Tail}(\delta, w) \in A^{\mathcal{I}}$ and, since $\mathcal{I} \models \mathcal{K}$, we have $\delta_w \in \{o^{\mathcal{I}}\}$. By condition (a) of unravelings, we then have that $(\delta, w) \in N_o^{\mathcal{J}}$ as required. For $N_o \sqsubseteq A$ with $o \in \mathsf{nom}(\mathcal{K})$, assume that $(\delta, w) \in N_o^{\mathcal{J}}$. By condition (a), we have $\delta_w = \mathsf{Tail}(\delta, w) \in \{o^{\mathcal{I}}\}$ and, since $\mathcal{I} \models \mathcal{K}$, we have $\delta_w \in A^{\mathcal{I}}$. By condition (b) of unravelings, we then have $(\delta, w) \in A^{\mathcal{J}}$ as required.

Let $\mathsf{Ax}$ be an axiom of the form $A \sqsubseteq \forall U.B$ and assume that $(\delta, w) \in A^{\mathcal{J}}$. By condition (b), we have $\delta_w = \mathsf{Tail}(\delta, w) \in A^{\mathcal{I}}$ and, since $\mathcal{I} \models \mathcal{K}$, we have each $\delta_{w'} \in \Delta^{\mathcal{I}}$ such that $\langle \delta_w, \delta_{w'} \rangle \in U^{\mathcal{I}}$ is such that $\delta_{w'} \in B^{\mathcal{I}}$. Let $(\delta', w')$ be such that $\langle (\delta, w), (\delta', w') \rangle \in U^{\mathcal{J}}$ and $(\delta', w') \notin B^{\mathcal{J}}$. By condition (c) of unravelings, we then have that $\langle \delta_w, \delta_{w'} \rangle \in U^{\mathcal{I}}$ for $\delta_{w'} = \mathsf{Tail}(\delta', w')$ and by condition (b) that $\delta_{w'} \notin B^{\mathcal{I}}$, which is a contradiction.

Let $\mathsf{Ax}$ be an axiom of the form $A \sqsubseteq \exists U.B$ and assume that $(\delta, w) \in A^{\mathcal{J}}$. By condition (b), we have $\delta_w = \mathsf{Tail}(\delta, w) \in A^{\mathcal{I}}$ and, since $\mathcal{I} \models \mathcal{K}$, we have there is at least one $\delta_{w'} \in \Delta^{\mathcal{I}}$ such that $\langle \delta_w, \delta_{w'} \rangle \in U^{\mathcal{I}}$ and $\delta_{w'} \in B^{\mathcal{I}}$. In case there is more than one such element, let $\delta_{w'}$ be such that $\delta_{w'} = \mathsf{choose}(C, \delta_w)$. Then, by definition of sequences, there is some neighbor $(\delta, w')$ of $(\delta, w)$ with $\mathsf{Tail}(\delta, w') = \delta_{w'}$. Let $\lambda(\delta, w) = \delta_1 \cdots \delta_n$, i.e., $\delta_n = \delta_w$. We distinguish two cases:

1. The element $\delta_{w'}$ is such that $\delta_{w'} = \delta_{n-1}$. By definition of the bijection $\lambda$, $w = w' \cdot c$, by definition of $\mathcal{J}$ from $\mathcal{I}$ (condition (c)) and since $\langle \delta_w, \delta_{w'} \rangle \in U^{\mathcal{I}}$, we have $\langle (\delta, w), (\delta, w') \rangle \in U^{\mathcal{J}}$. Then, since $B$ is a concept name and $\delta_{w'} \in B^{\mathcal{I}}$, we have by condition (b) that $(\delta, w') \in B^{\mathcal{J}}$, which proves the claim.

2. The element $\delta_{w'}$ is such that $\delta_{w'} \neq \delta_{n-1}$. By definition of sequences and the bijection $\lambda$, we have that $\lambda(\delta, w') = \delta_1 \cdots \delta_n \cdot \delta_{w'}$. Now, by definition of $\mathcal{J}$ from $\mathcal{I}$ (in particular properties (b) and (c)), we have $\langle (\delta, w), (\delta, w') \rangle \in U^{\mathcal{J}}$ and, again since $B$ is a concept name, $(\delta, w') \in B^{\mathcal{J}}$, which proves the claim.

Let $\mathsf{Ax}$ be an axiom of the form $\mathsf{func}(r)$ for $r \in \mathsf{rol}(\mathcal{K})$. Assume, to the contrary of what is to be shown, that $(\delta, w)$ has two distinct neighbors $(\delta, w_1), (\delta, w_2)$ such that $\langle (\delta, w), (\delta, w_1) \rangle, \langle (\delta, w), (\delta, w_2) \rangle \in r^{\mathcal{J}}$. Since the function $\lambda$ introduced in the unraveling is a bijection, there are two distinct sequences $s_1$ and $s_2$ such that $\lambda(\delta, w_1) = s_1$ and $\lambda(\delta, w_2) = s_2$ and $\mathsf{Tail}(\delta, w_1) = \delta_1, \mathsf{Tail}(\delta, w_2) = \delta_2$ with $\delta_1 \neq \delta_2$. Since $\langle (\delta, w), (\delta, w_1) \rangle, \langle (\delta, w), (\delta, w_2) \rangle \in r^{\mathcal{J}}$ we get, due to condition (c), that $\langle \mathsf{Tail}(\delta, w), \delta_1 \rangle, \langle \mathsf{Tail}(\delta, w), \delta_2 \rangle \in r^{\mathcal{I}}$, which is a contradiction since $\mathcal{I} \models \mathcal{K}$.

Since $(\delta, w)$ was arbitrarily chosen, we have that each element in the domain of $\mathcal{J}$ is locally $\mathcal{K}'$-consistent as required and $\mathcal{J} \models \mathcal{K}'$ by Lemma 10. □

**Lemma 14.** *Let $\mathcal{K}$ be a consistent $\mathcal{ALCOIF}b$ knowledge base, $\mathcal{I} = (\Delta^{\mathcal{I}}, \cdot^{\mathcal{I}})$ a model of $\mathcal{K}$, and $\mathcal{J} = (\Delta^{\mathcal{J}}, \cdot^{\mathcal{J}}) = \downarrow(\mathcal{I})$ an unraveling for $\mathcal{I}$. Then $\mathcal{J}$ has a branching degree bounded in $|\mathsf{cl}(\mathcal{K})|$.*

*Proof.* Let $m$ be the number of axioms in $\mathcal{K}$. Each axiom of a simplified knowledge base can contain at most one existential restriction and, due to the definition of the function $\mathsf{choose}$ used in the unraveling, there are, for each sequence $s \in S$, at most $m$ elements $\delta_1, \ldots, \delta_m \in \Delta^{\mathcal{I}}$ such that $s \cdot \delta_i$ with $1 \leq i \leq m$ is a sequence in $S$. Since the mapping $\lambda$ from the forest $\Delta^{\mathcal{J}}$ to sequences is a bijection, $\Delta^{\mathcal{J}}$ is a forest with branching degree $m$. □

In the following steps, we traverse a forest quasi-model in an order in which elements with smaller tree depth are always of smaller order than elements with greater tree depth. Elements with the same tree depth are ordered lexicographically. The bounded branching degree of unravelings then guarantees that, after a finite number of steps, we go on to the next level in the forest and process all nodes eventually. Further, we can merge nodes such that, finally, all nominal placeholders (in the extension of some $N_o$) can be interpreted as nominals without violating functionality restrictions. In fact, we do not only have to merge nominal placeholders, but also elements that are related to a nominal placeholder by an inverse functional role since, by definition of the semantics, these elements have to correspond to the same element in a model. In order to identify such elements, we define the notion of *backwards counting paths* as follows:

**Definition 15** (Paths and BCPs)**.** Let $\mathcal{I} = (\Delta^{\mathcal{I}}, \cdot^{\mathcal{I}})$ be an interpretation. We call $\delta_1 \cdot \ldots \cdot \delta_n$ a *path* from $\delta_1$ to $\delta_n$ if, for each $i$ with $1 \leq i < n$, $\langle \delta_i, \delta_{i+1} \rangle \in r_i^{\mathcal{I}}$ for some role $r_i \in \mathsf{rol}(\mathcal{K})$. The *length* $|p|$ of a path $p = \delta_1 \cdot \ldots \cdot \delta_n$ is $n-1$. Each element $\delta \in \Delta^{\mathcal{I}}$ is a path of length 0. We write $\delta_1 \overset{U_1}{\to} \delta_2 \ldots \overset{U_{n-1}}{\to} \delta_n$ to denote a path from $\delta_1$ to $\delta_n$ such that $\langle \delta_i, \delta_{i+1} \rangle \in U_i^{\mathcal{I}}$ for each $1 \leq i < n$ and $U_i$ a safe Boolean role expression.

Let $\mathcal{K}$ be an $\mathcal{ALCOIF}b$ knowledge base and $\mathcal{I} = (\Delta^{\mathcal{I}}, \cdot^{\mathcal{I}})$ a forest model (a forest quasi-model) of $\mathcal{K}$. A path $p = \delta_1 \cdot \ldots \cdot \delta_n$ in $\mathcal{I}$ is a *descending* path if there is some root $(\rho, \varepsilon) \in \Delta^{\mathcal{I}}$ such that, for each $i$ with $1 \leq i \leq n, \delta_i = (\rho, w_i)$ and, for $1 \leq i < n, |w_i| < |w_{i+1}|$. The path $p$ is a *backwards counting path* (BCP) in $\mathcal{I}$ if $\delta_n = o^{\mathcal{I}}$ ($\delta_n \in N_o^{\mathcal{I}}$) for some nominal $o \in \mathsf{nom}(\mathcal{K})$ and, for each $i$ with $1 \leq i < n$, $\langle \delta_i, \delta_{i+1} \rangle \in f_i^{\mathcal{I}}$ for some inverse functional role $f_i \in \mathsf{rol}(\mathcal{K})$. The path $p$ is a *descending BCP* if it is a BCP and a descending path. Given a BCP $p = \delta_1 \overset{f_1}{\to} \delta_2 \ldots \overset{f_n}{\to} \delta_{n+1}$ with $\delta_{n+1} \in o^{\mathcal{J}}$ ($\delta_{n+1} \in N_o^{\mathcal{J}}$), we call the sequence $f_1 \cdots f_n o$ a *path sketch* of $p$. △

Please note that an element $\delta$ in the domain of $\mathcal{J}$ already counts as a (descending) BCP if $\delta \in o^{\mathcal{J}}$ ($N_o^{\mathcal{J}}$) for some $o \in \mathsf{nom}(\mathcal{K})$.

We now define the order that guarantees that in an iterative parsing process, we not only process all nodes, but also that we can merge nodes as required so that, finally, all nominal placeholders can be interpreted as nominals without violating functionality restrictions.

**Definition 16** (Ordering)**.** For convenience and without loss of generality, we assume that the set of individual names $N_I$ is ordered. Let $\mathcal{K}$ be a consistent $\mathcal{ALCOIF}b$ knowledge base and $\mathcal{J}$ a forest quasi-interpretation for $\mathcal{K}$. We extend the order to elements in $\Delta^{\mathcal{J}}$ as follows: let $w_1 = w_p \cdot c_1^1 \cdots c_1^n, w_2 = w_p \cdot c_2^1 \cdots c_2^m \in \mathbb{N}^*$ where $w_p \in \mathbb{N}^*$ is the longest common prefix of $w_1$ and $w_2$, then $w_1 < w_2$ if either $n < m$ or both $n = m$ and $c_1^1 < c_2^1$. For $(\rho_1, \varepsilon), (\rho_2, \varepsilon) \in \Delta^{\mathcal{J}}$, let $o_1 \in \mathsf{nom}(\mathcal{K})$ be the smallest nominal such that $(\rho_1, \varepsilon) \in N_{o_1}^{\mathcal{J}}$ and $o_2 \in \mathsf{nom}(\mathcal{K})$ the smallest nominal such that $(\rho_2, \varepsilon) \in N_{o_2}^{\mathcal{J}}$. Now $(\rho_1, w_1) < (\rho_2, w_2)$ if either (i) $|w_1| < |w_2|$ or (ii) $|w_1| = |w_2|$ and $o_1 < o_2$ or (ii) $|w_1| = |w_2|, o_1 = o_2$ and $w_1 < w_2$.

In the following, we are merging elements in an unraveling and, in this process, create new roots of the form $(\rho w, \varepsilon)$ from elements of the form $(\rho, w)$ and elements of the form $(\rho w, w')$ from $(\rho, ww')$. We extend, therefore, the order to elements of this form as follows: $(\rho_1 w_1, w_1') < (\rho_2 w_2, w_2')$ if $(\rho_1, w_1 w_1') < (\rho_2, w_2 w_2')$. △

Roughly speaking, we proceed as follows in order to transform a quasi-forest model $\mathcal{J}$ into a forest model $\mathcal{I}$: we work our way downwards the trees level by level along the descending BCPs and use the above defined order for this purpose. By definition of the semantics, elements that start the same descending BCP or, more precisely, that start BCPs with identical path sketches, have to correspond to the same element in the forest model $\mathcal{I}$ that we produce. During the traversal of the forest quasi-model, we distinguish two situations: (i) we encounter an element $(\rho, w)$ that starts a descending BCP and we have not seen another element before that starts a descending BCP with the same path sketch. In this case, we promote $(\rho, w)$ to become a new root node of the form $(\rho w, \varepsilon)$ and we shift the subtree rooted in $(\rho, w)$ with it; (ii) we encounter a node $(\rho, w)$ that starts a descending BCP, but we have already seen a node $(\rho', w')$ that starts a descending BCP with that path sketch and which is now a root of the form $(\rho' w', \varepsilon)$. In this case, we delete the subtree rooted in $(\rho, w)$ and identify $(\rho, w)$ with $(\rho' w', \varepsilon)$. If $(\rho, w)$ is an $f$-successor of its predecessor for some inverse functional role $f$, we delete all $f^-$-successors of $(\rho' w', \varepsilon)$ and their subtrees in order to satisfy the functionality restriction. We use a notion of collapsing admissibility to characterize models in which the predecessor of $(\rho, w)$ satisfies the same atomic concepts as the deleted successor of $(\rho', w')$, which ensures that local consistency is preserved. By virtue of this notion, we can characterize forest quasi-models that can be collapsed into proper models irrespective of whether they have been obtained by unraveling of a model or not.

In order to keep the domain as a forest when promoting an element $(\rho, w)$ to a new root, we build the new domain with elements of the form $(\rho w, \varepsilon)$ for $(\rho, w)$ and elements of the form $(\rho w, w')$ for descendants $(\rho, ww')$ of $(\rho, w)$.

**Definition 17** (Equivalence Relation $\sim$ and Collapsings)**.** Let $\mathcal{K}$ be an $\mathcal{ALCOIF}b$ knowledge base, $\mathcal{K}' = \mathsf{nomFree}(\mathcal{K})$, and $\mathcal{J} = (\Delta^{\mathcal{J}}, \cdot^{\mathcal{J}})$ a forest quasi-interpretation of $\mathcal{K}$. We define $\sim$ as the smallest equivalence relation on $\Delta^{\mathcal{J}}$ that satisfies $\delta_1 \sim \delta_2$ if $\delta_1, \delta_2$ start descending BCPs with identical path sketches.

Let $\mathcal{J}$ be a strict forest quasi-interpretation for $\mathcal{K}$, $\mathcal{J}_0 = (\Delta^{\mathcal{J}_0}, \cdot^{\mathcal{J}_0}) = \mathcal{J}$ and $(\rho_0, w_0) \in \Delta^{\mathcal{J}_0}$ the smallest element with $w_0 \neq \varepsilon$ that starts a descending BCP. We call $\mathcal{J}_0$ an *initial collapsing* for $\mathcal{J}$ and $(\rho_0, w_0)$ the *focus* of $\mathcal{J}_0$.

Let $\mathcal{J}_i$ be a collapsing for $\mathcal{J}$ and $(\rho_i, w_i) \in \Delta^{\mathcal{J}_i}$ the focus of $\mathcal{J}_i$. We obtain a collapsing $\mathcal{J}_{i+1} = (\Delta^{\mathcal{J}_{i+1}}, \cdot^{\mathcal{J}_{i+1}})$ with focus $(\rho_{i+1}, w_{i+1})$ for $\mathcal{J}$ from $\mathcal{J}_i$ according to the following two cases:

1. There is no element $(\rho, \varepsilon) \in \Delta^{\mathcal{J}_i}$ such that $(\rho, \varepsilon)$ is smaller than the focus $(\rho_i, w_i)$ and $(\rho, \varepsilon) \sim (\rho_i, w_i)$. Then $\mathcal{J}_{i+i}$ is obtained from $\mathcal{J}_i$ by renaming each element $(\rho_i, w_i w_i') \in \Delta^{\mathcal{J}_i}$ to $(\rho_i w_i, w_i')$.

2. There is an element $(\rho, \varepsilon) \in \Delta^{\mathcal{J}_i}$ such that $(\rho, \varepsilon)$ is smaller than the focus $(\rho_i, w_i)$ and $(\rho, \varepsilon) \sim (\rho_i, w_i)$. Let $(\rho, \varepsilon)$ be the smallest such element.

(a) $\Delta^{\mathcal{J}_{i+1}} = \Delta^{\mathcal{J}_i} \setminus (\{(\rho_i, w_i w'_i) \mid w'_i \in \mathbb{N}^*\} \cup \{(\rho, w) \mid w = c \cdot w', c \in \mathbb{N}, w' \in \mathbb{N}^*, (\rho_i, w_i)$ has a predecessor $(\rho_i, w'_i)$ such that $\langle(\rho_i, w'_i), (\rho_i, w_i)\rangle \in f^{\mathcal{J}_i}$ for an inverse functional role $f$ in $\mathsf{rol}(\mathcal{K})$ and $\langle(\rho, c), (\rho, \varepsilon)\rangle \in f^{\mathcal{J}_i}\})$;

(b) for each concept name $A \in \mathsf{con}(\mathcal{K})$ and $(\rho, w) \in \Delta^{\mathcal{J}_{i+1}}$, $(\rho, w) \in A^{\mathcal{J}_{i+1}}$ iff $(\rho, w) \in A^{\mathcal{J}_i}$;

(c) for each role name $r \in \mathsf{rol}(\mathcal{K})$ and $(\rho_1, w_1), (\rho_2, w_2) \in \Delta^{\mathcal{J}_{i+1}}$, $\langle(\rho_1, w_1), (\rho_2, w_2)\rangle \in r^{\mathcal{J}_{i+1}}$ iff

  i. $\langle(\rho_1, w_1), (\rho_2, w_2)\rangle \in r^{\mathcal{J}_i}$ or

  ii. $(\rho_1, w_1)$ is the predecessor of $(\rho_i, w_i)$ in $\mathcal{J}_i$ (i.e., $\rho_1 = \rho_i$ and $w_i = w_1 \cdot c$ for some $c \in \mathbb{N}$), $(\rho_2, w_2) = (\rho, \varepsilon)$, and $\langle(\rho_1, w_1), (\rho_i, w_i)\rangle \in r^{\mathcal{J}_i}$.

The focus $(\rho_{i+1}, w_{i+1})$ in $\mathcal{J}_{i+1}$ is the smallest descending BCP with $(\rho_i, w_i) < (\rho_{i+1}, w_{i+1})$.

For a collapsing $\mathcal{J}_i$, let $\mathsf{safe}(\mathcal{J}_i)$ be the restriction of $\mathcal{J}_i$ to elements $(\rho, w)$ such that $(\rho, w) \in \mathcal{J}_j$ for all $j \geq i$. With $\mathcal{J}_\omega$ we denote the non-disjoint union of all interpretations $\mathsf{safe}(\mathcal{J}_i)$ obtained from subsequent collapsings $\mathcal{J}_i$ for $\mathcal{J}$. The interpretation obtained from $\mathcal{J}_\omega$ by interpreting each $o \in \mathsf{nom}(\mathcal{K})$ as $(\rho, \varepsilon) \in N_o^{\mathcal{J}_\omega}$ is denoted by $\mathsf{collapse}(\mathcal{J})$ and called a *purified interpretation with respect to* $\mathcal{J}$. If $\mathsf{collapse}(\mathcal{J}) \models \mathcal{K}$, we call $\mathsf{collapse}(\mathcal{J})$ a *purified model of* $\mathcal{K}$. △

In Figures 4 to 7 we illustrate the first collapsing steps for the unraveling depicted in Figure 3. Apart from the nominal placeholder concepts, the concept interpretations are not shown in the figures, but are assumed to be as indicated in Figure 3. The edges for descending BCPs are shown in red color, and the dashed lines in Figure 4 indicate the levels of the tree because, within a tree, the order in which the nodes are processed depends firstly on their level. Within a level, we assume that the order increases from left to right. The numbers next to nodes in Figure 4 indicate, which elements are used as focus element in a collapsing step and their order. For the initial collapsing (Figure 4) the focus is on the first non-root element that starts a BCP, which we indicate with a black border around the node and a black triangle pointing to the focus.

In the first collapsing step we just rename elements to promote the focus from Figure 4 to a root. Because the focus element highlighted in Figure 5 starts a BCP with path sketch different from the ones started by smaller elements, we again only rename elements to obtain a new root (Figure 6). Now, the focus is on a nominal placeholder and since nominal placeholder are BCPs, we have a root with the same path sketch and use the second case of Definition 17. The resulting collapsing is depicted in Figure 7.

Finally, we obtain a collapsing for the unraveling shown in Figure 3 as the one depicted in Figure 8.

We can now show that the collapsing of an unraveling results in a forest model for $\mathcal{K}$. Our aim is, however, to show something more general. We want to collapse not only unravelings into forest models, but also forest quasi-models which have been obtained in another way. Unfortunately, it is not the case that the collapsing of any forest quasi-model results in a forest model for $\mathcal{K}$ since the elements that we merge in the collapsing process do not necessarily satisfy the same atomic concepts. We define, therefore, the following admissibility criterion that characterizes forest quasi-models that can be collapsed into forest models.

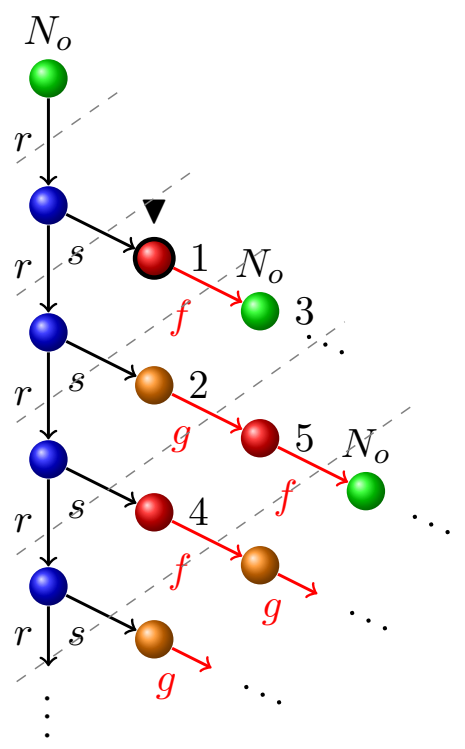

Figure 4: An initial collapsing for the unraveling depicted in Figure 3.

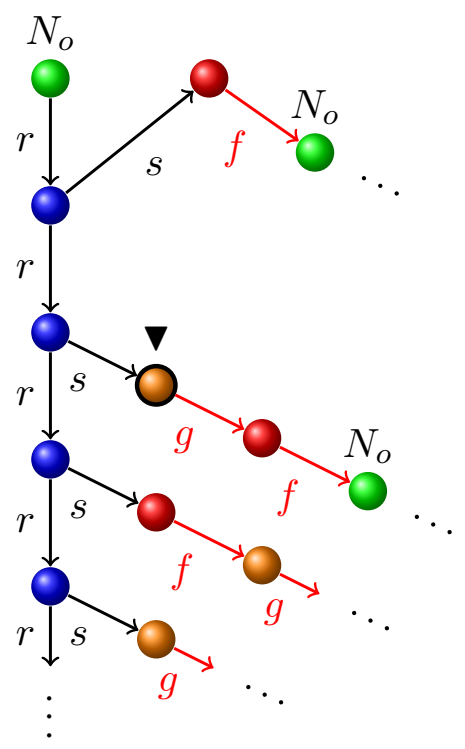

Figure 5: In the first collapsing step we just rename elements to promote the focus from Figure 4 to a root.

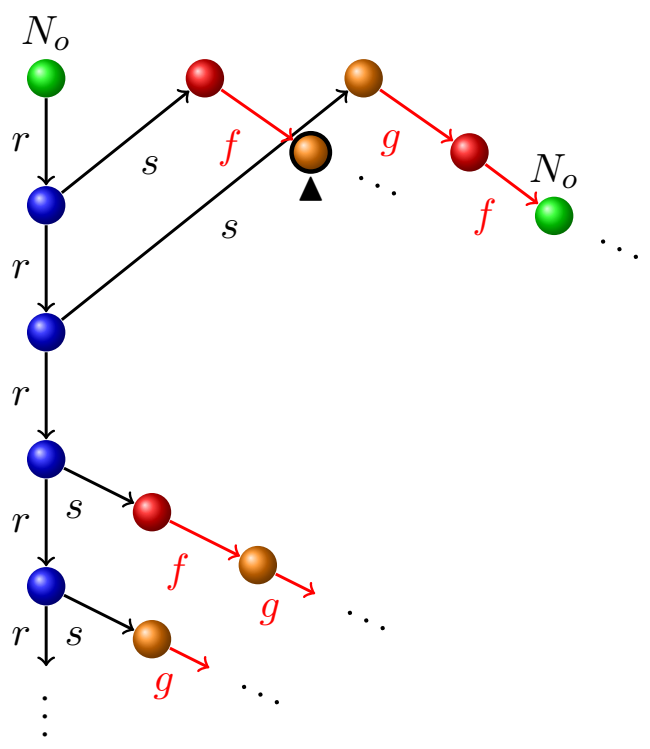

Figure 6: A collapsing obtained from the one depicted in Figure 5.

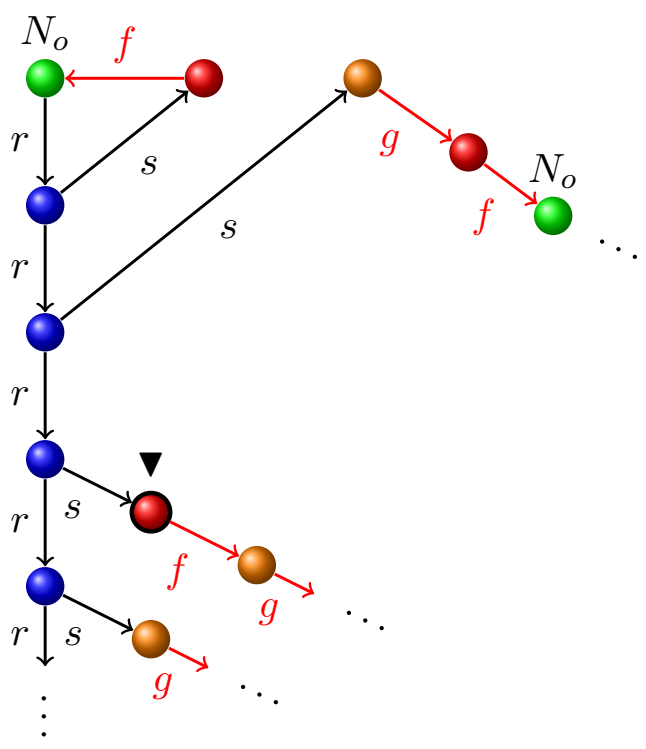

Figure 7: A collapsing obtained by using the second case of Definition 17 on the collapsing from Figure 6.

**Definition 18** (Collapsing-admissibility)**.** Let $\mathcal{J}$ be a forest quasi-interpretation for some $\mathcal{ALCOIF}b$ knowledge base $\mathcal{K}$. Then $\mathcal{J}$ is *collapsing-admissible* if there exists a function $\mathsf{ch}\colon (\mathsf{cl}(\mathcal{K}) \times \Delta^{\mathcal{J}}) \to \Delta^{\mathcal{J}}$ such that

1. for each concept $C = \exists U.B \in \mathsf{cl}(\mathcal{K})$ and each $\delta \in C^{\mathcal{J}}$, we have $\langle \delta, \mathsf{ch}(C, \delta)\rangle \in U^{\mathcal{J}}$ and $\mathsf{ch}(C, \delta) \in B^{\mathcal{J}}$. Moreover, if there is no functional role $f$ for which $\langle \delta, \mathsf{ch}(C, \delta)\rangle \in f^{\mathcal{J}}$ then $\mathsf{ch}(C, \delta)$ is a successor of $\delta$,
2. for each concept $C = \exists U.B \in \mathsf{cl}(\mathcal{K})$ and elements $\delta, \delta' \in C^{\mathcal{J}}$ that start descending BCPs with identical path sketches, we have $\langle \delta, \mathsf{ch}(C, \delta)\rangle \cong \langle \delta', \mathsf{ch}(C, \delta')\rangle$.

△

**Lemma 19.** *Let $\mathcal{K}$ be an $\mathcal{ALCOIF}b$ knowledge base. Any unraveling $\mathcal{J}$ of a model $\mathcal{I}$ for $\mathcal{K}$ is collapsing-admissible.*

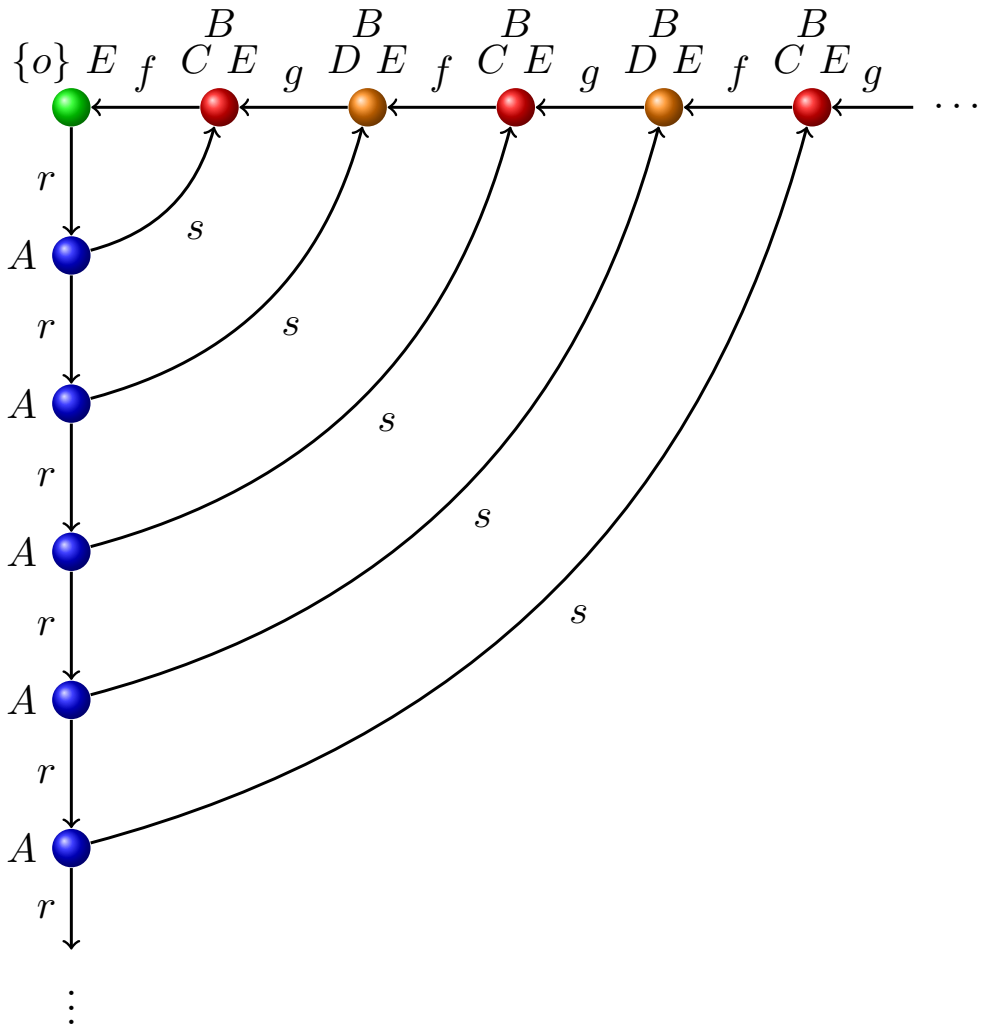

Figure 8: Result of collapsing the unraveling from Fig. 3. The infinitely many new root elements are displayed in the top line.

*Proof.* We define a function $\mathsf{ch}$ directly from the function $\mathsf{choose}$ used in the unraveling as follows: for each $C \in \mathsf{cl}(\mathcal{K})$ and $(\rho, w) \in \Delta^{\mathcal{J}}$ with $\lambda(\rho, w) = \delta_1 \cdot \ldots \cdot \delta_n$ and $\mathsf{choose}(C, \mathsf{Tail}(\rho, w)) = \{\delta'\}$, we set $\mathsf{ch}(C, (\rho, w)) = (\rho, w')$ for $(\rho, w') = \lambda^-(\delta_1 \cdot \ldots \cdot \delta_n \cdot \delta')$ if $\delta_1 \cdot \ldots \cdot \delta_n \cdot \delta'$ is a sequence in $S$ and $(\rho, w') = \lambda^-(\delta_1 \cdot \ldots \cdot \delta_{n-1})$ otherwise. This is well-defined since the function $\lambda$ in unravelings is total and bijective and it is as required for admissibility since elements that start BCPs with identical path sketches are always generated from the same element in $\mathcal{I}$. The first condition of collapsing-admissibility holds since in unravelings, we always add $\delta_1 \cdot \ldots \cdot \delta_n \cdot \delta'$ to the set of sequences unless the pair $\langle \delta_n, \delta_{n-1} \rangle$ is in the interpretation of some functional role. In that case, the function $\mathsf{ch}$ uses the predecessor instead of the successor, which is still admissible. □

**Lemma 20.** *Let $\mathcal{K}$ be a consistent $\mathcal{ALCOIF}b$ knowledge base, $\mathcal{J} = (\Delta^{\mathcal{J}}, \cdot^{\mathcal{J}})$ a strict forest quasi-model for $\mathcal{K}$ with branching degree $b$ that is collapsing-admissible. Then $\mathsf{collapse}(\mathcal{J})$ is a forest model for $\mathcal{K}$ with branching degree $b$.*

*Proof.* Let $\mathcal{K}' = \mathsf{nomFree}(\mathcal{K})$. Since $\mathcal{J}$ is a forest quasi-model of $\mathcal{K}$ by assumption, $\mathcal{J} \models \mathcal{K}'$.

We first show that each collapsing $\mathcal{J}_i$ for $\mathcal{J}$ is a forest quasi-model for $\mathcal{K}$, i.e., $\mathcal{J}_i \models \mathcal{K}'$. We then show that each collapsing $\mathcal{J}_{i+1}$ produced from an admissible collapsing $\mathcal{J}_i$ is again collapsing-admissible. Finally, we show that, for each $o \in \mathsf{nom}(\mathcal{K})$, there is exactly one node in $\mathcal{J}_\omega$ of the form $(\rho, \varepsilon)$ such that $(\rho, \varepsilon) \in N_o^{\mathcal{J}_\omega}$, which implies by Proposition 11 that $\mathsf{collapse}(\mathcal{J})$ is a forest model for $\mathcal{K}$.

We start with the first claim: For the initial collapsing this is immediate since $\mathcal{J}$ is a forest quasi-model for $\mathcal{K}$. In particular, $\mathcal{J}_0$ is locally $\mathcal{K}'$-consistent. Assume that $\mathcal{J}_i$ is a locally $\mathcal{K}'$-consistent collapsing and $(\rho_i, w_i)$ is the focus in $\mathcal{J}_i$. We show that $\mathcal{J}_{i+1}$ is locally $\mathcal{K}'$-consistent. Since $\mathcal{K}'$ is simplified by assumption, we only have to consider axioms of form (1).

If $\mathcal{J}_{i+1}$ is obtained according to the first case of Definition 17, we only rename elements of the domain in order to create a new root node and local $\mathcal{K}'$-consistency is immediate. We assume, thus, that $\mathcal{J}_{i+1}$ is obtained according to the second case of Definition 17.

Axioms of the form $\bigsqcap A_i \sqsubseteq \bigsqcup B_j$ and $A \equiv \{o\}$ (rewritten into $A \equiv N_o$ in $\mathcal{K}'$) hold immediately due to condition 2.b of collapsings.

Let $\mathsf{Ax}$ be an axiom of the form $A \sqsubseteq \forall U.B$ and assume that $(\rho, w) \in A^{\mathcal{J}}$. The only interesting elements are the predecessor $(\rho_i, w_i')$ of the focus $(\rho_i, w_i)$ and $(\rho, \varepsilon)$. However, $(\rho_i, w_i) \sim (\rho, \varepsilon)$ and, since $\mathcal{J}$ is collapsing-admissible, $(\rho_i, w_i)$ and $(\rho, \varepsilon)$ satisfy the same atomic concepts with respect to $\mathsf{con}(\mathcal{K})$. Further, the interpretation of atomic concepts is not changed due to 2.b, which again implies local $\mathcal{K}'$-consistency for this kind of axioms.

Let $\mathsf{Ax}$ be an axiom of the form $A \sqsubseteq \exists U.B$ and assume that $(\rho, w) \in A^{\mathcal{J}_{i+1}}$. We concentrate on the three interesting cases where the direct neighborhoods of elements change:

1. We start with the case where the focus $(\rho_i, w_i)$ is the corresponding $U$-successor of $(\rho, w)$, i.e., $\rho_i = \rho, w_i = w \cdot c$ for some $c \in \mathbb{N}$, $\langle(\rho, w), (\rho_i, w_i)\rangle \in U^{\mathcal{J}_i}$, and $(\rho_i, w_i) \in B^{\mathcal{J}_i}$. Since $(\rho, \varepsilon)$ and $(\rho_i, w_i)$ are in the same equivalence class for $\sim$ by assumption, $(\rho, \varepsilon)$ starts a BCP with the same path sketch as $(\rho_i, w_i)$ and both $(\rho, \varepsilon)$ and $(\rho_i, w_i)$ satisfy the same atomic concepts with respect to $\mathsf{con}(\mathcal{K})$. Then condition 2.(c)ii. ensures that $(\rho, \varepsilon)$ is the required $U$-successor for $(\rho, w)$ in $\mathcal{J}_{i+1}$.

2. Assume that $(\rho, w) = (\rho, \varepsilon), \langle(\rho, \varepsilon), (\rho, c)\rangle \in U^{\mathcal{J}_i}, (\rho, c) \in B^{\mathcal{J}_i}$, $(\rho, c) \notin \Delta^{\mathcal{J}_{i+1}}$, and $(\rho, \varepsilon) \notin (\exists U.B)^{\mathcal{J}_{i+1}}$. Due to 2.a, the focus $(\rho_i, w_i)$ has a predecessor $(\rho_i, w_i')$ such that $\langle(\rho_i, w_i'), (\rho_i, w_i)\rangle \in f^{\mathcal{J}_i}$ for an inverse functional role $f \in \mathsf{rol}(\mathcal{K})$ and $\langle(\rho, \varepsilon), (\rho, c)\rangle \in (f^-)^{\mathcal{J}_i}$. Since $f$ is inverse functional and $\mathcal{J}_i$ is, by assumption, locally $\mathcal{K}'$-consistent, there is no successor $(\rho_i, w_i \cdot c_i)$ of $(\rho_i, w_i)$ such that $\langle(\rho_i, w_i), (\rho_i, w_i \cdot c_i)\rangle \in (f^-)^{\mathcal{J}_i}$. Similarly, there is no element $(\rho', w')$ such that $\langle(\rho, \varepsilon), (\rho', w')\rangle \in (f^-)^{\mathcal{J}_i}$. Then, since $\mathcal{J}_i$ is collapsing-admissible, we have that $(\rho_i, w_i') \in \mathsf{ch}(\exists U.B, (\rho_i, w_i)), (\rho, c) \in \mathsf{ch}(\exists U.B, (\rho, \varepsilon))$, and $\langle(\rho_i, w_i), (\rho_i, w_i')\rangle \cong \langle(\rho, \varepsilon), (\rho, c)\rangle$ since $(\rho_i, w_i)$ and $(\rho, \varepsilon)$ start descending BCPs with identical path sketches. In particular, $\langle(\rho_i, w_i), (\rho_i, w_i')\rangle \in U^{\mathcal{J}_i}$ and $(\rho_i, w_i') \in B^{\mathcal{J}_i}$. Then, by condition 2.(c)ii., $\langle(\rho, \varepsilon), (\rho_i, w_i')\rangle \in U^{\mathcal{J}_{i+1}}$, by condition 2.b, $(\rho_i, w_i') \in B^{\mathcal{J}_{i+1}}$, and, thus, $(\rho, \varepsilon) \in (\exists U.B)^{\mathcal{J}_{i+1}}$ as required.

3. We assume that $(\rho_i, w_i)$ has a predecessor $(\rho_i, w_i')$ such that $\langle(\rho_i, w_i'), (\rho_i, w_i)\rangle \in f^{\mathcal{J}_i}$ for an inverse functional role $f$ in $\mathsf{rol}(\mathcal{K})$ and $\langle(\rho, c), (\rho, \varepsilon)\rangle \in f^{\mathcal{J}_i}$, causing the deletion of $(\rho, c)$ and its descendants, one of which, say $(\rho, v)$ is connected to some $(\varsigma, \varepsilon)$, such that $\langle(\varsigma, \varepsilon), (\rho, v)\rangle \in U^{\mathcal{J}_i}$ and $(\rho, v) \in B^{\mathcal{J}_i}$. Now, if there is no inverse functional role $g$ for which $\langle(\varsigma, \varepsilon), (\rho, v)\rangle \in g^{\mathcal{J}_i}$, then the strictness and collapsing-admissibility of $\mathcal{J}_i$ ensure the existence of a $c' \in \mathbb{N}$ for which $\langle(\varsigma, \varepsilon), (\varsigma, c)\rangle \in U^{\mathcal{J}_i}$ and $(\varsigma, c) \in B^{\mathcal{J}_i}$ and, consequently, also $\langle(\varsigma, \varepsilon), (\varsigma, c)\rangle \in U^{\mathcal{J}_{i+1}}$ and $(\varsigma, c) \in B^{\mathcal{J}_{i+1}}$. If $\langle(\varsigma, \varepsilon), (\rho, v)\rangle \in g^{\mathcal{J}_i}$ for some inverse functional role $g$, then strictness of the initial collapsing implies that $(\rho, v)$ itself started a descending BCP and, due to the defined order, it must have been a focus before and now be a root itself. This contradicts, however, the initial assumption that $(\rho, v)$ is a descendant of $(\rho, \varepsilon)$ and we are done.

In all cases, the elements in $\Delta^{\mathcal{J}_{i+1}}$ have the required successors.

Let $\mathsf{Ax}$ be an axiom of the form $\mathsf{func}(f)$ for $f \in \mathsf{rol}(\mathcal{K})$. We concentrate on relations between the predecessor $(\rho_i, w_i')$ of the focus and $(\rho, \varepsilon)$ since otherwise local $\mathcal{K}'$-consistency is immediate. A predecessor exists for the focus since we process elements in ascending order starting with non-root nodes. Assume $\langle(\rho_i, w_i'), (\rho_i, w_i)\rangle \in f^{\mathcal{J}_i}$, in which case $\langle(\rho_i, w_i'), (\rho, \varepsilon)\rangle \in f^{\mathcal{J}_{i+1}}$ due to 2.(c)ii. and assume $(\rho, \varepsilon)$ has a successor $(\rho, c)$ in $\mathcal{J}_i$ such that $\langle(\rho, c), (\rho, \varepsilon)\rangle \in f^{\mathcal{J}_i}$. In this case, $(\rho, c) \notin \Delta^{\mathcal{J}_{i+1}}$ according to 2.a and together with local $\mathcal{K}'$-consistency of $\mathcal{J}_i$, this implies that $(\rho_i, w_i')$ is the only element in $\Delta^{\mathcal{J}_{i+1}}$ such that $\langle(\rho_i, w_i'), (\rho, \varepsilon)\rangle \in f^{\mathcal{J}_{i+1}}$.

We now show that each $\mathcal{J}_{i+1}$ produced from an admissible collapsing $\mathcal{J}_i$ is again admissible for collapsing. By assumption, the initial collapsing is admissible, so let $\mathcal{J}_i$ be an admissible collapsing and $\mathsf{ch}_i$ the required function. We distinguish two cases:

1. Let $\mathcal{J}_{i+1}$ be produced according to the first case of collapsings. We define a function $\mathsf{ch}_{i+1}$ for $\mathcal{J}_{i+1}$ as follows: For each $C \in \mathsf{cl}(\mathcal{K})$ and $\delta \in \Delta^{\mathcal{J}_{i+1}}$, we set $\mathsf{ch}_{i+1}(C, \delta) = \delta'$ for $\delta' = (\rho_i w_i, w_1')$ if $\mathsf{ch}_i(C, \delta) = (\rho_i, w_i w_i')$ for $(\rho_i, w_i)$ the focus in $\mathcal{J}_i$ and $\delta' = \mathsf{ch}_i(C, \delta)$ otherwise. Since we just change the names of the elements and leave the interpretation of concepts and roles as before, the function is as required for admissibility.

2. Let $\mathcal{J}_{i+1}$ be produced according to the second case of collapsings. We define a function $\mathsf{ch}_{i+1}$ for $\mathcal{J}_{i+1}$ as follows: For each $C \in \mathsf{cl}(\mathcal{K})$,

   (a) for each $\delta \in \Delta^{\mathcal{J}_{i+1}} \notin \{(\rho_i, w_i'), (\rho, \varepsilon)\}$ with $(\rho_i, w_i')$ the predecessor of the focus $(\rho_i, w_i)$, we set $\mathsf{ch}_{i+1}(C, \delta) = \delta'$ for $\delta'$ such that $\delta' \in \Delta^{\mathcal{J}_{i+1}}$ and $\delta' = \mathsf{ch_i}(C, \delta)$; this is well-defined since only successors of $(\rho, \varepsilon)$ and $(\rho_i, w_i')$ are deleted in $\mathcal{J}_{i+1}$.

   (b) for $\delta = (\rho, \varepsilon)$ and $(\rho_i, w_i')$ the predecessor of the focus, $\mathsf{ch}_{i+1}(C, \delta) = \delta'$ for $\delta' = (\rho_i, w_i')$ if $\mathsf{ch_i}(C, \delta) = (\rho, c)$ and $(\rho, c) \notin \Delta^{\mathcal{J}_{i+1}}$ and $\delta' = \mathsf{ch_i}(C, \delta)$ otherwise;

   (c) for $\delta = (\rho_i, w_i')$ the predecessor of the focus, we set $\mathsf{ch}_{i+1}(C, \delta) = \delta'$ for $\delta' = (\rho, \varepsilon)$ if $\mathsf{ch_i}(C, \delta) = (\rho_i, w_i)$ and $\delta' = \mathsf{ch_i}(C, \delta)$ otherwise.

   For elements apart from the predecessor of the focus $(\rho_i, w_i')$ and the root $(\rho, \varepsilon)$ that replaces $(\rho_i, w_i)$, the interpretation of concepts and roles remains as before by properties 2.b and 2.c and the function is as required. For $(\rho_i, w_i')$, we change the function so that in cases where $(\rho_i, w_i)$ was returned, $(\rho, \varepsilon)$ is returned. Since $(\rho_i, w_i) \sim (\rho, \varepsilon)$, this is admissible. Similarly, if a successor $(\rho, c)$ of $(\rho, \varepsilon)$ is not contained in $\Delta^{\mathcal{J}_{i+1}}$, then $(\rho_i, w_i')$ is used instead. This is admissible since, in this case, $\langle(\rho_i, w_i), (\rho_i, w_i')\rangle \cong \langle(\rho, \varepsilon), (\rho, c)\rangle$ as argued above for axioms of the form $A \sqsubseteq \exists U.B$.

We now show that, for each $o \in \mathsf{nom}(\mathcal{K})$, there is exactly one node in $\mathcal{J}_\omega$ of the form $(\rho, \varepsilon)$ such that $(\rho, \varepsilon) \in N_o^{\mathcal{J}_\omega}$. Nominal placeholders are descending BCPs by definition and, when a nominal placeholder becomes the focus, it is merged into a root that is in the same equivalence class of $\sim$ and which is by definition of lower order. Such a root exists because of property $\mathsf{FQ2}$ of forest quasi-interpretations.

The interpretation $\mathcal{J}_\omega$ is obtained by building the non-disjoint union of the "safe" parts for all collapsings, which contain only elements which will neither be renamed nor deleted. Thus, $\mathcal{J}_\omega$ does not contain nominal placeholders as required. Considering any one element $(\rho, w) \in \Delta^{\mathcal{J}_\omega}$, we find that there is an $i \in \mathbb{N}$ such that all successors and all root neighbors of $(\rho, w)$ in $\mathcal{J}_i$ are the same as in $\mathcal{J}_\omega$. As we have shown, $\mathcal{J}_i$ is locally $\mathcal{K}'$-consistent and therefore $(\rho, w)$ has a consistent neighborhood. Hence $\mathcal{J}_\omega$ is a forest quasi-model of $\mathcal{K}$.

Now, when interpreting each $o \in \mathsf{nom}(\mathcal{K})$ as $\{(\rho, \varepsilon) \in \Delta^{\mathcal{J}_\omega} \mid (\rho, \varepsilon) \in N_o^{\mathcal{J}_\omega}\}$ in $\mathsf{collapse}(\mathcal{J})$, we obtain a forest model for $\mathcal{K}$, where the set of roots is $\{(\rho, \varepsilon) \mid (\rho, \varepsilon) \in \Delta^{\mathcal{J}_\omega}\}$.

The bounded branching degree is an immediate consequence of the construction since we never add successors during the construction and the starting forest quasi-interpretation $\mathcal{J}$ has a bounded branching degree by assumption. □

Since unravelings of a model $\mathcal{I}$ for $\mathcal{K}$ are strict forest quasi-models of $\mathcal{K}$ with branching degree bounded in $|\mathsf{cl}(\mathcal{K})|$ by Lemma 13, and unravelings are collapsing-admissible by Lemma 19, it is an immediate consequence of Lemma 20 that the collapsing of an unraveling yields a forest model branching degree bounded in $|\mathsf{cl}(\mathcal{K})|$.

**Corollary 21.** *Let $\mathcal{K}$ be an $\mathcal{ALCOIF}b$ knowledge base and $\mathcal{I}$ an interpretation such that $\mathcal{I} \models \mathcal{K}$, then the purified interpretation $\mathsf{collapse}(\downarrow(\mathcal{I}))$ is a forest model for $\mathcal{K}$ with branching degree $b$ bounded in $|\mathsf{cl}(\mathcal{K})|$.*

Since the number of roots might still be infinite in purified models, we could, up to now, have obtained the same result by unraveling an arbitrary model, where we take all elements on BCPs as roots instead of taking just the nominals and creating new roots in the collapsing process. In the next sections, however, we show how we can transform an unraveling of a counter-model for the query such that it remains collapsing-admissible and such that it can in the end be collapsed into a forest model with a finite number of roots that is still a counter model for the query. For this transformation it is much more convenient to work with real (strict) trees and forests, which is why we use (strict) forest quasi-interpretations.

In the next sections, we also use the following alternative characterization of the result of a collapsing, which comes in handy for the subsequent proofs.

We start by defining the so-called *pruning* of a forest quasi-interpretation, which is, roughly speaking, the structure obtained by just deleting all the nodes, which will be erased in the course of the collapsing process anyway.

**Definition 22** (Pruning)**.** Let $\mathcal{J}$ be a strict forest quasi-model for an $\mathcal{ALCOIF}b$ knowledge base $\mathcal{K}$ that is collapsing-admissible and let $\mathcal{J}_0, \mathcal{J}_1, \ldots, \mathcal{J}_\omega$ be as defined in Definition 17. The *pruning* of $\mathcal{J}$ (written $\mathsf{prune}(\mathcal{J})$) is obtained by restricting $\mathcal{J}$ to a set $\Gamma \subseteq \Delta^{\mathcal{J}}$ which is defined as follows: $\Gamma$ contains all $\langle \rho w_1, w_2 w_3 \rangle \in \Delta^{\mathcal{J}}$ for which $\langle \rho w_1 w_2, w_3 \rangle \in \Delta^{\mathcal{J}_\omega}$ or $\langle \rho w_1 w_2, w_3 \rangle$ is the focus in some $\mathcal{J}_i$. △

We again use the equivalence relation $\sim$ for elements that start descending BCPs with identical path sketches from Definition 17 and construct an interpretation from a pruning by identifying equivalent nodes, also known as factorization.

**Definition 23** (Factorization)**.** Let $\mathcal{K}$ be an $\mathcal{ALCOIF}b$ knowledge base, $\mathcal{J}$ a strict forest quasi-interpretation for $\mathcal{K}$ that is collapsing-admissible, and $\mathcal{L} = \mathsf{prune}(\mathcal{J})$.

The factorization of $\mathcal{L}$ by $\sim$ (denoted by $\mathcal{L}_{/\sim}$) is now defined as the forest quasi-interpretation $\mathcal{M} = (\Delta^{\mathcal{M}}, \cdot^{\mathcal{M}})$ with

- $\Delta^{\mathcal{M}} = \{[\delta]_\sim \mid \delta \in \Delta^{\mathcal{L}}\}$;

- for each $A \in \mathsf{con}(\mathcal{K})$, $A^{\mathcal{M}} = \{[\delta]_\sim \mid \delta \in A^{\mathcal{L}}\}$,
- for each $r \in \mathsf{rol}(\mathcal{K})$, $r^{\mathcal{M}} = \{\langle [\delta]_\sim, [\delta']_\sim \rangle \mid \langle \delta, \delta' \rangle \in r^{\mathcal{L}}\}$, and
- for each $o \in \mathsf{nom}(\mathcal{K})$, $o^{\mathcal{M}} = [\delta]_\sim$ for $\delta \in N_o^{\mathcal{L}}$.

△

Note that the interpretation of nominals in $\mathcal{M}$ is well defined as, by definition, all $N_o$-instances are in the same $\sim$-equivalence class.

Now we are ready to establish the wanted correspondence: the collapsing of a forest quasi-interpretation can essentially be obtained by first pruning and then factorizing it.

**Lemma 24.** *Let $\mathcal{J}$ be a strict forest quasi-model for an $\mathcal{ALCOIF}b$ knowledge base $\mathcal{K}$ and let $\mathcal{J}$ be collapsing-admissible. Then $\mathsf{collapse}(\mathcal{J}) \cong \mathsf{prune}(\mathcal{J})_{/\sim}$. Moreover the new roots in $\mathsf{collapse}(\mathcal{J})$ correspond to those $\sim$-equivalence classes that contain $\mathcal{J}$-elements which start descending BCPs in $\mathcal{J}$.*

*Proof.* Considering the first claim, note that by definition of the collapsing procedure, for every $(\rho w, w') \in \Delta^{\mathsf{collapse}(\mathcal{J})}$ there is exactly one pair $w_1, w_2$ with $w = w_1 w_2$ such that $(\rho w_1, w_2 w') \in \Delta^{\mathsf{prune}(\mathcal{J})}$. Moreover, case 1 of the construction assures that $\Delta^{\mathsf{collapse}(\mathcal{J})}$ contains one element from every $\sim$-equivalence class from $\Delta^{\mathsf{prune}(\mathcal{J})/\sim}$. Hence the mapping $\varphi\colon \Delta^{\mathsf{collapse}(\mathcal{J})} \to \Delta^{\mathsf{prune}(\mathcal{J})/\sim}$ with $\varphi(\rho w, w') = [(\rho w_1, w_2 w')]_\sim$ is a bijection and, as a consequence of the construction, an isomorphism.

The second claim is also a direct consequence of the construction of the collapsing. □

## 5. Quasi-Entailment in Quasi-Models

In this section, we will provide a characterization for forest quasi-models that mirrors query entailment for the corresponding "proper models". In our further argumentation, we will talk about the initial part of a tree, i.e., the part that is left if branches are cut down to a fixed length. For a forest interpretation $\mathcal{I} = (\Delta^{\mathcal{I}}, \cdot^{\mathcal{I}})$ and an $n \in \mathbb{N}$, we therefore denote with $\mathsf{cut}_n(\mathcal{I})$ the interpretation obtained from $\mathcal{I}$ by restricting $\Delta^{\mathcal{I}}$ to those pairs $(\rho, w)$ for which $|w| \leq n$.

The following lemma ensures that in the case of purified models, we find only finitely many unraveling trees of depth $n$ that "look different".

**Lemma 25.** *Let $\mathcal{K}$ be a consistent $\mathcal{ALCOIF}b$ knowledge base. Then there is a purified interpretation $\mathcal{I}$ such that $\mathcal{I} \models \mathcal{K}$ and, for every $n \in \mathbb{N}$, there are only finitely many non-isomorphic trees of depth $n$.*

*Proof.* Since $\mathcal{K}$ has a model by assumption, Corollary 21 guarantees that there is some purified model $\mathcal{I}$ of $\mathcal{K}$. In particular, $\mathcal{I}$ is a forest model with the branching degree bounded in the size of $\mathsf{cl}(\mathcal{K})$.

We now compute the maximal number of non-isomorphic trees of depth $n$ over the domain of $\mathcal{I}$. We denote this bound by $T_n$. The argumentation is close to the one used by Levy and Rousset (1998) for their definition of tree blocking.

Let $c = |\mathsf{cl}(\mathcal{K})|$ and $r = |\mathsf{rol}(\mathcal{K})|$. We first consider trees of depth $n = 0$. We have $2^c$ choices for the different subsets of concepts in $\mathsf{cl}(\mathcal{K})$. For $n > 0$, each concept at level 0 can trigger the generation of a new successor and we can have any number of successors between 0 and $c$. Assume, for now, that we have only a single role name $r \in \mathsf{rol}(\mathcal{K})$ and that each node in a level smaller than $n$ is the root of a tree with depth $n-1$ with exactly $c$ successors for each node. In this case, there are $O(2^c T_{n-1}^c)$ non-isomorphic sub-trees of depth $n$. Taking into account that a node does not necessarily have $c$ successors, but we can choose any number between 0 and $c$, we get a bound of $O(2^c c T_{n-1}^c)$ for the number of non-isomorphic sub-trees of depth $n$. Finally, since we have not only one but a choice of $r$ roles, we get a bound of $O(2^c(cT_{n-1}^c)^r)$. We now abbreviate $2^c c^r$ with $x$ and $rc$ with $a$ and rewrite the obtained bound as $T_n = O(x(T_{n-1})^a)$. Unfolding yields $T_n = O((x^{1+a+\ldots+a^{n-1}})(T_0)^{a^n})$ which is bounded by $O((x^{a^n})(2^c)^{a^n}) = O((x2^c)^{a^n})$. By expanding the abbreviated symbols, we obtain a bound for $T_n$ of $O((2^c c^r)^{(rc)^n})$. □

For our further considerations, we introduce the notion of "anchored $n$-components". These are meant to be certain substructures of forest quasi-interpretations. In the first place, these substructures contain a connected set of nodes $W'$ which are situated "closely together" in the original structure, this closeness being witnessed by the fact that all elements of $W'$ are descendants of some node $\delta$ and have a distance $\leq n$ to $\delta$. Moreover for each of those nodes $\delta'$ from $W'$, the anchored $n$-component may (but does not need to) contain a finite number of descending BCPs starting from $\delta'$.

**Definition 26** (Anchored Components)**.** Let $\mathcal{J}$ be a forest quasi-interpretation and $\delta \in \Delta^{\mathcal{J}}$. An interpretation $\mathcal{C}$ will be called *anchored $n$-component of $\mathcal{J}$ with witness $\delta$* if $\mathcal{C}$ can be created by restricting $\mathcal{J}$ to a set $W \subseteq \Delta^{\mathcal{J}}$ obtained as follows:

- Let $\mathcal{J}_\delta$ be the subtree of $\mathcal{J}$ that is started by $\delta$ and let $\mathcal{J}_{\delta,n} := \mathsf{cut}_n(\mathcal{J}_\delta)$. Select a subset $W' \subseteq \Delta^{\mathcal{J}_{\delta,n}}$ that is closed under predecessors.
- For every $\delta' \in W'$, let $P$ be a finite set (possibly empty) of descending BCPs $\mathbf{p}$ starting from $\delta'$ and let $W_{\delta'}$ contain all nodes from all $\mathbf{p} \in P$.
- Set $W = W' \cup \bigcup_{\delta' \in W'} W_{\delta'}$.

△

Now think of a forest quasi-model $\mathcal{J}$ and some query $q$. The following definition and lemma employ the notion of anchored $n$-components to come up with the notion of *quentailment* (short for quasi-entailment), a criterion that reflects query-entailment in the world of forest quasi-models.

**Definition 27** (Quentailment)**.** Let $q$ be a conjunctive query with $\sharp(q) = n$ and $\mathcal{J}$ some forest quasi-model of an $\mathcal{ALCOIFb}$ knowledge base $\mathcal{K}$. We say that $\mathcal{J}$ *quentails* $q$ (written $\mathcal{J} \approx\!\!\!| q$) if, for $V = \mathsf{var}(q)$, $\mathcal{J}$ contains connected anchored $n$-components $\mathcal{C}_1, \ldots, \mathcal{C}_\ell$ and there are according functions $\mu_i \colon V \to 2^{\Delta^{\mathcal{C}_i}}$ such that the following hold:

Q1 For every $x \in V$, there is at least one $\mathcal{C}_i$, such that $\mu_i(x) \neq \emptyset$

Q2 For all $A(x) \in q$, we have $\mu_i(x) \subseteq A^{\mathcal{J}}$ for all $i$.

Q3 For every $r(x, y) \in q$ there is a $\mathcal{C}_i$ such that there are $\delta_1 \in \mu_i(x)$ and $\delta_2 \in \mu_i(y)$ such that $\langle \delta_1, \delta_2 \rangle \in r^{\mathcal{J}}$.

Q4 If, for some $x \in V$, there are connected anchored $n$-components $\mathcal{C}_i$ and $\mathcal{C}_j$ with $\delta \in \mu_i(x)$ and $\delta' \in \mu_j(x)$, then there is

- a sequence $\mathcal{C}_{n_1}, \dots, \mathcal{C}_{n_k}$ with $n_1 = i$ and $n_k = j$ and
- a sequence $\delta_1, \dots, \delta_k$ with $\delta_1 = \delta$ and $\delta_k = \delta'$ as well as $\delta_m \in \mu_{n_m}(x)$ for all $1 \leq m < k$,

such that, for every $m$ with $1 \leq m < k$, we have that

- $\mathcal{C}_{n_m}$ contains a descending BCP $p_1$ started by $\delta_m$,
- $\mathcal{C}_{n_{m+1}}$ contains a descending BCP $p_2$ started by $\delta_{m+1}$,
- $p_1$ and $p_2$ have the same path sketch.

For a union of conjunctive queries $u = q_1 \vee \ldots \vee q_h$, we say that $\mathcal{J}$ *quentails* $u$ (written $\mathcal{J} \approx\!\!\models u$) if $\mathcal{J} \approx\!\!\models q$ for a $q \in \{q_1, \dots, q_h\}$. △

Note that an anchored component may contain none, one or several instantiations of a variable $x \in V$. Intuitively, the definition ensures, that we find matches of query parts which when fitted together by identifying BCP-equal elements yield a complete query match.

**Lemma 28.** *Let $u = q_1 \vee \ldots \vee q_h$ be a union of conjunctive queries and $\mathcal{K}$ an $\mathcal{ALCOIFb}$ knowledge base. Then*

1. *For any model $\mathcal{I}$ of $\mathcal{K}$, $\downarrow(\mathcal{I}) \approx\!\!\models u$ implies $\mathcal{I} \models u$.*
2. *For any strict forest quasi-model $\mathcal{J}$ of $\mathcal{K}$ that is collapsing-admissible,* $\mathsf{collapse}(\mathcal{J}) \models u$ *implies $\mathcal{J} \approx\!\!\models u$.*

*Proof.* 1. Let $q$ be a disjunct of $u$ such that $\downarrow(\mathcal{I}) \approx\!\!\models q$, $V = \mathsf{var}(q)$, and $\mathcal{C}_1, \dots, \mathcal{C}_\ell$ the connected anchored $n$-components witnessing the quentailment. We use the function $\mathsf{Tail}$ from Definition 12 and exploit its properties as a homomorphism. Note that $\mathsf{Tail}$ maps nodes in $\downarrow(\mathcal{I})$ to the same individual in $\mathcal{I}$, if they start descending BCPs with the same path sketches. Due to condition Q4 from Definition 27, this implies, for every $x \in V$ and every $\delta_1 \in \mu_i(x)$ and $\delta_2 \in \mu_j(x)$, that $\mathsf{Tail}(\delta_1) = \mathsf{Tail}(\delta_2)$. Hence, the total function $\mu\colon V \to \Delta^{\mathcal{I}}$ defined by letting $\mu(x) = \gamma$ whenever $\mathsf{Tail}(\delta) = \gamma$ for some $\delta \in \mu_i(x)$ and some $i$ with $1 \leq i \leq \ell$, is well-defined. We now show that $\mu$ is a query match for $q$ in $\mathcal{I}$ by examining the atoms of $q$:

- For every unary atom $A(x)$, the definition of quentailment ensures that there exist a $\mathcal{C}_i$ and a $\delta \in \Delta^{\mathcal{C}_i}$ with $\delta \in \mu_i(x)$ and $\delta \in A^{\mathcal{J}}$. Then, by definition of $\mathsf{Tail}$, it follows that $\mu(x) = \mathsf{Tail}(\delta) \in A^{\mathcal{I}}$.
- Likewise, for every binary atom $r(x, y)$, the definition of quentailment ensures that there exists a $\mathcal{C}_i$ and $\delta_1, \delta_2 \in \Delta^{\mathcal{C}_i}$ such that $\delta_1 \in \mu_i(x)$ and $\delta_2 \in \mu_i(y)$ as well as $\langle \delta_1, \delta_2 \rangle \in r^{\mathcal{J}}$. Again, by definition of $\mathsf{Tail}$, it follows that $\langle \mu(x), \mu(y) \rangle = \langle \mathsf{Tail}(\delta_1), \mathsf{Tail}(\delta_2) \rangle \in r^{\mathcal{I}}$.

2. To prove this, we employ the alternative characterization of collapsings as established in Lemma 24. Let $\mathcal{I}' = (\Delta^{\mathcal{I}'}, \cdot^{\mathcal{I}'}) = \mathsf{prune}(\mathcal{J})_{/\sim}$ and let $\mu\colon V \to \Delta^{\mathcal{I}'}$ be a match for $q$ in $\mathcal{I}'$. We use $\mu$ to construct the anchored $n$-components and functions needed to show that $\mathcal{J}$ quentails $q$.

   Let $V^* \subseteq V$ contain those variables that $\mu$ maps to a singleton $\sim$-equivalence class. We now define $\mathfrak{V} = \{V_1, \ldots, V_m\}$ as the finest partitioning on $V^*$ such that, for any $x, y \in V^*$, $x$ and $y$ are in the same partition whenever $r(x, y) \in q$ for some $r \in \mathsf{rol}(\mathcal{K})$. Next, we assign to every partition $V' \in \mathfrak{V}$ the set $Q_{V'}$ of query atoms containing variables from $V'$. We now construct for every $V'$ an anchored $n$-component $\mathcal{C}_{V'}$ and a function $\mu_{V'}$ (initialized as yielding $\emptyset$ for all inputs) as follows:

   - For every $x \in V'$, let $\mathcal{C}_{V'}$ contain the $\mathcal{J}$-element $\delta$ for which $\mu(x) = \{\delta\}$. Note that $\Delta^{\mathcal{I}'}$ consists of the $\sim$-equivalence classes over elements from $\mathcal{J}$, i.e., $\{\delta\}$ is one of the $\sim$-equivalence classes from $\Delta^{\mathcal{I}'}$. Moreover, set $\mu_{V'}(x) = \mu_{V'}(x) \cup \{\delta\}$.
   - For every $r(x, y) \in Q_{V'}$ with $y \notin V'$ and $\mu(x) = \{\delta\}$, let $\mathcal{C}_{V'}$ contain an additional element $\delta' \in \mu(y)$ for which $\langle \delta, \delta' \rangle \in r^{\mathcal{J}}$ (existence assured by definition of collapsing via factorization) and all elements from some descending BCP in $\mathsf{prune}(\mathcal{J})$ starting from $\delta$ (existence assured since $[\delta']_\sim$ is not a singleton class). Moreover set $\mu_{V'}(y) = \mu_{V'}(y) \cup \{\delta'\}$.
   - Likewise, for every $r(x, y) \in Q_{V'}$ with $x \notin V'$ and $\mu(y) = \{\delta\}$, let $\mathcal{C}_{V'}$ contain an additional element $\delta' \in \mu(x)$ for which $\langle \delta', \delta \rangle \in r^{\mathcal{J}}$ (existence assured by definition of collapsing via factorization) and all elements from the shortest descending BCP in $\mathsf{prune}(\mathcal{J})$ starting from $\delta'$ (existence assured since $[\delta']_\sim$ is not a singleton class). Moreover set $\mu_{V'}(x) = \mu_{V'}(x) \cup \{\delta'\}$.

   We furthermore construct, for each query atom $a$ that contains no variables from $V^*$, its own anchored $n$-component $\mathcal{C}_a$ and function $\mu_a$ (again initialized to always return $\emptyset$) as follows:

   - If $a = r(x, y)$, let $\mathcal{C}_a$ contain two nodes $\delta_1$ and $\delta_2$ for which $\delta_1 \in \mu(x)$ and $\delta_2 \in \mu(y)$ and $\langle \delta_1, \delta_2 \rangle \in r^{\mathcal{J}}$ (existence assured by definition via factorization) as well as some $\mathsf{prune}(\mathcal{J})$-descending BCP starting from $\delta_1$ and the shortest $\mathsf{prune}(\mathcal{J})$-descending BCP starting from $\delta_2$.
   - If $a = A(x)$, let $\mathcal{C}_a$ contain a node $\delta$ for which $\delta \in \mu(x)$ and $\delta \in A^{\mathcal{J}}$ (existence assured by definition via factorization) as well as the shortest $\mathsf{prune}(\mathcal{J})$-descending BCP starting from $\delta$.

   Let $\mathfrak{C}$ contain all $\mathcal{C}_{V'}$ and $\mathcal{C}_a$ defined so far. Note that $\mathfrak{C}$ already satisfies the conditions Q1-Q3 of Definition 27. We now have to add some more anchored $n$-components in order to satisfy condition Q4 as well. Let $\mathfrak{C}'$ be initially empty. For any $x \in V$ where $\mu(x)$ is a non-singleton equivalence class and any two $\mathcal{C}_\alpha, \mathcal{C}_\beta \in \mathfrak{C}$ with $\delta \in \mu_\alpha(x)$ and $\delta' \in \mu_\beta(x)$, we have that, since $\delta$ and $\delta'$ are in the same $\sim$-equivalence class $\mu(x)$, there is a sequence $\delta_1, \ldots, \delta_k$ of $\mathcal{J}$-nodes with $\delta = \delta_1$ and $\delta' = \delta_k$ and every $\delta_i$ and $\delta_{i+1}$ start a descending BCP having the same path sketch. We enhance $\mathfrak{C}'$ by one anchored component per $\delta_i$ which contains just $\delta_i$ and the corresponding descending

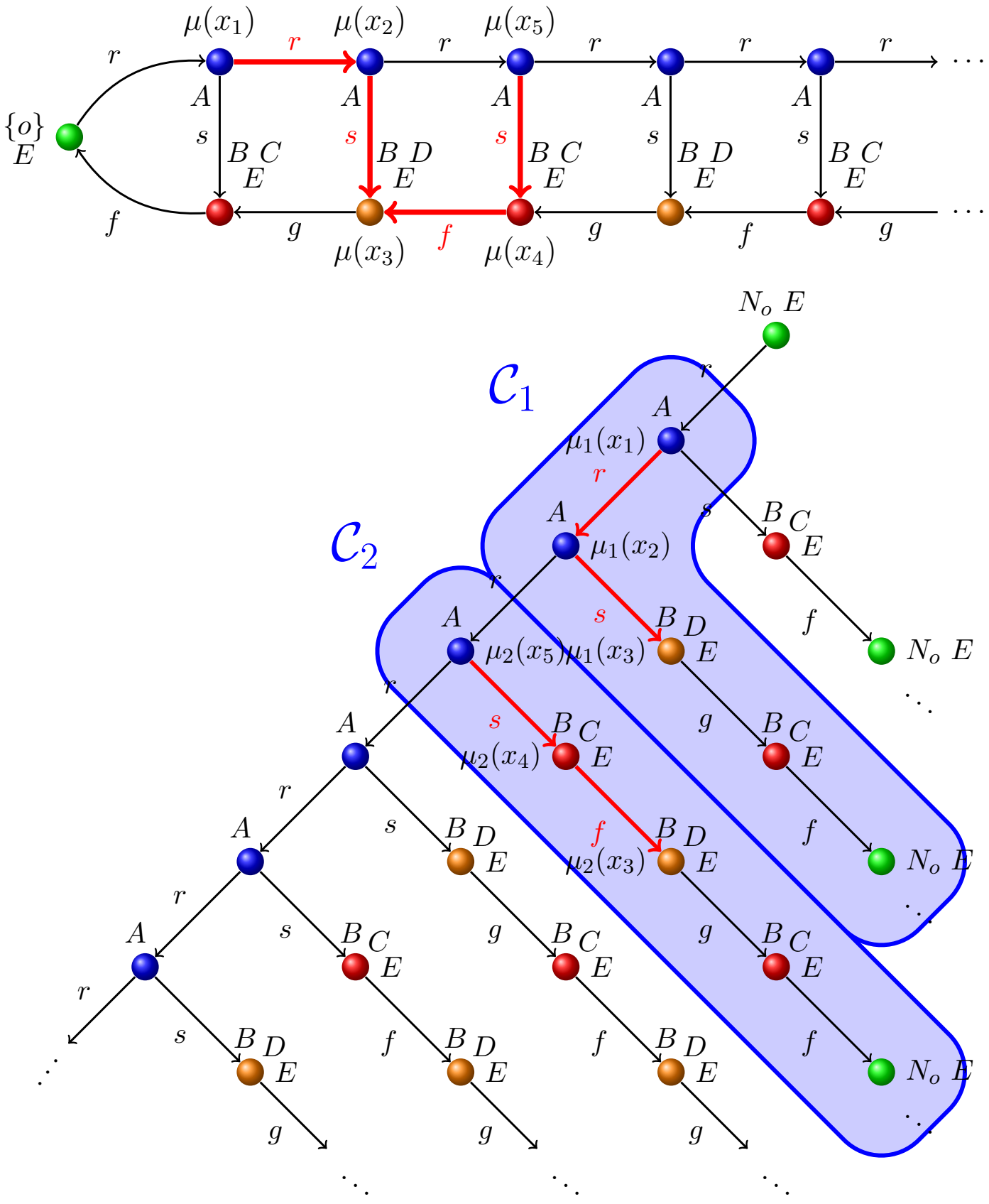

Figure 9: Correspondence between entailment and quentailment.

BCPs. Then, by construction, the elements of $\mathfrak{C} \cup \mathfrak{C}'$ constitute the necessary anchored $n$-components to justify that $\mathcal{J}$ quentails $q$ and, thus, $\mathcal{J}$ quentails $u$.

□

As an example for the correspondence between (query) entailment and quentailment, consider the conjunctive query

$$q = \{r(x_1, x_2), s(x_2, x_3), f(x_4, x_3), s(x_5, x_4)\}.$$

A match $\mu$ for this query into our example model from Figure 2 is displayed in the upper part of Figure 9, which witnesses $\mathcal{I} \models q$. In the lower part, the anchored components $\mathcal{C}_1$ and $\mathcal{C}_2$ and according functions $\mu_1$ and $\mu_2$ establish $\downarrow(\mathcal{I}) \approx q$.

## 6. Limits and Forest Transformations

Before introducing the following constructions in detail, we will try to provide some high-level explanation to convey the intuition behind the subsequent steps. As mentioned before, one of the major obstacles for a decision procedure for conjunctive query entailment is that for DLs including inverses, nominals, and cardinality restrictions (or alternatively functionality), there are potentially infinitely many so-called "new nominals": domain elements which can be identified by being linked to a "proper nominal" via a BCP.

However, we will show that for every model of a knowledge base that does not satisfy a conjunctive query (i.e., for every countermodel), there is a "nice" countermodel with only finitely many new nominals (in the subsequent section, we will then argue that this ensures the existence of a procedure that terminates when the query is not entailed by the knowledge base in question). We provide a construction which transforms an arbitrary countermodel into a "nice" one by first unraveling it into a quasi forest model, then substituting new nominals by uncritical elements and finally collapsing the result back into a proper model. For doing this, we have to find appropriate substitutes for most of the new nominals. Those substitutes have to fit in their environment without themselves introducing new nominals.

Due to the global cardinality constraints that BCPs impose on their elements, the existence of infinitely many new nominals implies that their "witnessing" BCPs must get longer and longer, such that by just looking at some finite-distance neighborhood, most of those new nominals just look like non-nominal domain elements. This state of affairs can be exploited by essentially constructing new domain elements as "environment-limits". In a way, those new domain elements are characterized by the property that they can be approximated with arbitrary precision by already present domain elements – possibly without themselves being present in the domain.[3] We will see in the following that those new domain elements can serve as the substitutes we are looking for.

**Definition 29** (Limits of a Model)**.** Let $\mathcal{I} = (\Delta^{\mathcal{I}}, \cdot^{\mathcal{I}})$ with $\delta \in \Delta^{\mathcal{I}}$ be some model of an $\mathcal{ALCOIF}b$ knowledge base $\mathcal{K}$. A tree interpretation $\mathcal{J}$ is said to be *generated by* $\delta$ (written: $\mathcal{J} \lhd \delta$), if it is isomorphic to the restriction of $\downarrow(\mathcal{I}, \delta)$ to elements of $\{(\delta, cw) \mid (\delta, cw) \in \Delta^{\downarrow(\mathcal{I},\delta)}, c \notin H\}$ for some $H \subseteq \mathbb{N}$.

The set of *limits* of $\mathcal{I}$ (written $\lim \mathcal{I}$) is the set of all tree interpretations $\mathcal{J}$ such that for every $k \in \mathbb{N}$, there are infinitely many $\delta \in \Delta^{\mathcal{I}}$ such that $\mathsf{cut}_k(\mathcal{L}) \cong \mathsf{cut}_k(\mathcal{J})$ for some $\mathcal{L} \lhd \delta$. △

Figure 10 displays one limit element of our example model.

The following lemma gives some useful properties of limits. Besides some rather obvious compatibility considerations with respect to knowledge base satisfaction, claim 3 of Lemma 30 provides us with the as pleasant as useful insight that the root node of a limit can never be part of a BCP at all.

**Lemma 30.** *Let $\mathcal{K}$ be an $\mathcal{ALCOIF}b$ knowledge base, $\mathcal{K}' = \mathsf{nomFree}(\mathcal{K})$, $\mathcal{I}$ a purified model of $\mathcal{K}$, and $n$ some fixed natural number. Then the following hold:*

1. *Let $\mathcal{L}'$ be a tree interpretation such that there are infinitely many $\delta \in \Delta^{\mathcal{I}}$ with $\mathcal{L}' \cong \mathsf{cut}_n(\mathcal{L})$ for some $\mathcal{L} \lhd \delta$. Then, there is at least one limit $\mathcal{J} \in \lim \mathcal{I}$ such that $\mathsf{cut}_n(\mathcal{J}) \cong \mathcal{L}'$.*

2. *Every $\mathcal{J} \in \lim \mathcal{I}$ is locally $\mathcal{K}'$-consistent apart from its root $(\rho, \varepsilon)$.*

3. As an analogy, consider the fact that any real number can be approximated by a sequence of rational numbers, even if it is itself irrational.

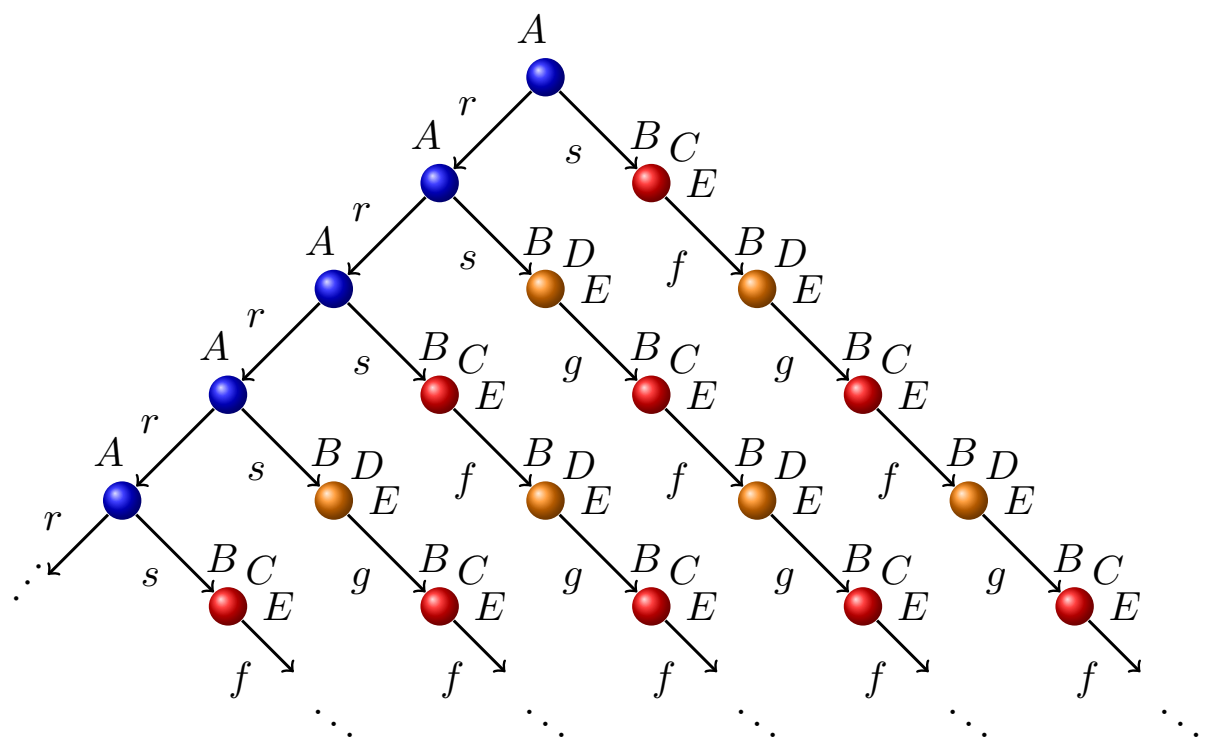

Figure 10: One limit for the model from Fig. 2

3. *For every $\mathcal{J} \in \lim \mathcal{I}$ it holds that its root $(\rho, \varepsilon)$ has no BCP to any $(\rho, w) \in \Delta^{\mathcal{J}}$.*

4. *If $\mathcal{J} \in \lim \mathcal{I}$ contains a node $\delta$ starting two backwards counting paths with path sketches $s_1$ and $s_2$, then for any element $\delta^*$ in any unraveling holds: if the direct neighborhood of $\delta^*$ is isomorphic to that of $\delta$ and $\delta^*$ starts a descending BCP with path sketch $s_1$ then it also starts a descending BCP with path sketch $s_2$.*

5. *Every $\mathcal{J} \in \lim \mathcal{I}$ is collapsing-admissible.*

*Proof.* 1. Let $b$ be the branching degree of $\mathcal{I}$, let $D_n$ be the (by assumption infinite) set of all $\delta \in \Delta^{\mathcal{I}}$ such that $\mathcal{L}' \cong \mathsf{cut}_n(\mathcal{L})$ for some $\mathcal{L} \lhd \delta$, and let $\mathfrak{I}_n$ contain all those $\mathcal{L}$. Starting with $k = n$, we now iteratively increase $k$ and construct sets $\mathfrak{I}_k$ and $D_k$ and tree interpretations $\mathcal{L}_k$. On our way, we inductively prove some properties.

By induction hypothesis we know that $D_k$ is infinite and there is some $\mathcal{L}_k$ with $\mathcal{L}_k \cong \mathsf{cut}_k(\mathcal{M})$ for all $\mathcal{M} \in \mathfrak{I}_k$. By Lemma 25, there are only finitely many non-isomorphic tree interpretations of depth $k+1$ with branching degree $b$, and we can partition $\mathfrak{I}_k$ into finitely many sets $\mathfrak{I}_{k,1}, \ldots, \mathfrak{I}_{k,m}$ such that every two $\mathcal{M}, \mathcal{M}'$ from any $\mathfrak{I}_{k,i}$ satisfy $\mathsf{cut}_{k+1}(\mathcal{M}) \cong \mathsf{cut}_{k+1}(\mathcal{M}')$. Likewise, we can define classes $D_{k,1}, \ldots, D_{k,m}$ with $D_k = D_{k,1} \cup \ldots \cup D_{k,m}$ such that $\delta \in D_{k,i}$ if there is an $\mathcal{L} \lhd \delta$ with $\mathcal{L} \in \mathfrak{I}_{k,i}$. Now, as $D_k$ is infinite, one of the $D_{k,i}$ must be infinite as well and we can set $D_{k+1} = D_{k,i}$ and $\mathfrak{I}_{k+1} = \mathfrak{I}_{k,i}$. Hence, we know that $D_{k+1}$ is infinite and there is some $\mathcal{L}_{k+1}$ with $\mathcal{L}_{k+1} \cong \mathsf{cut}_{k+1}(\mathcal{M})$ for all $\mathcal{M} \in \mathfrak{I}_{k+1}$.

Thus, we have established an infinite sequence $\mathcal{L}_n, \mathcal{L}_{n+1}, \ldots$ with $\mathcal{L}_i \cong \mathsf{cut}_i(\mathcal{L}_j)$ for all $j > i$. Without loss of generality, we can assume that isomorphic nodes are named identically, i.e., we even have $\mathcal{L}_i = \mathsf{cut}_i(\mathcal{L}_j)$ for all $j > i$. Now we can define $\mathcal{J}$ as the (non-disjoint) union of all $\mathcal{L}_i$. This way we have established the structure $\mathcal{J}$ for which $\mathsf{cut}_k(\mathcal{J}) = \mathcal{L}_k$ and we know that for every $k$ there are infinitely many $\delta$ (namely all elements from $D_k$) such that $\mathsf{cut}_k(\mathcal{L}) \cong \mathcal{L}_k$ for some $\mathcal{L} \lhd \delta$. Hence this $\mathcal{J}$ is the limit element with the desired properties.

2. Let $(\rho, w)$ with $w \neq \varepsilon$ be a node in $\mathcal{J}$. Now choose a $\delta \in \Delta^{\mathcal{I}}$ such that $\mathsf{cut}_{|w|+1}(\mathcal{L}) \cong \mathsf{cut}_{|w|+1}(\mathcal{J})$ for some $\mathcal{L} \lhd \delta$ (by definition, there are even infinitely many such $\delta$s to

choose from). Then $\mathcal{L}$ contains a node $\delta^*$ whose direct neighborhood is isomorphic to that of $(\rho, w)$. However, as $\mathcal{L}$ is contained in $\downarrow(\mathcal{I}, \delta)$ and $\mathcal{I} \models \mathcal{K}$ by assumption, it is locally $\mathcal{K}'$-consistent and hence $\delta^*$ is. Therefore $(\rho, w)$ is locally $\mathcal{K}'$-consistent in $\mathcal{J}$.

3. Assume the contrary, i.e., that some $\mathcal{J} \in \lim \mathcal{I}$ has a BCP from the root $(\rho, \varepsilon)$ to some $(\rho, w) \in \Delta^{\mathcal{J}}$ with $(\rho, w) \in N_o^{\mathcal{J}}$ for some $o \in \mathsf{nom}(\mathcal{K})$. Since we have only functionality and by definition of BCPs, a BCP uniquely identifies one domain individual. By definition of $\lim \mathcal{I}$, however, there are infinitely many $\delta \in \Delta^{\mathcal{I}}$ satisfying $\mathsf{cut}_{|w|}(\mathcal{L}) \cong \mathsf{cut}_{|w|}(\mathcal{J})$ for some $\mathcal{J} \lhd \delta$ and we have an infinite number of individuals with the same counting path to $o^{\mathcal{I}}$. This is a contradiction.

4. Choose $k$ to be the maximum length of the two BCPs. By definition of the limit, $\mathcal{I}$ contains an element $\gamma$ such that $\mathsf{cut}_{|w|+k}(\mathcal{L}) \cong \mathsf{cut}_{|w|+k}(\mathcal{J})$ for some $\mathcal{L} \lhd \gamma$. Now, let $\delta' \in \Delta^{\mathcal{L}}$ be the element that (with respect to this isomorphism) corresponds to $\delta \in \Delta^{\mathcal{J}}$. Then, $\delta'$ is the origin of two descending BCPs with path sketches $s_1$ and $s_2$. Let $\mathsf{Tail}(\delta') = \gamma'$. Since path sketches of descending BCPs uniquely identify one domain individual, every node $\delta^*$ in any unraveling that starts a descending BCP with path sketch $s_1$ must have been caused by $\gamma'$ as well. Furthermore (as their direct neighborhoods are isomorphic and by the specific design of the $\mathsf{choose}$ function from Definition 12 which renders all successors non-isomorphic), all successors of $\delta^*$ uniquely correspond to neighbors of $\gamma'$ and in turn to successors of $\delta'$.

   This in turn implies that, for every successor of $\delta^*$, one finds a successor of $\delta'$ with isomorphic direct neighbourhood. Yet, this synchronicity argument can be inductively applied and thereby iterated down the BCP. Thus, we obtain that $\delta^*$ also starts a descending BCP with path sketch $s_2$.

5. We define the function $\mathsf{ch}\colon (\mathsf{cl}(\mathcal{K}) \times \Delta^{\mathcal{J}}) \to \Delta^{\mathcal{J}}$ essentially like in the proof of Lemma 19, namely by referring to the function $\mathsf{choose}$. For a given element $\delta \in \Delta^{\mathcal{J}}$ that starts a BCP of length $\ell$ in $\mathcal{J}$, choose a $\delta' \in \Delta^{\mathcal{I}}$ such that $\mathsf{cut}_{|\delta|+\ell}(\mathcal{L}) \cong \mathsf{cut}_{|\delta|+\ell}(\mathcal{J})$ for some $\mathcal{L} \lhd \delta'$. As $\mathcal{L}$ is contained in $\downarrow(\mathcal{I}, \delta')$, we can now proceed and define $\mathsf{ch}(C, \delta)$ as demonstrated in the proof of Lemma 19.

□

Having defined limit elements as convenient building blocks for restructuring forest quasi-interpretations, the following definition provides the first hints on *where* inside such a structure one existing node (and all its successors) can safely be exchanged by a limit element.

**Definition 31** ($n$-Secure Replacement)**.** Let $\mathcal{K}$ be an $\mathcal{ALCOIF}b$ knowledge base, $\mathcal{I}$ a model for $\mathcal{K}$, $\mathcal{J}$ some forest quasi-model for $\mathcal{K}$ with $\delta \in \Delta^{\mathcal{J}}$. A strict tree quasi-interpretation $\mathcal{J}' \in \lim \mathcal{I}$ is called an *$n$-secure replacement* for $\delta$ if

- $\mathsf{cut}_n(\downarrow(\mathcal{J}, \delta))$ is isomorphic to $\mathsf{cut}_n(\mathcal{J}')$,
- for every anchored $n$-component of $\mathcal{J}'$ with witness $\delta'$, there is an isomorphic anchored $n$-component of $\mathcal{J}$ with witness $\delta$.

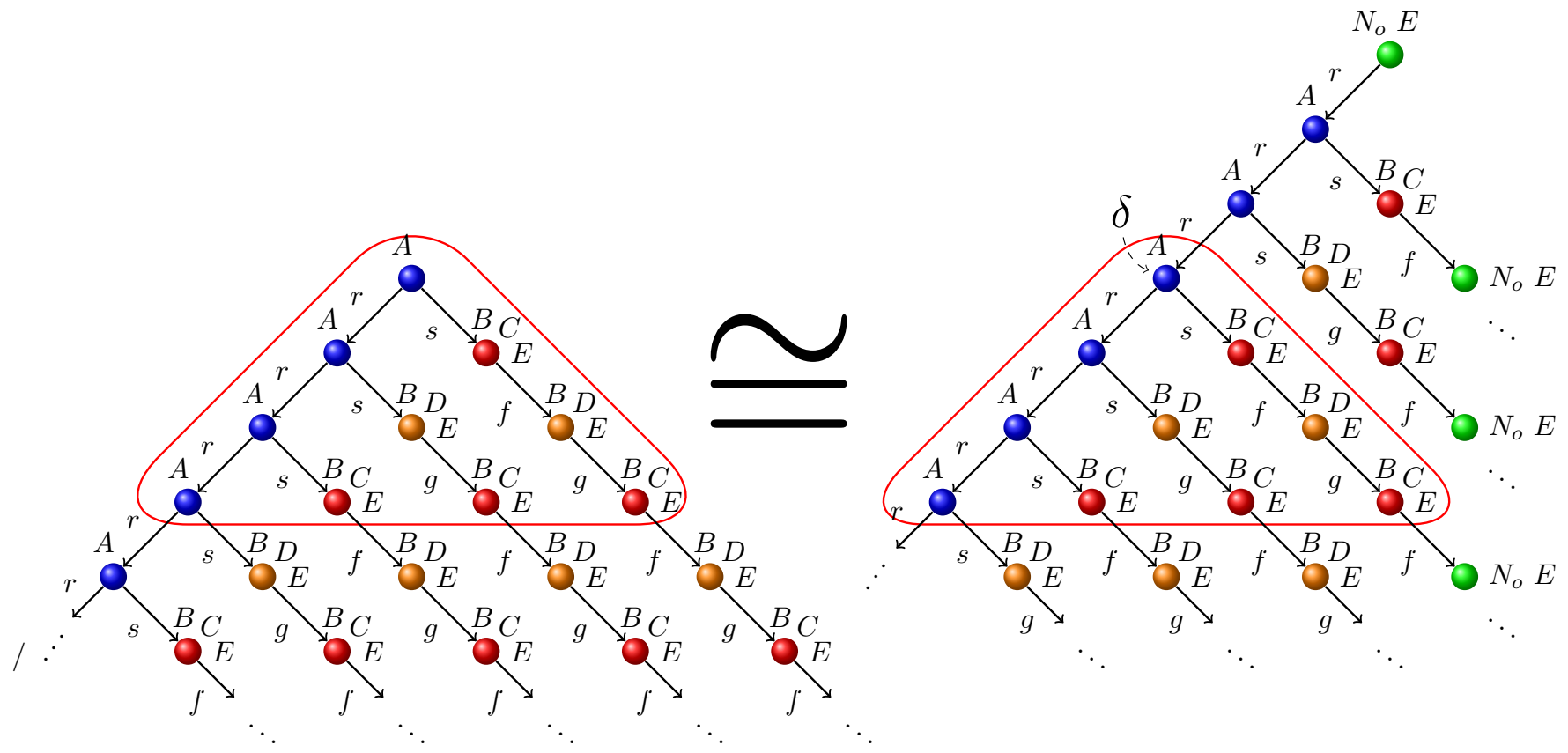

Figure 11: Forest quasi-model (right) and according 3-secure replacement for $\delta$ (left).

If a $\delta \in \Delta^{\mathcal{J}}$ has an $n$-secure replacement in $\lim \mathcal{I}$, we call $\delta$ *n-replaceable* w.r.t. $\mathcal{I}$, otherwise we call $\delta$ *n-irreplaceable* w.r.t. $\mathcal{I}$. △

Figure 11 displays a 3-secure replacement in the considered unraveling of our example model.

After having defined which elements of a forest quasi-model are eligible for being replaced by a limit element, we have to make sure that not too many elements (actually defined in terms of the original model) are exempt from being replaced.

**Lemma 32.** *Every purified model $\mathcal{I}$ of an $\mathcal{ALCOIFb}$ knowledge base $\mathcal{K}$ contains only finitely many distinct elements that start a BCP and are the cause for $n$-irreplaceable nodes in the unraveling of $\mathcal{I}$.*

*Proof.* Assume the converse: let a purified model $\mathcal{I}$ of $\mathcal{K}$ contain an infinite set $D$ of elements giving rise to $n$-irreplaceable nodes in $\downarrow(\mathcal{I})$. Then there must be an $\mathcal{L}'$ such that there is an infinite set $D' \subseteq D$ such that every $d' \in D'$ generates an $\mathcal{L}$ for which $\mathsf{cut}_n(\mathcal{L}) \cong \mathcal{L}$ (since by Lemma 25, there are only finitely many non-isomorphic choices for $\mathcal{L}'$). This set $D'$ can be used to guide the construction of a specific limit element $\mathcal{J} \in \lim \mathcal{I}$ according to Lemma 30.1. Now, for an element $(\rho, w)$ from $\mathcal{J}$ starting a BCP, let $l_{(\rho,w)} \in \mathbb{N}$ be the length of the shortest such BCP starting from $(\rho, w)$. Then, let $k$ be the maximum over all $l_{(\rho,w)}$ of individuals $(\rho, w)$ from $\mathcal{J}$ that start a BCP and for which $|w| \leq n$. By construction, $D'$ contains one element $d''$ generating an $\mathcal{L}$ with $\mathsf{cut}_k(\mathcal{L}) \cong \mathsf{cut}_k(\mathcal{J})$ (actually infinitely many). By the choice of $k$ and Lemma 30.4, we can conclude that $\mathcal{J}$ is an $n$-secure replacement for the irreplaceable $\downarrow(\mathcal{I})$-node caused by $d''$ which contradicts the fact that $d'' \in D$. □

Now we know, which elements of a forest quasi-model can be replaced by a suitable limit element. The following definition exactly tells us, how such a replacement is carried out: the respective element and all its successors are deleted and the limit element (together with its successors) is inserted at the same position.

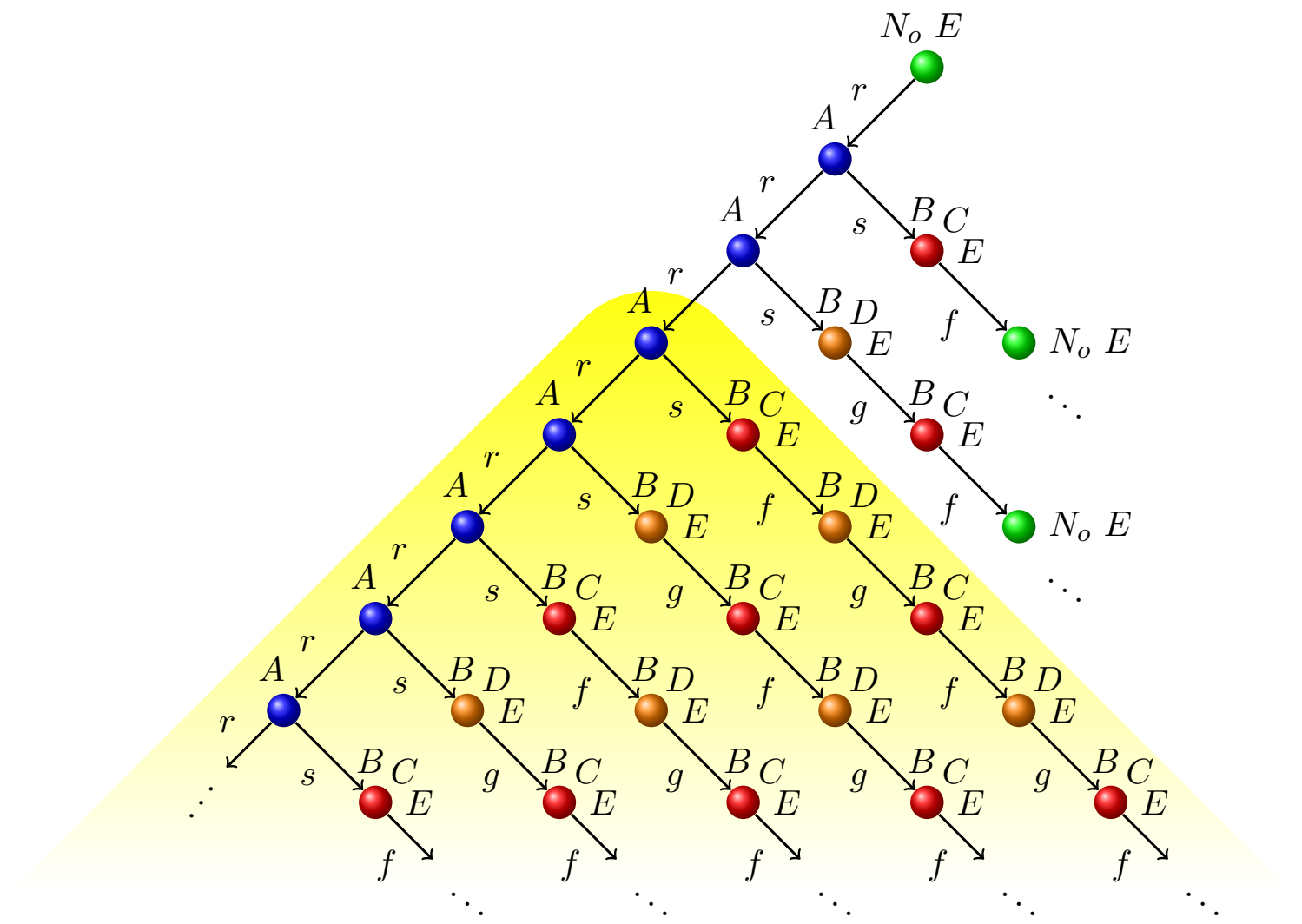

Figure 12: Result of replacing the element $\delta$ by the 3-secure replacement depicted in Figure 11. The inserted component is highlighted.

**Definition 33** (Replacement Step)**.** Let $\mathcal{K}$ be an $\mathcal{ALCOIF}b$ knowledge base, $\mathcal{I}$ a model of $\mathcal{K}$, and $\mathcal{J}$ a forest quasi-model of $\mathcal{K}$, i.e., $\mathcal{J} \models \mathcal{K}' = \mathsf{nomFree}(\mathcal{K})$. Let $(\rho, w) \in \Delta^{\mathcal{J}}$ be $n$-replaceable w.r.t. $\mathcal{I}$ and $\mathcal{J}'$ an according $n$-replacement for $(\rho, w)$ from $\lim \mathcal{I}$ with root $(\varsigma, \varepsilon)$.

We define the result of replacing $(\rho, w)$ by $\mathcal{J}'$ as the interpretation $\mathcal{R}$ where

- $\Delta^{\mathcal{R}} = \Delta^{\mathcal{J}}_{\mathsf{red}} \cup \{(\rho, ww'') \mid (\varsigma, w'') \in \Delta^{\mathcal{J}'}\}$ with $\Delta^{\mathcal{J}}_{\mathsf{red}} = (\Delta^{\mathcal{J}} \setminus \{(\rho, ww') \mid |w'| > 1\})$
- for each $A \in \mathsf{con}(\mathcal{K}')$, $A^{\mathcal{R}} = (A^{\mathcal{J}} \cap \Delta^{\mathcal{J}}_{\mathsf{red}}) \cup \{(\rho, ww') \mid (\varsigma, w') \in A^{\mathcal{J}'}\}$
- for each $r \in \mathsf{rol}(\mathcal{K}')$, $r^{\mathcal{R}} = (r^{\mathcal{J}} \cap \Delta^{\mathcal{J}}_{\mathsf{red}} \times \Delta^{\mathcal{J}}_{\mathsf{red}}) \cup \{\langle(\rho, ww'), (\rho, ww'')\rangle \mid \langle(\varsigma, w'), (\varsigma, w'')\rangle \in r^{\mathcal{J}'}\}$

$\triangle$

Figure 12 displays the result of carrying out this replacement step on our example.

The following lemma assures that during a replacement as described above, no new anchored $n$-components are introduced, instead all anchored $n$-components present after an $n$-secure transformation were present before or completely contained in the inserted limit element.

**Lemma 34.** *Let $\mathcal{K}$ be an $\mathcal{ALCOIF}b$ knowledge base, $\mathcal{I}$ a model for $\mathcal{K}$, $\mathcal{J}$ a forest quasi-model for $\mathcal{K}$, i.e., $\mathcal{J} \models \mathcal{K}' = \mathsf{nomFree}(\mathcal{K})$, and let $(\rho, w) \in \Delta^{\mathcal{J}}$ be $n$-replaceable w.r.t. $\mathcal{I}$. Let $\mathcal{J}'$ be an $n$-replacement for $(\rho, w)$ with root $(\varsigma, \varepsilon)$ and $\mathcal{R}$ be the result of replacing $(\rho, w)$ by $\mathcal{J}'$. Then the following hold:*

1. $\mathsf{cut}_n(\downarrow(\mathcal{J}, (\rho, w)))$ *is isomorphic to* $\mathsf{cut}_n(\downarrow(\mathcal{R}, (\rho, w)))$.

2. *If $n \geq 1$, then $\mathcal{R}$ is locally $\mathcal{K}'$-consistent.*

3. *Whenever $\mathcal{R}$ contains an anchored $n$-component $\mathcal{C}$, then one of $\mathcal{J}$ or $\mathcal{J}'$ contains an anchored $n$-component isomorphic to $\mathcal{C}$.*

*Proof.* 1. This is a direct consequence from Definitions 31 and 33.

2. We make a case distinction when element-wise investigating local consistency of $\mathcal{R}$ (note that $\mathcal{K}$ and $\mathcal{K}'$ are simplified and that local consistency of a node $(\rho, v) \in \Delta^{\mathcal{R}}$ depends only on this node and its direct neighbors):

   - $v = ww'$ for some $w' \neq \varepsilon$: then the direct neighborhood of $(\rho, v)$ in $\mathcal{R}$ is isomorphic to the direct neighborhood of $(\varsigma, w')$ in $\mathcal{J}'$ (recall that $(\varsigma, \varepsilon)$ is the root of $\mathcal{J}'$). By Lemma 30.2, $\mathcal{J}'$ is locally $\mathcal{K}'$-consistent except possibly for $(\varsigma, \varepsilon)$. Hence also $(\rho, v)$ is locally $\mathcal{K}'$-consistent in $\mathcal{R}$.
   - $v \neq ww'$ for any $w'$, i.e., $(\rho, v)$ was not affected by the replacement: then the direct neighborhood of $(\rho, v)$ has not changed by the replacement and, therefore, the neighborhoods of $(\rho, v)$ in $\mathcal{J}$ and $\mathcal{R}$ coincide. As $\mathcal{J}$ is locally $\mathcal{K}'$-consistent by assumption, so is $(\rho, v)$ in $\mathcal{R}$.
   - $v = w$: in that case, the direct neighborhood of $(\rho, v)$ has changed but remained isomorphic. This follows from the preceding statement (34.1).

3. Let $(\rho', w')$ be the witness of $\mathcal{C}$. We distinguish three cases:

   - $\rho' = \rho$ and $w$ is a prefix of $w'$. Then, clearly $\mathcal{C}$ is completely contained in $\mathcal{J}'$.
   - $\rho' = \rho$ and $w'$ is a prefix of $w$. Let $\mathcal{C}'$ be the structure obtained by restricting $\mathcal{C}$ to all elements of the form $(\rho, ww'')$ and then renaming every element $(\rho, ww'')$ to $(\varsigma, w'')$, where $(\varsigma, \varepsilon)$ is the root of $\mathcal{J}'$. Then $\mathcal{C}'$ is an anchored $n$-component in $\mathcal{J}'$ with witness $(\varsigma, \varepsilon)$. Now, by definition of replacing, $\mathcal{J}$ must contain an isomorphic copy of $\mathcal{C}'$ with witness $(\rho, w)$. Since the other part of $\mathcal{C}$ (consisting of those nodes $(\rho', w')$ such that $w$ is not a prefix of $w'$) has not been altered by the replacement, we can conclude that $\mathcal{J}$ must contain an isomorphic copy of $\mathcal{C}$.
   - Neither of the above. Then, $(\rho', w')$ and the subtree rooted in $(\rho', w')$ is contained in $\mathcal{J}$ as this part of $\mathcal{J}$ has not been affected by the replacement. Then, clearly also $\mathcal{C}$ is contained in $\mathcal{J}$.

□

We are now ready for defining the whole process of restructuring a forest quasi-model essentially by substituting as many nodes as possible by appropriate limit elements.

**Definition 35** ($n$-Secure Transformation)**.** Let $\mathcal{I}$ be a model of some $\mathcal{ALCOIF}b$ knowledge base $\mathcal{K}$ and $\mathcal{J}$ an unraveling for $\mathcal{I}$. An interpretation $\mathcal{J}'$ is called an *$n$-secure transformation* of $\mathcal{J}$ if it is obtained by (possibly infinitely) repeating the following step:

Choose one unvisited and w.r.t. tree-depth minimal node $(\rho, w)$ that is $n$-replaceable w.r.t. $\mathcal{I}$. Replace $(\rho, w)$ with one of its $n$-secure replacements from $\lim \mathcal{I}$ and mark $(\rho, w)$ as visited. △

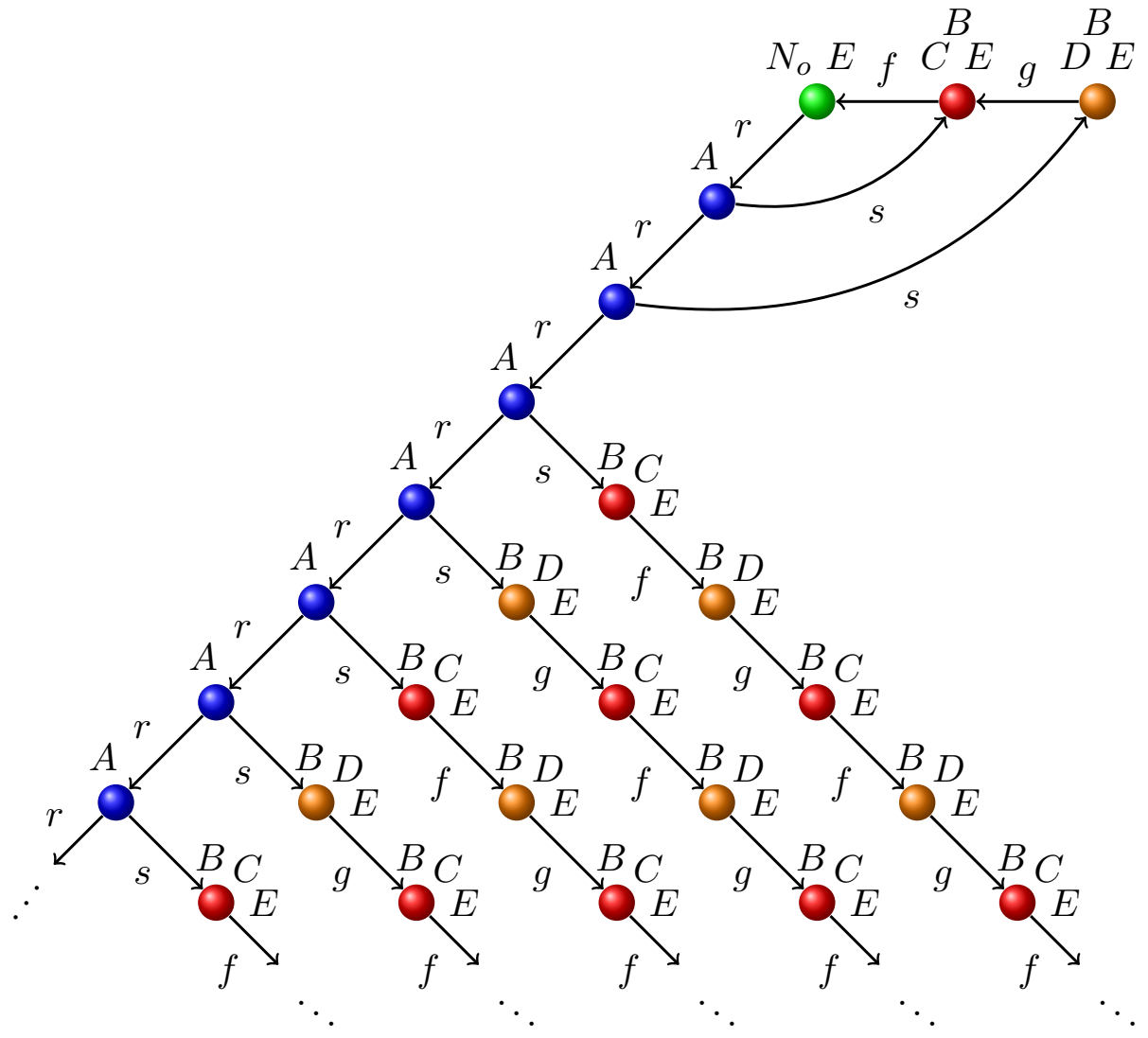

Figure 13: Result of collapsing the forest quasi-model displayed in Figure 12.

Note that this is well-defined as every node is visited at most once and no formerly irreplaceable node ever becomes replaceable. Hence for every $k \in \mathbb{N}$, the "initial segment" $\mathsf{cut}_k(\overline{\mathcal{J}})$ of the current intermediate structure $\overline{\mathcal{J}}$ is already isomorphic to the initial segment $\mathsf{cut}_k(\mathcal{J}')$ of $\mathcal{J}'$ after a bounded number of replacement steps, due to the fact that all involved structures have bounded branching degree.

By now, the whole effort might still look a bit contrived and pointless, however, the following lemma establishes a bunch of properties that in the end allow us to deduce the existence of a very well-behaved countermodel whenever there is any at all.

We show that the process of unraveling, $n$-secure transformation and collapsing preserves the property of being a model of a knowledge base and (with the right choice of $n$) also preserves the property of *not* entailing a conjunctive query. Moreover, this model conversion process ensures that the resulting model contains only finitely many new nominals (witnessed by a bound on the length of BCPs). Figure 13 illustrates these properties for our example model. Note that only two new nominals are left whereas collapsing the original unraveling yields infinitely many.

**Lemma 36.** *Let $\mathcal{I}$ be a purified model of some $\mathcal{ALCOIF}b$ knowledge base $\mathcal{K}$, $\mathcal{J}$ an unraveling of $\mathcal{I}$, and $\mathcal{J}'$ an $n$-secure transformation of $\mathcal{J}$. Then the following hold:*

1. *$\mathcal{J}'$ is a strict forest quasi-model for $\mathcal{K}$.*
2. *$\mathcal{J}'$ is collapsing-admissible.*
3. $\mathsf{collapse}(\mathcal{J}')$ *is a model of $\mathcal{K}$.*
4. *There is a natural number $m$ such that $\mathcal{J}'$ does not contain any node whose shortest descending BCP has a length greater than $m$.*

5. *If $\mathcal{J}'$ contains an anchored $n$-component $\mathcal{C}$, then $\mathcal{J}$ contains an anchored $n$-component isomorphic to $\mathcal{C}$.*

6. *If, for some union of conjunctive queries $u = q_1 \vee \ldots \vee q_h$, we have $\mathcal{J} \not\approx u$ and $n > \max_{q \in \{q_1,\ldots,q_h\}} \sharp(q)$, then $\mathcal{J}' \not\approx u$.*

7. *If, for some union of conjunctive queries $u = q_1 \vee \ldots \vee q_h$, we have $\mathcal{I} \not\models u$ and $n > \max_{q \in \{q_1,\ldots,q_h\}} \sharp(q)$, then collapse$(\mathcal{J}') \not\models u$.*

*Proof.* 1. Let $\mathcal{K}' = \mathsf{nomFree}(\mathcal{K})$. Due to Lemma 13, $\mathcal{J}$ is a strict forest quasi-model for $\mathcal{K}$. By Lemma 34.2, each replacement step preserves local $\mathcal{K}'$-consistency and results, thus, in a forest quasi-model for $\mathcal{K}$. Since each $n$-replacement is a strict tree quasi-interpretation also strictness is preserved. By induction it follows that every interpretation produced in the $n$-secure transformation procedure is a strict forest quasi-model for $\mathcal{K}$. For every node in $\mathcal{J}'$, its direct predecessor and direct successors have not changed any more after finitely many replacement steps and local $\mathcal{K}'$-consistency depends solely on those neighbors. Hence $\mathcal{J}'$ is also locally $\mathcal{K}'$-consistent.

2. By Lemma 19, $\mathcal{J}$ is collapsing-admissible, by Lemma 30.5 every limit of $\mathcal{I}$ is. Moreover, as is obvious from the proofs of both propositions, it is possible to define the respective $\mathsf{ch}$-functions recurring to the original $\mathsf{choose}$-function on $\mathcal{I}$, hence every two elements from (even different) unravelings or limits that start descending BCPs with identical path sketches correspond to the very same element in $\mathcal{I}$ whence the separate $\mathsf{ch}$-functions are compatible with each other. Therefore, replacing an element in the unraveling yields a strict forest quasi-model that is collapsing-admissible. Applying the same argument inductively yields that every intermediate strict forest quasi-model during the $n$-secure transformation is collapsing-admissible. Finally, as the according $\mathsf{ch}$-function stabilizes after finitely many replacement steps (together with the neighborhood of the considered elements), also $\mathcal{J}'$ is collapsing-admissible.

3. This follows from the two previous facts (36.1 and 36.2) together with Lemma 20.

4. Consider the set $D$ of all $\delta \in \Delta^{\mathcal{I}}$ causing $n$-irreplaceable nodes in $\mathcal{J}$. By Lemma 32, $D$ is finite. We obtain $D'$ by removing all $\delta$ from $D$ that do not start any descending BCPs.

   For $\delta \in D'$, let dBCP$(\delta)$ denote the set of descending BCPs starting in $\delta$ and choose

$$m := \max_{\delta \in D'} \left( \min_{p \in \mathrm{dBCP}(\delta)} |p| \right)$$

   Now assume there were a $\delta' \in \Delta^{\mathcal{I}}$ having a shortest descending BCP of length greater than $m$. Obviously, as $\delta' \notin D'$, there must be a $(\rho, w)$ generated by $\delta'$ that is $n$-replaceable. However, during the $n$-secure transformation all $n$-replaceable elements have been replaced by elements that do not start any descending BCPs at all due to Lemma 30.3.

5. We prove this by induction on the replacement steps of the $n$-secure transformation process by showing that this is true for every intermediate replacement result $\mathcal{R}'$.

   The claim for $\mathcal{J}'$ then follows from the fact that, for every considered $\mathcal{C}$ (which is always finite), only a finite part $\mathsf{cut}_\ell(\mathcal{J}')$ is relevant and that for every $\ell$, there is a bounded number of replacement steps after which we have $\mathsf{cut}_\ell(\mathcal{R}') = \mathsf{cut}_\ell(\mathcal{J}')$ for every further intermediate $\mathcal{R}'$.

   As base case (zero replacement steps carried out), we find that for $\mathcal{R}' = \mathcal{J}$, the claim is trivially true.

   Now assume that the claim has been established for $\mathcal{R}$ and has to be shown for $\mathcal{R}'$ that is created by replacing $(\rho, w)$ in $\mathcal{R}$ with some $\mathcal{J}''$. By Lemma 34.3, we then know that one of the following is the case:

   - $\mathcal{R}$ contains $\mathcal{C}$. Yet, we can apply the induction hypothesis and conclude that also $\mathcal{J}$ contains $\mathcal{C}$ as claimed.
   - $\mathcal{J}''$ contains $\mathcal{C}$. But, since $\mathcal{C}$ is finite, it is already contained in $\mathsf{cut}_k(\mathcal{J}'')$ for some $k \in \mathbb{N}$ and, as $\mathcal{J}''$ is a limit element, we find one $\delta \in \Delta^\mathcal{I}$ with $\mathsf{cut}_k(\downarrow(\mathcal{I}, \delta)) = \mathsf{cut}_k(\mathcal{J}'')$. Since $\mathcal{I}$ is purified, we find a $(\rho, w) \in \Delta^\mathcal{J}$ that corresponds to $\delta$, i.e., $\mathcal{J}$ contains an isomorphic copy of $\downarrow(\mathcal{I}, \delta)$ which in turn contains an isomorphic copy of $\mathcal{C}$.

6. This is actually a straightforward consequence from the preceding proposition and the definition of quentailment.

   For the indirect proof, we suppose $\mathcal{J} \not\approx u$ and $n > \max_{q \in \{q_1,\ldots,q_h\}} \sharp(q)$ and $\mathcal{J}' \approx u$, the latter witnessed by $\mathcal{J}' \approx q$ for a $q \in \{q_1, \ldots, q_h\}$. By definition, the latter assures the existence of adequate anchored $n$-components in $\mathcal{J}'$. Then, applying the preceding proposition (36.5), we obtain that isomorphic copies of all those anchored $n$-components are contained in $\mathcal{J}$ which, by definition, just means $\mathcal{J} \approx q$ and, therefore, $\mathcal{J} \approx u$. Hence, we have a contradiction, which proves the claim.

7. We prove this indirectly, so assume $\mathcal{I} \not\models u$, $n > \max_{q \in \{q_1,\ldots,q_h\}} \sharp(q)$, and $\mathsf{collapse}(\mathcal{J}') \models u$, witnessed by $\mathsf{collapse}(\mathcal{J}') \models q$ for a $q \in \{q_1, \ldots, q_h\}$.

   Then, from Lemma 28.2, it follows that $\mathcal{J}' \approx q$. By the previous proposition (36.6), we conclude $\mathcal{J} \approx q$, which in turn implies $\mathcal{I} \models q$ by Lemma 28.1. This implies $\mathcal{I} \models u$, a contradiction.

□

Now we are able to establish our first milestone on the way to showing finite representability of countermodels.

**Theorem 37.** *For every $\mathcal{ALCOIF}b$ knowledge base $\mathcal{K}$ with $\mathcal{K} \not\models u$, there is a forest model $\mathcal{I}$ of $\mathcal{K}$ with finitely many roots such that $\mathcal{I} \not\models u$. Moreover, $\mathcal{I}$ has bounded branching degree.*

*Proof.* Let $u = q_1 \vee \ldots \vee q_h$. Since an inconsistent knowledge base entails every query, we can assume that $\mathcal{K}$ is consistent and, since $\mathcal{K} \not\models u$, there is a model $\mathcal{I}$ of $\mathcal{K}$ with $\mathcal{I} \not\models u$. Choose an $n > \max_{q \in \{q_1,\ldots,q_h\}} \sharp(q)$ and let $\mathcal{J}'$ be obtained by carrying out an $n$-secure transformation on $\downarrow(\mathcal{I})$ and let $\mathcal{I}' = \mathsf{collapse}(\mathcal{J}')$. We know that $\mathcal{I}'$ is a model of $\mathcal{K}$ (via Lemma 36.3) and that $\mathcal{I}' \not\models u$ (by Lemma 36.7).

By Lemma 36.4, we know that there is a fixed natural number $m$ such that the shortest descending BCP started by any node in $\mathcal{J}'$ is shorter than $m$. Note that there are only finitely many path sketches of length $\leq m$. This means that every node in $\mathcal{J}'$ that starts a descending BCP at all can be assigned to one such path sketch. However, this entails that there are only finitely many elements (i.e., $\sim$-equivalence classes) in $\mathcal{I}'$ that contain $\mathcal{J}'$-elements starting descending BCPs in $\mathcal{J}'$. This implies, via Lemma 24, that $\mathcal{I}'$ contains only finitely many roots.

The fact that $\mathcal{I}'$ has bounded branching degree is a direct consequence from the fact that the initial unraveling has bounded branching degree, that replacement do not change the branching degree nor do collapsings as assured by Lemma 20. □

## 7. Finite Representations of Models

In this section, we show how we can construct a finite representation of a forest model of a knowledge base that has only a finite number of roots. We then show that these finite representations can be used to check query entailment. In order to do this, we use a technique that is very similar to the blocking techniques used in tableau algorithms (see, e.g., Horrocks & Sattler, 2007). A tableau algorithm builds a so-called completion graph that is a finite representation of a model. A completion graph has essentially the same structure as our forest quasi-models. It contains root nodes for the nominals occurring in the input knowledge base plus further root nodes for new nominals that start BCPs. Each (new and old) nominal is the root of a tree, and relations only occur between direct neighbors within a tree, between elements within a tree and a root, or between the roots. An initial completion graph contains only nodes for the nominals occurring in the input knowledge base. Concepts are expanded according to a set of expansion rules, and new nodes are added to the graph when expanding existential restrictions. New nominals are added by the so-called NN-rule whenever an element from within a tree has a relationship with an inverse functional role to a root node that represents a nominal from the input knowledge base, i.e., when a BCP is created. In order to obtain a finite representation, tableau algorithms usually employ some cycle detection mechanism, called blocking. Otherwise the depth of the trees and the number of new nominals might grow infinitely. For logics as expressive as $\mathcal{ALCOIF}b$, blocking usually requires two pairs of elements. In our notation, a (non-root) node $n$ with predecessor $n'$ blocks a node $m$ with predecessor $m'$, if $\langle n', n\rangle \cong \langle m', m\rangle$. In order to obtain a real model from the finite representation, the part between $n$ and $m$ is copied and appended infinitely often. We use a similar technique to obtain a finite representation for a forest model. Since we want to preserve non-entailment, working with just pairs of elements is not sufficient. Instead, we take the length $n$ of the query into account and use isomorphic trees of depth $n$ to define blocking. This technique has first been employed for deciding query entailment in $\mathcal{ALCN}$ with role conjunctions (Levy & Rousset, 1998) and has recently been extended to the logics $\mathcal{ALCHIQ}$, $\mathcal{ALCHOQ}$, and $\mathcal{ALCHOI}$ (Ortiz, 2008;

Ortiz et al., 2008a) and extends, as our result, to the DLs $\mathcal{SHIQ}$, $\mathcal{SHOQ}$, and $\mathcal{SHOI}$ (i.e., with transitivity) as long as the query contains only simple roles.

As for forest quasi-interpretations, we use isomorphisms between forest interpretations or parts of them.

**Definition 38** (Isomorphism between Forest Interpretations)**.** Let $\mathcal{K}$ be an $\mathcal{ALCOIFb}$ knowledge base and $\mathcal{I} = (\Delta^{\mathcal{I}}, \cdot^{\mathcal{I}}), \mathcal{I}' = (\Delta^{\mathcal{I}'}, \cdot^{\mathcal{I}'})$ two forest interpretations of $\mathcal{K}$. Without loss of generality, we assume from now on that each root $\delta = (\rho, \varepsilon) \in \Delta^{\mathcal{I}}$ is in the extension of a unique concept $N_\delta$ that does not occur in $\mathsf{con}(\mathcal{K})$. Then $\mathcal{I}$ and $\mathcal{I}'$ are called *isomorphic* w.r.t. $\mathcal{K}$, written: $\mathcal{I} \cong_{\mathcal{K}} \mathcal{I}'$, iff there is a bijection $\varphi : \Delta^{\mathcal{I}} \to \Delta^{\mathcal{I}'}$ such that:

- $\delta_1$ is a successor of $\delta_2$ iff $\varphi(\delta_1)$ is a successor of $\varphi(\delta_2)$ for all $\delta_1, \delta_2 \in \Delta^{\mathcal{I}}$,
- $\langle \delta_1, \delta_2 \rangle \in r^{\mathcal{I}}$ iff $\langle \varphi(\delta_1), \varphi(\delta_2) \rangle \in r^{\mathcal{I}'}$ for all $\delta_1, \delta_2 \in \Delta^{\mathcal{I}}$ and $r \in \mathsf{rol}(\mathcal{K})$,
- $\delta \in A^{\mathcal{I}}$ iff $\varphi(\delta) \in A^{\mathcal{I}'}$ for all $\delta \in \Delta^{\mathcal{I}}$ and $A \in \mathsf{con}(\mathcal{K}) \cup \{N_\delta \mid \delta = (\rho, \varepsilon) \in \Delta^{\mathcal{I}}\}$.
- $\delta = o^{\mathcal{I}}$ iff $\varphi(\delta) = o^{\mathcal{I}'}$ for all $\delta \in \Delta^{\mathcal{I}}$ and $o \in \mathsf{nom}(\mathcal{K})$.

△

Usually, we omit the subscript $\mathcal{K}$ from $\cong_{\mathcal{K}}$ and assume that it is clear from the context.

**Definition 39** ($n$-Blocking)**.** Let $n \in \mathbb{N}$ be a fixed natural number and $\mathcal{I} = (\Delta^{\mathcal{I}}, \cdot^{\mathcal{I}})$ with $(\delta, w) \in \Delta^{\mathcal{I}}, w \neq \varepsilon$ a forest interpretation for some $\mathcal{ALCOIFb}$ knowledge base $\mathcal{K}$. An *$n$-blocking-tree* w.r.t. $(\delta, w)$, denoted $\mathsf{block}^n_{\mathcal{I}}(\delta, w)$, is the interpretation obtained from $\mathcal{I}$ by restricting $\mathcal{I}$ to elements in $\{(\delta, ww') \mid |w'| \leq n\} \cup \{(\rho, \varepsilon) \mid (\rho, \varepsilon) \in \Delta^{\mathcal{I}}\}$. An $n$-blocking-tree $\mathsf{block}^n_{\mathcal{I}}(\delta, w)$ $n$-blocks an $n$-blocking-tree $\mathsf{block}^n_{\mathcal{I}}(\delta, ww')$ if

1. $\mathsf{block}^n_{\mathcal{I}}(\delta, w)$ and $\mathsf{block}^n_{\mathcal{I}}(\delta, ww')$ have disjoint domains except for root elements,
2. there is a bijection $\varphi$ from elements in $\mathsf{block}^n_{\mathcal{I}}(\delta, w)$ to elements in $\mathsf{block}^n_{\mathcal{I}}(\delta, ww')$ that witnesses $\mathsf{block}^n_{\mathcal{I}}(\delta, w) \cong \mathsf{block}^n_{\mathcal{I}}(\delta, ww')$, and
3. for each descendant $(\delta, wv)$ of $(\delta, w)$, there is no inverse functional role $f$ and root $(\rho, \varepsilon) \in \Delta^{\mathcal{I}}$ such that $\langle (\delta, wv), (\rho, \varepsilon) \rangle \in f^{\mathcal{I}}$.

A node $(\delta, v) \in \Delta^{\mathcal{I}}$ is *$n$-blocked*, if $(\delta, v)$ is either *directly* or *indirectly $n$-blocked*; $(\delta, v)$ is indirectly $n$-blocked, if one of its ancestors is $n$-blocked; $(\delta, v)$ is directly $n$-blocked if none of its ancestors is $n$-blocked and $(\delta, v)$ is a leaf of some $n$-blocking-tree $\mathsf{block}^n_{\mathcal{I}}(\delta, ww')$ in $\mathcal{I}$ that is $n$-blocked; in this case we say that $(\delta, v)$ is (directly) $n$-blocked by $\varphi^{-}(\delta, ww')$ for $\varphi$ the bijection witnessing $\cong$.

Without loss of generality, we assume that the $n$-blocking-trees used above are minimal w.r.t. the order of elements in $\Delta^{\mathcal{I}}$ (cf. Definition 16).

A forest interpretation $\mathcal{I} = (\Delta^{\mathcal{I}}, \cdot^{\mathcal{I}})$ for $\mathcal{K}$ is an *$n$-representation* of $\mathcal{K}$ if

1. $\Delta^{\mathcal{I}}$ is finite,
2. $\Delta^{\mathcal{I}}$ contains no indirectly $n$-blocked nodes,
3. for each $o \in \mathsf{nom}(\mathcal{K})$, there is one element of the form $(\rho, \varepsilon) \in \Delta^{\mathcal{I}}$ such that $o^{\mathcal{I}} = (\rho, \varepsilon)^{\mathcal{I}}$,

4. each element that is not directly $n$-blocked is locally $\mathcal{K}$-consistent.

$\triangle$

Note that $n = 1$ is more restrictive than standard pairwise blocking since two trees of depth one need to be isomorphic before blocking occurs, whereas standard blocking already occurs for two isomorphic pairs of nodes. For DLs as expressive as $\mathcal{ALCOIF}b$, however, $n$ has to be greater than 0 (at least trees of depth 1) if we want to transform $n$-representations into models of the knowledge base. We now show that each knowledge base has an $n$-representation for some fixed $n \in \mathbb{N}$ and, afterwards, that we can use an $n$-representation to build a model for the knowledge base.

**Lemma 40.** *Let $\mathcal{K}$ be a consistent $\mathcal{ALCOIF}b$ knowledge base and $u = q_1 \vee \ldots \vee q_h$ a union of conjunctive queries and $n$ a fixed natural number greater than $max_{1 \leq i \leq h}|q_i|$. If $\mathcal{K} \not\models u$, then there is an $n$-representation of $\mathcal{K}$ that does not satisfy $u$.*

*Proof.* By assumption, $\mathcal{K}$ is consistent and $\mathcal{K} \not\models u$. Then, by Theorem 37, there is a forest model $\mathcal{I}$ of $\mathcal{K}$ with finitely many roots and branching degree bounded in $|\mathsf{cl}(\mathcal{K})|$, and for all $q \in \{q_1, \ldots, q_h\}$ holds $\mathcal{I} \not\models q$. We show that we can find an $n$-representation $\mathcal{R}$ for $\mathcal{I}$.

We use a similar argumentation as in Lemma 25 to show that there are only finitely many non-isomorphic $n$-blocking trees. We again denote this bound by $T_n$. Let $c = |\mathsf{cl}(\mathcal{K})|, r = |\mathsf{rol}(\mathcal{K})|$, and $m$ the (finite) number of roots in $\mathcal{I}$. Each root $\rho \in \Delta^{\mathcal{I}}$ is annotated with a special concept $N_\rho$ by assumption. For $n = 0$, we again have $2^c$ choices. For $n > 0$, each element can have between 0 and $c$ successors and between 0 and $m$ relations with roots. For roots we have $2^{c+m}$ choices for the concepts. We use $2^{cm}$ as bound for the choice of concepts for roots and this clearly bounds the choice for non-roots as well. Each non-root node in a level smaller than $n$ is the root of a tree with depth $n-1$ and each node in the sub-tree can again have up to $m$ relations to a root. Assuming that we have only a single role name $r \in \mathsf{rol}(\mathcal{K})$, we get a bound of $O(2^c cm T_{n-1}^{cm})$ for the number of non-isomorphic sub-trees of depth $n$ with relations to the at most $m$ roots. Since we have not only one but a choice of $r$ roles, we get a bound of $O(2^c (cm T_{n-1}^{cm})^r)$. We now abbreviate $2^c (cm)^r$ with $x$ and $cmr$ with $a$ and rewrite the obtained bound as $T_n = O(x(T_{n-1})^a)$. Unfolding yields $T_n = O((x^{1+a+\ldots+a^{n-1}})(T_0)^{a^n})$ which is bounded by $O((x^{a^n})(2^c)^{a^n}) = O((x2^c)^{a^n})$. By expanding the abbreviated symbols, we obtain a bound for $T_n$ of $O((2^c(cm)^r)^{(cmr)^n})$.

Together with the fact that $\mathcal{I}$ is obtained from a collapsing and relations from elements within a tree to a root in collapsings are never for inverse functional roles, this shows that there is an $n$-representation of $\mathcal{I}$ because for each tree rooted in a node $(\delta, \varepsilon) \in \Delta^{\mathcal{I}}$ with depth greater than $T_n$, there are two nodes $(\delta, w)$ and $(\delta, ww')$ such that $\mathsf{block}_{\mathcal{I}}^n(\delta, w)$ $n$-blocks $\mathsf{block}_{\mathcal{I}}^n(\delta, ww')$, and we can simply discard indirectly $n$-blocked nodes from $\mathcal{I}$ to obtain the desired $n$-representation.

Since $\mathcal{I} \not\models q$ and the $n$-representation is a restriction of $\mathcal{I}$, non-entailment of $q$ is clearly preserved. $\square$

Please note that we would not obtain such a bound if we had not fixed a bound on the number of new nominals (roots) beforehand and that we cannot use the standard tableau algorithms to obtain this result. The reason for this is that the number of new nominals (roots) in the tableau algorithms depends on the length of the longest path before blocking occurs. For our $n$-blocking-trees, however, we also have to consider relations back to the roots, which means that blocking occurs the later the more roots we have. On the other hand, delaying blocking may lead to the introduction of more and more new roots. Due to this cyclic argument, termination cannot be guaranteed for the tableau algorithms unless we have fixed a bound on the number of new nominals beforehand. This is also the reason why the tableau algorithm for entailment of conjunctive queries with only simple roles in the query of Calvanese et al. (2009) is sound, complete, and terminating on $\mathcal{SHIQ}$, $\mathcal{SHOQ}$, and $\mathcal{SHOI}$ knowledge bases, but is not guaranteed to terminate on $\mathcal{SHOIQ}$ knowledge bases (transitivity, i.e., having a DL with $\mathcal{S}$ instead of $\mathcal{ALC}$ does not have any impact on this).

We now show, how we can obtain a model for a knowledge base $\mathcal{K}$ from some $n$-representation of $\mathcal{K}$. We use a technique that is directly inspired from tableau algorithms and resembles the process of building a tableau from a complete and clash-free completion graph. In particular the tableau algorithm by Ortiz et al. (Ortiz, 2008; Ortiz et al., 2008a) is very similar as it also uses tree blocking.

**Definition 41** (Models for $n$-Representations)**.** Let $\mathcal{R} = (\Delta^{\mathcal{R}}, \cdot^{\mathcal{R}})$ be an $n$-representation of some $\mathcal{ALCOIF}b$ knowledge base $\mathcal{K}$. Let $\mathbf{s} = \frac{\delta'_1}{\delta_1}, \ldots, \frac{\delta'_m}{\delta_m}$ be a sequence of pairs of elements from $\Delta^{\mathcal{R}}$. With $|\mathbf{s}|$ we denote the *length* $m$ of $\mathbf{s}$. For such a sequence $\mathbf{s}$, we set $\mathsf{last}^*(\mathbf{s}) = \delta'_m$ and $\mathsf{last}_*(\mathbf{s}) = \delta_m$. By $\mathbf{s} \mid \frac{\delta'_{m+1}}{\delta_{m+1}}$ we denote the sequence $\frac{\delta'_1}{\delta_1}, \ldots, \frac{\delta'_m}{\delta_m}, \frac{\delta'_{m+1}}{\delta_{m+1}}$.

The set of $\mathcal{R}$-induced elements, denoted $\mathsf{elem}(\mathcal{R})$, is inductively defined as follows:

- If $\delta = (\rho, \varepsilon) \in \Delta^{\mathcal{R}}$, then $\frac{\delta}{\delta} \in \mathsf{elem}(\mathcal{R})$.
- If $\mathbf{s} \in \mathsf{elem}(\mathcal{R}), \delta = (\rho, w) \in \Delta^{\mathcal{R}}, \delta$ is not $n$-blocked, and $\delta$ is a successor of $\mathsf{last}^*(\mathbf{s})$, then $\mathbf{s} \mid \frac{\delta}{\delta} \in \mathsf{elem}(\mathcal{R})$.
- If $\mathbf{s} \in \mathsf{elem}(\mathcal{R}), \delta = (\rho, w) \in \Delta^{\mathcal{R}}, \delta$ is directly $n$-blocked by some $\delta' \in \Delta^{\mathcal{R}}$, and $\delta$ is a successor of $\mathsf{last}^*(\mathbf{s})$, then $\mathbf{s} \mid \frac{\delta'}{\delta} \in \mathsf{elem}(\mathcal{R})$.

We define the interpretation $\mathcal{I} = (\Delta^{\mathcal{I}}, \cdot^{\mathcal{I}})$ induced by $\mathcal{R}$ as follows:

- $\Delta^{\mathcal{I}} = \mathsf{elem}(\mathcal{R})$,
- for each $\mathbf{s} \in \Delta^{\mathcal{I}}$ and $A \in \mathsf{con}(\mathcal{K})$, $\mathbf{s} \in A^{\mathcal{I}}$ iff $\mathsf{last}^*(\mathbf{s}) \in A^{\mathcal{R}}$,
- for each $\mathbf{s} \in \Delta^{\mathcal{I}}$ and $o \in \mathsf{nom}(\mathcal{K})$, $\mathbf{s} = o^{\mathcal{I}}$ iff $\mathsf{last}^*(\mathbf{s}) = o^{\mathcal{R}}$,
- for each $\mathbf{s}, \mathbf{s}' \in \Delta^{\mathcal{I}}$ and $r \in \mathsf{rol}(\mathcal{K})$, $r^{\mathcal{I}} =$

$$\{\langle \mathbf{s}, \mathbf{s}' \rangle \mid \mathbf{s}' = \mathbf{s} \mid \tfrac{\delta'}{\delta} \text{ and } \langle \mathsf{last}^*(\mathbf{s}), \mathsf{last}_*(\mathbf{s}') \rangle \in r^{\mathcal{R}}\} \cup$$
$$\{\langle \mathbf{s}, \mathbf{s}' \rangle \mid \mathbf{s} = \mathbf{s}' \mid \tfrac{\delta'}{\delta} \text{ and } \langle \mathsf{last}_*(\mathbf{s}), \mathsf{last}^*(\mathbf{s}') \rangle \in r^{\mathcal{R}}\} \cup$$
$$\{\langle \mathbf{s}, \mathbf{s}' \rangle \mid \mathbf{s}' = \tfrac{\delta}{\delta} \text{ and } \langle \mathsf{last}^*(\mathbf{s}), \delta \rangle \in r^{\mathcal{R}}\} \cup$$
$$\{\langle \mathbf{s}, \mathbf{s}' \rangle \mid \mathbf{s} = \tfrac{\delta}{\delta} \text{ and } \langle \delta, \mathsf{last}^*(\mathbf{s}') \rangle \in r^{\mathcal{R}}\}.$$

△

The interpretation of nominals is well-defined since $n$-representations are forest interpretations for $\mathcal{K}$ (hence, there is a unique root for each nominal) and pairs $\frac{\delta}{\delta}$ with $\delta = (\rho, \varepsilon)$ are never appended to sequences in $\mathsf{elem}(\mathcal{R})$.

**Lemma 42.** *Let $\mathcal{K}$ be a consistent $\mathcal{ALCOIF}b$ knowledge base, $u = q_1 \vee \ldots \vee q_h$ a union of conjunctive queries, and $n \geq 1$ a fixed natural number greater than $max_{1 \leq i \leq h}|q_i|$. If $\mathcal{R}$ is an $n$-representation of $\mathcal{K}$ such that $\mathcal{R} \not\models u$, then there is a model $\mathcal{I}$ of $\mathcal{K}$ such that $\mathcal{I} \not\models u$.*

The proof is essentially as the one by Ortiz et al. (2008a), but adapted to our case, where we work completely on interpretations. Our $n$-representations correspond to completion graphs and our models to tableaux in their case.

*Proof.* Let $\mathcal{I}$ be an interpretation induced by $\mathcal{R}$. Since $n$-representations do not contain relations from an element within a tree to a root for an inverse functional role by definition, functionality restrictions are not violated in $\mathcal{I}$. Further, since $\mathcal{K}$ is simplified and $\mathcal{R}$ is a forest interpretation for $\mathcal{K}$ such that all elements apart from (directly) $n$-blocked ones are locally $\mathcal{K}$-consistent, it is quite straightforward that each element in the induced interpretation is locally $\mathcal{K}$-consistent. Together with the restriction on nominals (property 3), this implies that $\mathcal{I}$ is a model for $\mathcal{K}$. This is essentially the same principle as the one used to prove that tableaux constructed from completion graphs are proper representations of models of the input knowledge base.

Assume, to the contrary of what is to be shown, that $\mathcal{I} \models u$. Then there is a disjunct $q \in \{q_1, \ldots, q_h\}$ and a match $\mu$ for $q$ such that $\mathcal{I} \models^\mu q$. We use $\mu$ to construct a match $\pi$ for $q$ in $\mathcal{R}$ by "shifting" the mapping for variables into parts that have no direct counterpart in $\mathcal{R}$ upwards.

We define the match graph $G$ for $q$ in $\mathcal{I}$ as an undirected graph containing a node $\mathsf{s}$ for each $\mathsf{s} \in \Delta^{\mathcal{I}}$ such that $\mu(x) = \mathsf{s}$ for some $x \in \mathsf{var}(q)$ and containing an edge $\langle \mathsf{s}, \mathsf{s}' \rangle$ for each $\mathsf{s}, \mathsf{s}' \in \Delta^{\mathcal{I}}$ such that there is an atom $r(x, y) \in q, \mu(x) = \mathsf{s}$, and $\mu(y) = \mathsf{s}'$. We call nodes of $G$ that correspond to roots in $\Delta^{\mathcal{I}}$ root nodes of $G$ (i.e., nodes $\mathsf{s}$ such that $\mathsf{s} = \frac{\delta}{\delta}$) and we call all other nodes tree nodes. Note that the restriction of $G$ to tree nodes is a set of trees that we refer to as $G_1, \ldots, G_k$ and that each such tree has a depth smaller than $n$.

For each $x \in \mathsf{var}(q)$ such that $\mu(x) = \frac{\delta}{\delta}$ ($\frac{\delta}{\delta}$ is a root node in $G$), we set $\pi(x) = \mathsf{last}^*(\frac{\delta}{\delta})$. Note that $\delta$ is a root node in $\mathcal{R}$.

For each $G_i \in \{G_1, \ldots, G_k\}$, we distinguish two situations:

1. $G_i$ contains a node $\mathsf{s}$ such that $\mathsf{last}^*(\mathsf{s}) \neq \mathsf{last}_*(\mathsf{s})$ (i.e., $G_i$ contains a path from within an $n$-blocking tree to a copy of the path starting from the node that blocks). Due to the use of $n$-blocking, a single tree $G_i$ can never cover more than one $n$-blocking tree and it can use at most nodes from two $n$-blocking trees (leaving one and then entering the next one in less than $n$ steps). For each node $\mathsf{s}'$ in $G_i$ such that $|\mathsf{s}'| < |\mathsf{s}|$ and $x \in \mathsf{var}(q)$ such that $\mu(x) = \mathsf{s}'$, we set $\pi(x) = \varphi(\mathsf{last}^*(\mathsf{s}'))$. For each $\mathsf{s}'$ in $G_i$ with $|\mathsf{s}'| \geq |\mathsf{s}|$ and $x \in \mathsf{var}(q)$ such that $\mu(x) = \mathsf{s}'$, we set $\pi(x) = \mathsf{last}^*(\mathsf{s}')$.

2. $G_i$ contains no node $\mathsf{s}$ such that $\mathsf{last}^*(\mathsf{s}) \neq \mathsf{last}_*(\mathsf{s})$ (i.e., $G_i$ contains a path that lies completely within an $n$-blocking tree or from a path outside of an $n$-blocking-tree into an $n$-blocking-tree). For each node $\mathsf{s}$ in $G_i$ and $x \in \mathsf{var}(q)$ such that $\mu(x) = \mathsf{s}$, we set $\pi(x) = \mathsf{last}^*(\mathsf{s})$.

By definition of $\pi$, $\mathcal{I}$ as an induced model of $\mathcal{R}$, and $n$-blocking, we immediately have that, for each $A(x) \in q$, $\pi(x) \in A^{\mathcal{R}}$. We show that, for each $r(x, y) \in q$, $\langle \pi(x), \pi(y) \rangle \in r^{\mathcal{R}}$, which proves $\mathcal{R} \models q$. We distinguish three cases:

1. $\mu(x) = \frac{\delta}{\delta}$ for some $\delta \in \Delta^{\mathcal{R}}$. Then $\pi(x) = \delta = (\rho, \varepsilon) \in \Delta^{\mathcal{R}}$. We distinguish three cases for $\mu(y)$:

   (a) $\mu(y) = \frac{\delta'}{\delta'}$ is also a root, then $\pi(y) = \delta' = (\rho', \varepsilon) \in \Delta^{\mathcal{R}}$ and, since $\mu$ is a match for $q$ in $\mathcal{I}$ and by definition of $\mathcal{I}$ as an induced interpretation of $\mathcal{R}$, we have that $\langle \pi(x), \pi(y) \rangle = \langle \delta, \delta' \rangle \in r^{\mathcal{R}}$.

   (b) $\mu(y)$ is a successor of $\mu(x)$ in $\mathcal{I}$, i.e., $\mu(y) = \mathbf{s} = \frac{\delta}{\delta}|\frac{\delta'}{\delta'}$. Then $\mathbf{s}$ is not $n$-blocked and $\pi(y) = \delta' = (\rho', c) \in \Delta^{\mathcal{R}}$ for $c \in \mathbb{N}$. Again, since $\mu$ is a match for $q$ in $\mathcal{I}$ and by definition of $\mathcal{I}$ as an induced interpretation of $\mathcal{R}$, we have that $\langle \pi(x), \pi(y) \rangle = \langle \delta, \delta' \rangle \in r^{\mathcal{R}}$.

   (c) $\mu(y)$ is neither a root ($\mu(y) \neq \frac{\delta'}{\delta'}$ for any $\delta' \in \Delta^{\mathcal{R}}$) nor a successor of $\mu(x)$ in $\mathcal{I}$ ($\mu(y) \neq \frac{\delta}{\delta}|\frac{\delta'}{\delta'}$ for any $\delta' \in \Delta^{\mathcal{R}}$). Then $\mu(y)$ belongs to some graph match component $G_i$ and $\pi(y) = \mathsf{last}^*(\mu(y))$ or $\pi(y) = \varphi^-(\mathsf{last}^*(\mu(y)))$. Since the isomorphism between $n$-blocking trees also takes the relations to root nodes into account and other parts have direct counterparts in $\mathcal{R}$, we have that $\langle \pi(x), \pi(y) \rangle \in r^{\mathcal{R}}$.

2. $\mu(x) = \mathbf{s} \neq \frac{\delta}{\delta}$ for any $\delta \in \Delta^{\mathcal{R}}$. The cases when $\mu(y) = \frac{\delta'}{\delta'}$ for some $\delta' \in \Delta^{\mathcal{R}}$ is as above. We assume, therefore, that $\mu(y) = \mathbf{s}'$ with $|\mathbf{s}'| > 1$. By definition of $\mathcal{I}$, this means that either $\mathbf{s} = \mathbf{s}'|\frac{\delta}{\delta'}$ or $\mathbf{s}' = \mathbf{s}|\frac{\delta}{\delta'}$ for some $\delta, \delta' \in \Delta^{\mathcal{R}}$. We assume $\mathbf{s}' = \mathbf{s}|\frac{\delta}{\delta'}$. The opposite case is analogous. By definition of the match graph $G$, there is a component $G_i$ of $G$ that contains both $\mathbf{s}$ and $\mathbf{s}'$. We distinguish two cases:

   (a) The component $G_i$ contains a node $\bar{\mathbf{s}}$ such that $\mathsf{last}^*(\bar{\mathbf{s}}) \neq \mathsf{last}_*(\bar{\mathbf{s}})$. The most interesting case is when $\mathsf{last}^*(\mu(y)) \neq \mathsf{last}_*(\mu(y))$, i.e., $\bar{\mathbf{s}} = \mathbf{s}'$. Then $\pi(x) = \varphi^-(\mathsf{last}^*(\mathbf{s}))$ and $\pi(y) = \mathsf{last}^*(\mathbf{s}')$. Since $\mathsf{last}^*(\mathbf{s}') \neq \mathsf{last}_*(\mathbf{s}')$, we have that $\mathsf{last}^*(\mathbf{s}')$ is the node that directly $n$-blocks $\mathsf{last}_*(\mathbf{s}')$ and, by definition of the bijection $\varphi$, which witnesses the isomorphism, we have that $\pi(x) = \varphi^-(\mathsf{last}^*(\mathbf{s}))$ is the predecessor of $\pi(y) = \mathsf{last}^*(\mathbf{s}')$ and, by definition of $\mathcal{I}$ from $\mathcal{R}$, that $\langle \pi(x), \pi(y) \rangle \in r^{\mathcal{R}}$.

   (b) The component $G_i$ contains no node $\bar{\mathbf{s}}$ such that $\mathsf{last}^*(\bar{\mathbf{s}}) \neq \mathsf{last}_*(\bar{\mathbf{s}})$. Then $\pi(x) = \mathsf{last}^*(\mu(x))$ and $\pi(y) = \mathsf{last}^*(\mu(y))$. By definition of $\mathcal{I}$ from $\mathcal{R}$, we immediately have that $\langle \pi(x), \pi(y) \rangle \in r^{\mathcal{R}}$.

In any case, we have that $\langle \pi(x), \pi(y) \rangle \in r^{\mathcal{R}}$, which implies $\mathcal{R} \models^{\pi} q$ contradicting the initial assumption. □

Now Lemma 40 guarantees that, in case $\mathcal{K} \not\models q$, there is always a finite $n$-representation $\mathcal{R}$ for $\mathcal{K}$ such that $\mathcal{R} \not\models q$ and Lemma 42 guarantees that $\mathcal{R}$ can be transformed into a model $\mathcal{I}$ of $\mathcal{K}$ such that $\mathcal{I} \not\models q$. This suffices to show that we can enumerate all (finite) $n$-representations for $\mathcal{K}$ and check whether they entail a disjunct of the union of conjunctive queries. Together with the semi-decidability result for FOL, we get the following theorem.

**Theorem 43.** *Let $\mathcal{K}$ be an $\mathcal{ALCOIF}b$ knowledge base and $u = q_1 \vee \ldots \vee q_h$ a union of conjunctive queries. The question whether $\mathcal{K} \models u$ is decidable.*

## 8. Conclusions

We have solved the long-standing open problem of deciding conjunctive query entailment in the presence of nominals, inverse roles, and qualified number restrictions. We have shown that the problem is decidable by providing a decision procedure and proving its correctness. Since the approach is purely a decision procedure, the computational complexity of the problem remains open.

Our result also shows decidability of entailment of unions of conjunctive queries in $\mathcal{SHOIQ}$ and $\mathcal{SROIQ}$ (underlying OWL DL and OWL 2) if we disallow non-simple roles as binary query predicates. We thereby have reached a first important milestone towards tackling the problem of conjunctive queries for OWL 1 DL and OWL 2 DL.

Entailment of unions of conjunctive queries is also closely related to the problem of adding rules to a DL knowledge base, e.g., in the form of Datalog rules. Augmenting a DL KB with an arbitrary Datalog program easily leads to undecidability (Levy & Rousset, 1998). In order to ensure decidability, the interaction between the Datalog rules and the DL knowledge base is usually restricted by imposing a safeness condition. The $\mathcal{DL}$+*log* framework (Rosati, 2006a) provides the least restrictive integration proposed so far and Rosati presents an algorithm that decides the consistency of a $\mathcal{DL}$+*log* knowledge base by reducing the problem to entailment of unions of conjunctive queries. Notably, Rosati's results (2006a, Thm. 11) imply that the consistency of an $\mathcal{ALCHOIQ}b$ knowledge base extended with (weakly-safe) Datalog rules is decidable if and only if entailment of unions of conjunctive queries in $\mathcal{ALCHOIQ}b$ is decidable, which we have established.

**Corollary 44.** *The consistency of $\mathcal{ALCHOIQ}b$+log-knowledge bases (both under FOL semantics and under non-monotonic semantics) is decidable.*

Another related reasoning problem is *query containment*. Given a schema (or TBox) $\mathcal{S}$ and two queries $q$ and $q'$, we have that $q$ is contained in $q'$ w.r.t. $\mathcal{S}$ iff every interpretation $\mathcal{I}$ that satisfies $\mathcal{S}$ and $q$ also satisfies $q'$. It is well known that query containment w.r.t. a TBox can be reduced to deciding entailment for unions of conjunctive queries w.r.t. a knowledge base (Calvanese et al., 1998a). Decidability of unions of conjunctive query entailment in $\mathcal{ALCHOIQ}b$ implies, therefore, also decidability of query containment w.r.t. an $\mathcal{ALCHOIQ}b$ TBox.

There are two obvious avenues for future work. We will embark on extending our results in order to allow non-simple roles as query predicates. This is a non-trivial task as our current approach heavily relies on a certain locality of query matches, which has to be relinquished when considering non-simple roles. On the other hand, we are eager to determine the associated computational complexities and provide techniques that can form the basis for implementable algorithms.

## Acknowledgments

During his stay in Oxford where our collaboration started, Sebastian Rudolph was supported by a scholarschip of the German Academic Exchange Service (DAAD). Continuative work on the subject was enabled by funding through the ExpresST project of the German Research Foundation (DFG).

Birte Glimm was supported by EPSRC in the project *HermiT: Reasoning with Large Ontologies*.

We thank the three anonymous reviewers for their numerous helpful comments.

We thank Ian Pratt-Hartmann for (unknowingly) smashing our graphomata, Maria Magdalena Ortiz de la Fuente for establishing the competitive atmosphere, Yevgeny Kazakov for breath-taking discussions on black holes, Boris Motik for his motivating considerations on the value of the academic life, and last not least God for providing us with extraordinary weather and – most notably – infinity.